\documentclass[aps, prd, showkeys,superscriptaddress, nofootinbib, floatfix]{revtex4-2}

\usepackage{amsmath}
\usepackage{graphicx}
\usepackage{dcolumn}
\usepackage{bm}
\usepackage{ulem}

\usepackage{amssymb}
\usepackage{dsfont}
\usepackage{url}
\usepackage{caption,subcaption}
\usepackage[colorlinks=true,allcolors=blue]{hyperref}

\def\pp{\mathbf{p}}

\def\lan{\left\langle}
\def\ran{\right\rangle}

\begin{document}

\author{Gabriel S. Rocha}
\email{gabriel.soares.rocha@vanderbilt.edu}
\affiliation{Department of Physics and Astronomy, Vanderbilt University, 1221 Stevenson Center Lane,
Nashville, TN 37240, USA}
\affiliation{Instituto de F\'{\i}sica, Universidade Federal Fluminense, Niter\'{o}i, Rio de Janeiro, 24210-346,
Brazil}
\author{Gabriel S. Denicol}
\email{gsdenicol@id.uff.br}
\affiliation{Instituto de F\'{\i}sica, Universidade Federal Fluminense, Niter\'{o}i, Rio de Janeiro, 24210-346, Brazil}

\title{Transport coefficients of transient hydrodynamics for the hadron-resonance gas and thermal-mass quasiparticle models}

\begin{abstract}
We calculate all transport coefficients of second order transient hydrodynamics in two effective kinetic theory models: a hadron-resonance gas and a quasiparticle model with thermal masses tuned to reproduce QCD thermodynamics. We compare the corresponding results with calculations for an ultrarelativistic single-component gas,
that are widely employed in hydrodynamic simulations of heavy ion collisions. We find that both of these effective models display a qualitatively different normalized bulk viscosity, when compared to the calculation for the single-component gas. Indeed, $\zeta/[\tau_{\Pi}(\varepsilon_{0} + P_{0})] \simeq 16.91(1/3-c_{s}^{2})^{2}$, for the hadron-resonance gas model, and  $\zeta/[\tau_{\Pi}(\varepsilon_{0} + P_{0})] \simeq 5 (1/3-c_{s}^{2})$ for the quasiparticle model. Differences are also observed for many second-order transport coefficients, specially those related to the bulk viscous pressure. The transport coefficients derived are shown to be consistent with fundamental linear stability and causality conditions.
\end{abstract}

\maketitle

\section{Introduction}

Ultrarelativistic heavy-ion collisions produce a hot and dense fluid of nuclear matter in which the fundamental degrees of freedom of Quantum Chromodynamics (QCD), quarks and gluons, become directly manifest. Since the typical expansion speed of this so-called Quark-Gluon Plasma is comparable to that of light, relativistic dissipative hydrodynamic models are needed to describe these collisions \cite{Gale:2013da,Heinz:2013th,DerradideSouza:2015kpt} (see Refs.~\cite{Florkowski:2017olj,Rocha:2023ilf} for reviews on theories of relativistic dissipative hydrodynamics). For decades, it is known that relativistic extensions of Navier-Stokes theory, widely used in non-relativistic models \cite{struchtrup2005macroscopic,batchelor1967introduction}, are intrinsically inconsistent. Its equations of motion allow for the propagation of superluminal waves and perturbations around global equilibrium can be shown to grow indefinitely instead of being damped. These are the infamous acausality \cite{pichon:65etude} and instability \cite{hiscock:85generic,Hiscock:1983zz} problems of \textit{relativistic} Navier-Stokes theory. 

In the last decade, transient hydrodynamic models have been widely employed in Heavy-Ion Physics \cite{Gale:2013da, Heinz:2013th,Paquet:2023rfd}. In this case, the constitutive relations containing only space-like gradients, characteristic of Navier-Stokes theory, are substituted by relaxation-type partial differential equations for the dissipative currents. Such novel fluid-dynamical formalism was originally derived by Israel and Stewart \cite{israel1979jm,Israel:1979wp} and implements the idea that the dissipative currents relax exponentially to the Navier-Stokes constitutive relations (see Refs.~\cite{Rocha:2023ilf,Wagner:2023jgq} for a discussion on Israel-Stewart-like theories). This modification renders the equations of motion linearly causal and stable, as long as the viscosities and relaxation times appearing in the theory satisfy certain constrains \cite{Denicol:2021,Pu:2009fj,Brito:2020nou,PereiradeBrito:2021jul,Sammet:2023bfo}. References \cite{Gavassino:2021owo,Gavassino:2023mad,Gavassino:2021kjm} provide recent developments in assessing causality and its relation to stability.  

Recently, causality of transient theories in the full non-linear regime has been assessed making use of the theory of characteristic manifolds \cite{Bemfica:2019cop,Bemfica:2020xym,Disconzi:2023rtt}. In this case, the constrains involve the viscosities and relaxation times but also the remaining transport coefficients and the hydrodynamic fields themselves (e.g.~the components of the shear-stress tensor). This implies that causality cannot be assessed without solving the equations of motion. In Refs.~\cite{Plumberg:2021bme,Krupczak:2023jpa}, state-of-art codes that solve transient hydrodynamic equations of motion for Heavy-Ion collisions have been analyzed and it was shown that non-linear causality violations indeed occur. These violations were seen to take place predominantly at early times. This evidences that there is room for improvement in the hydrodynamic models employed, including a better understanding of the qualitative and quantitative behavior of the plethora of transport coefficients appearing in these theories. 

One instance of the aforementioned simulation codes is MUSIC \cite{Schenke:2010nt,Ryu:2015vwa,Paquet:2015lta,Ryu:2017qzn}. In this case, the expressions for the various transport coefficients required in the simulations were determined using calculations for an ultrarelativistic single-component gas \cite{Denicol:2014vaa}. The goal of this paper is to update these calculations considering systems that incorporate basic features of QCD thermodynamics. For this purpose, we calculate all second-order transport coefficients using two distinct effective kinetic descriptions: the hadron-resonance gas (HRG) \cite{Hagedorn:1965st,Venugopalan:1992hy,Karsch:2003vd,Huovinen:2009yb,Bazavov:2009zn,Borsanyi:2010cj,Ratti:2021ubw} and a quasiparticle model (QPM) \cite{Romatschke:2011qp,Albright:2015fpa,Alqahtani:2015qja,Chakraborty:2010fr,Tinti:2016bav,Rocha:2022fqz} with a temperature-dependent mass tuned to recover QCD thermodynamics \cite{Borsanyi:2010cj,Bazavov:2009zn}.
In both the QPM and HRG descriptions, we employ the relaxation time approximation to simplify the collision term, as was also the case in Ref.~\cite{Denicol:2014vaa}. 

The hadron-resonance gas is a very traditional model for strongly-interacting matter at sufficiently low temperatures, in which the system can be effectively treated as a multi-species gas consisting of the various strongly-interacting particles and unstable resonant states cataloged by the Particle Data Group \cite{ParticleDataGroup:2022pth}. At sufficiently high temperatures, after a cross-over transition \cite{Aoki:2006we}, the effective degrees of freedom change, quarks and gluons become deconfined and the equation of state computed from lattice QCD simulations provides the thermodynamic description of the system \cite{Ratti:2021ubw}. The HRG and lattice QCD descriptions are known to agree with each other for temperatures below that of the crossover \cite{Borsanyi:2010cj,Bazavov:2009zn}. In the high temperature regime, quasiparticle models have been employed in the literature to estimate the effects of QCD thermodynamics to transport properties. In fact, in Refs.~\cite{Romatschke:2011qp,Albright:2015fpa,Tinti:2016bav} expressions for transport coefficients are derived and, in order to obtain analytically accessible results, the Anderson-Witting \cite{andersonRTA:74} relaxation time approximation is employed. The present work is a continuation of Ref.~\cite{Rocha:2022fqz}, where only first-order transport coefficients were calculated -- here we shall fill this void and obtain expressions for all transport coefficients of transient relativistic fluid dynamics in a QPM employing an improved RTA prescription \cite{Rocha:2021zcw} and a judicious choice of matching conditions, which simplifies the dynamical constrain of the background field (see Sec.~\ref{sec:th-mass-QPM}).

We begin the discussion in Sec.~\ref{sec:Hydro}, with a general discussion of hydrodynamics in generic matching conditions. Then, we present the HRG model employed in Sec.~\ref{sec:HRG-model}, with the corresponding derivation of the transient hydrodynamic equations of motion in Landau matching conditions. Afterwards, in Sec.~\ref{sec:th-mass-QPM}, we present the QPM and derive the corresponding transient hydrodynamic equations in what we call the trace-anomaly matching condition. At the end of the section, in order to have a consistent comparison with the transport coefficients of the HRG model, we perform a change of matching conditions to Landau prescription in a manner that is consistent with the  hydrodynamic power-counting. In Sec.~\ref{sec:compar-HRG-QPM}, we, then, compare the HRG and QPM transport coefficients also with the expressions employed in MUSIC. In Section \ref{sec:concl} we make our concluding remarks.   

{\bf Notation:} We shall use units such that $\hbar = c = k_{B} = 1$ and the mostly minus $(+,-,-,-)$ metric signature.

\section{Hydrodynamics}
\label{sec:Hydro}

In the absence of conserved charges, which is a reasonable approximation in high energy heavy-ion collisions, the local conservation of energy and momentum are the most fundamental hydrodynamic equations of motion, 
\begin{equation}
\label{eq:consv-laws}
\begin{aligned}
&
\partial_{\mu}T^{\mu \nu} = 0,
\end{aligned}    
\end{equation}
where $T^{\mu \nu}$ is the energy-momentum tensor. The  tensor can be conveniently cast in terms of different components with respect to a time-like normalized vector $u^{\mu}$, $u_{\mu}u^{\mu}=1$ so that,
\begin{subequations}
\label{eq:decompos-tmunu}
\begin{align}    
\label{eq:decompos-tmunu-1}
T^{\mu \nu} &= \varepsilon u^{\mu} u^{\nu} - P \Delta^{\mu \nu} + h^{\mu} u^{\nu} + h^{\nu} u^{\mu} + \pi^{\mu \nu},
\\
\label{eq:decompos-tmunu-2}
\varepsilon & \equiv u_{\mu}u_{\nu}T^{\mu\nu}, P \equiv -\frac{1}{3}\Delta_{\mu\nu}T^{\mu\nu},  \\
    \label{eq:decompos-tmunu-3}
h^{\mu} & \equiv \Delta^{\mu}_{\nu} u_{\lambda} T^{\nu\lambda}, \, \, \pi^{\mu\nu} \equiv \Delta^{\mu\nu}_{\alpha\beta} T^{\alpha\beta}.
\end{align}    
\end{subequations}
where $\varepsilon$ is the total energy density in the local rest frame, $P$ is the total isotropic pressure, $\nu^\mu$ is the particle diffusion 4-current, $h^\mu$ is the energy diffusion 4-current, and $\pi^{\mu\nu}$ is the shear-stress tensor. We further introduced the projection operators
\begin{equation}
\begin{aligned}
&
\Delta^{\mu \nu} \equiv g^{\mu \nu} - u^{\mu} u^{\nu},   \\
&
\Delta^{\mu \nu \alpha \beta} \equiv \frac{1}{2}\left( \Delta^{\mu \alpha } \Delta^{\nu \beta} + \Delta^{\nu \alpha } \Delta^{\mu \beta} \right) - \frac{1}{3}\Delta^{\mu \nu} \Delta^{\alpha \beta},
\end{aligned}
\end{equation}
which render a given 4-vector in the 3-space orthogonal to $u^\mu$ and render a given second rank tensor symmetric and traceless and orthogonal to $u^{\mu}$ in both indices, respectively.

Now, we proceed to define a local equilibrium state, which shall allow the separation of the energy-momentum tensor in ideal and dissipative parts,
\begin{equation}
\label{eq:definitions2}
    \varepsilon \equiv \varepsilon_0(\beta) + \delta \varepsilon, \,\,\,
    P \equiv P_0(\beta) + \Pi,
\end{equation}
where $\beta \equiv 1/T$ is the inverse temperature of this fictitious local equilibrium state. The quantities $\varepsilon_0$ and $P_0$ are then related by an equilibrium equation of state, whereas the quantities  $\delta \varepsilon$, and $\Pi$ are the corresponding dissipative corrections. 

The definition of the parameters $\beta$ and $u^{\mu}$ as temperature and four-velocity is made by the so-called matching conditions. The most traditional matching condition employed in Heavy-Ion Collisions was put forward by Landau \cite{landau:59fluid} and shall be used in the Hadron-Resonance gas model. Landau's prescription defines the fluid 4-velocity as a time-like and normalized eigenvector of $T^{\mu \nu}$, such that $T^{\mu}_{\ \nu} u^{\nu} \equiv \varepsilon u^{\mu}$, hence implying in the condition $h^\mu=0$. Moreover, the inverse temperature is defined assuming that the particle number and energy densities in the local rest frame are given by their respective thermodynamic values, i.e., $\delta \varepsilon \equiv 0$. In the next section, an alternative matching condition shall be considered for the thermal mass quasi-particle model in order to simplify the computations. Overall, using the decomposition \eqref{eq:decompos-tmunu} into the conservation laws \eqref{eq:consv-laws}, and projecting them into their components parallel and orthogonal to $u^\mu$, we obtain the following equations of motion, 
\begin{subequations}
 \label{eq:basic-hydro-EoM}
\begin{align}
\label{eq:hydro-EoM-eps}
 D\varepsilon_{0}+D\delta \varepsilon + (\varepsilon_{0}+\delta \varepsilon + P_{0} + \Pi) \theta - \pi^{\mu \nu} \sigma_{\mu \nu} + \partial_{\mu}h^{\mu} + u_{\mu} Dh^{\mu} &= 0, \\
\label{eq:hydro-EoM-umu}
(\varepsilon_{0} + \delta \varepsilon + P_{0} + \Pi)Du^{\mu} - \nabla^{\mu}(P_{0} + \Pi) + h^{\mu} \theta + h^{\alpha} \Delta^{\mu \nu} \partial_{\alpha}u_{\nu} +  \Delta^{\mu \nu} Dh_{\nu} + \Delta^{\mu \nu} \partial_{\alpha}\pi^{\alpha}_{ \ \nu} &= 0,
\end{align}
\end{subequations}
where $D = u^\mu \partial_\mu$ is the comoving time derivative, $\nabla^\mu = \Delta^{\mu\nu} \partial_\nu$ is the 4-gradient operator, $\theta = \partial_\mu u^\mu$ is the expansion rate, and  $\sigma^{\mu \nu} = \Delta^{\mu \nu \alpha \beta} \partial_{\alpha} u_{\beta} $ is the shear tensor. Naturally, Eqs.~\eqref{eq:basic-hydro-EoM} do not form a closed system of equations and relations between the dissipative currents and other quantities appearing in the energy-momentum tensor. In the present text, we shall derive second-order transient hydrodynamic equations of motion from kinetic theory.

\section{Hadron resonance gas model}
\label{sec:HRG-model}

At low temperatures, nuclear matter can be modelled as a gas of weakly-interacting hadrons and resonances \cite{Venugopalan:1992hy,Karsch:2003vd,Huovinen:2009yb,Bazavov:2009zn,Borsanyi:2010cj,Ratti:2021ubw}. In this regime, an effective kinetic theory\footnote{Another hadronic effective kinetic theory, that is not the HRG, has been employed in Ref.~\cite{Albright:2015fpa}.} approach can also be employed, where the single-particle distribution function $f(x,\mathbf{p}_{i}) \equiv f_{{\bf p},i} \ (i = 1, \cdots, N_{\mathrm{spec}})$ of each particle species is determined by the relativistic Boltzmann equation, 
\begin{equation}
\label{eq:BoltzmannHRG}
p^{\mu} \partial_{\mu} f_{{\bf p},i}  =  
\sum_{j,a,b=1}^{N_{\mathrm{spec}}}\int dQ_{j} \ dQ^{\prime}_{a} \ dP^{\prime}_{b} W_{pp' \leftrightarrow qq'}^{ij \leftrightarrow ab} \left(  \Tilde{f}_{\pp, a}  \Tilde{f}_{\pp', b} f_{\mathbf{q},i}f_{\mathbf{q}',j}  
-
f_{\pp, a}f_{\pp', b} 
\Tilde{f}_{\mathbf{q},i} \Tilde{f}_{\mathbf{q}',j}
\right) 
\equiv 
C_{i}\left[ f_{\bf p}\right].
\end{equation}
Above, $C_{i}$ denotes the collision term for the $i$-th particle species, which contains the Lorentz-invariant transition rate $W_{qq' \leftrightarrow pp'}^{ab \leftrightarrow ij}$, that enforces the conservation of four-momentum. We also defined the integral measure $dP_{j} = d^{3}p_{j}/[(2 \pi)^{3} E_{{\bf p},j}]$, where $E_{{\bf p},j} = \sqrt{ {\bf p}_{j}^{2} + m^{2}_{j}}$ and $\Tilde{f}_{\mathbf{p},i} = 1 - (a_{i}/g_{i}) f_{\mathbf{q}',i}$, where $g_{i}$ is the spin-degeneracy of the species $i$ and $a_{i} = +1$ $(-1)$ if the particle is a fermion (boson) and $a_{i} \to 0$ recovers the classical particle statistics.  

The energy-momentum tensor is expressed in terms of the single-particle distribution of each particle species as,
\begin{equation}
\begin{aligned}
&
T^{\mu \nu} = \sum_{i = 1}^{N_{\mathrm{spec}}} \int dP_{i} p^{\mu}_{i} p^{\nu}_{i} f_{{\bf p},i}. 
\end{aligned}    
\end{equation}
As with all hydrodynamic models, the inverse temperature $\beta$ and fluid 4-velocity $u^{\mu}$ are defined by matching conditions. In this section, we employ the usual Landau matching conditions, which defines $\beta$ so that the total energy density follows the equilibrium equation of state and defines $u^{\mu}$ as the time-like normalized eigenvector of the energy-momentum tensor, so that we have the constrains
\begin{equation}
\begin{aligned}
&
\delta \varepsilon \equiv 0, \quad h^{\mu} \equiv 0.
\end{aligned}    
\end{equation}
With the restrictions imposed by Landau matching conditions, the energy-momentum tensor for the hadron-resonance gas model reads
\begin{equation}
\begin{aligned}
&
T^{\mu \nu}_{L} = \varepsilon_{0,L} u^{\mu}_{L} u^{\nu}_{L} - (P_{0,L} + \Pi_{L}) \Delta^{\mu \nu}_{L} + \pi^{\mu \nu}_{L},
\end{aligned}    
\end{equation}
in which the hydrodynamic fields defining energy density, pressure, bulk viscous pressure and shear-stress tensor are defined, respectively, as
\begin{equation}
\begin{aligned}
& 
\varepsilon_{0,L} = \sum_{i = 1}^{N_{\mathrm{spec}}} \int dP_{i} E_{{\bf p},i}^{2}  f_{0{\bf p},i}, 
\quad P_{0,L} = - \frac{1}{3} \sum_{i = 1}^{N_{\mathrm{spec}}} \int dP_{i} \left(\Delta _{\alpha \beta} p^{\alpha}_{i} p^{\beta}_{i}\right) f_{0{\bf p},i},
\\
&
\Pi_{L} = - \frac{1}{3} \sum_{i = 1}^{N_{\mathrm{spec}}}\int dP_{i} \left(\Delta _{\alpha \beta} p^{\alpha}_{i} p^{\beta}_{i}\right) \delta f_{{\bf p},i} , \quad \pi^{\mu \nu}_{L} = \sum_{i = 1}^{N_{\mathrm{spec}}} \int dP p^{\langle \mu}_{i} p^{\nu \rangle}_{i} \delta f_{{\bf p},i},
\end{aligned}    
\end{equation}
which are moments of the local equilibrium distribution and the deviation function, respectively, 
\begin{equation}
\begin{aligned}
&    
f_{0 {\bf p},i} = \frac{g_{i}}{\exp{\left( \beta u_{\mu}p^{\mu}_{i}\right)}+a_{i}}, 
\quad
\delta f_{{\bf p},i} = f_{{\bf p},i} - f_{0 {\bf p},i}.
\end{aligned}    
\end{equation}
In the remainder of the present section, for the sake of convenience, we shall omit the $L$ subscripts in the hydrodynamic variables for the sake of compactness.

\subsection{Moment equations for the hadron-resonance gas}
\label{sec:moments-HRG}

Transient hydrodynamic equations of motion are commonly derived from kinetic theory employing the method of moments \cite{Denicol:2012cn,Denicol:2021}. In this formalism, exact equations of motion for irreducible moments of the non-equilibrium distribution function are derived \cite{deBrito:2024vhm}. In the present case, we now derive the equations of motion for
\begin{equation}
\label{eq:irreducible_moments-cap6}
\begin{aligned}
\rho^{\mu_{1} \cdots \mu_{\ell}}_{i,r} = \int dP  E_{\mathbf{p},i}^{r} p^{\langle \mu_{1}} \cdots p^{\mu_{\ell} \rangle}_{i} \delta f_{\mathbf{p},i},
\quad
i = 1, \cdots, N_{\mathrm{spec}},
\end{aligned}    
\end{equation}
where $p^{\langle \mu_{1}} \cdots p^{\mu_{\ell} \rangle}_{i} \equiv \Delta^{\mu_{1} \cdots \mu_{\ell}}_{\ \ \nu_{1} \cdots \nu_{\ell}} p^{\nu_{1}} \cdots p^{\nu_{\ell}}_{i}$, and $\Delta^{\mu_{1} \cdots \mu_{\ell}}_{\ \ \nu_{1} \cdots \nu_{\ell}}$ denotes the $2 \ell$-rank tensor projector that is traceless, fully-orthogonal to $u^{\mu}$, and double-symmetric. Here, we only derive the equations of motion for the scalar and rank-2 tensor moments, which are relevant for the derivation of a fluid-dynamical theory (since we consider systems with a vanishing chemical potential, rank-1 irreducible moments will be of a higher-order in the hydrodynamic power-counting and can be neglected). We then have the following equations of motion for the scalar irreducible moments,
\begin{equation}
\label{eq:transient-l=0}
\begin{aligned}
&
D\rho _{i,r} 
- 
r Du_{\mu} \rho_{i,r-1}^{\mu} 
+
\nabla_{\mu} \rho^{\mu}_{i,r-1}
+  \frac{\theta}{3} \left[ -m^{2}_{i} (r-1) \rho_{i,r-2} +(r+2) \rho_{i,r} \right] - (r-1) \rho_{i,r-2}^{\mu \nu} \sigma_{\mu \nu}\\
& 
-
\frac{J_{r+1,0}^{(i)}}{\sum_{j} J_{3,0}^{(j)}}[D\delta \varepsilon +  (\delta \varepsilon + \Pi) \theta - \pi^{\mu \nu} \sigma_{\mu \nu} +  \partial_{\mu}h^{\mu} + u_{\mu} Dh^{\mu}]
+
\alpha_{i,r}^{(0)}  \theta
=
\int dP E_{\mathbf{p}}^{r-1} C_{i}[f_{\mathbf{p}}]
\equiv
\mathcal{C}_{i,r-1}.
\end{aligned}
\end{equation}
And the following equations of motion for the rank-2 irreducible moments,
\begin{equation}
\label{eq:transient-l=2}
\begin{aligned}
&  D\rho ^{\langle \alpha \beta \rangle}_{i,r}
-
r Du_{\mu} \rho^{\mu \alpha \beta}_{i,r-1} 
+
\frac{2}{5}  Du^{ \langle \alpha} \left[ \left( r + 5\right) \rho_{i,r+1}^{\beta\rangle} - r m^{2}_{i} \rho_{r-1}^{\beta\rangle} \right]
+\Delta^{\alpha \beta}_{ \ \ \alpha' \beta'}
 \nabla_{\mu} \rho_{i,r-1}^{\alpha' \beta' \mu} 
\\
& 
-
(r-1) \sigma_{\mu \nu} \rho^{\mu \alpha \beta \nu}_{i,r-2}
+
\frac{2}{5} 
 \nabla^{\langle \alpha}\left( m^{2}_{i} \rho_{i,r-1}^{\beta \rangle} - \rho_{i,r+1}^{\beta \rangle} \right)
+
\frac{\theta}{3} \left[ (r+4) \rho_{i,r}^{\alpha \beta} - (r-1) m^{2}_{i} \rho_{i,r-2}^{\alpha \beta}\right]
+
2 
 \omega_{\mu}^{ \ \langle \alpha \vert} \rho^{\mu \vert \beta \rangle}_{i,r} 
\\
& 
+ \frac{2}{7} 
\sigma_{\mu}^{\ \langle \alpha}  \left[ (2r+5) \rho_{i,r}^{\beta \rangle \mu} - 2(r-1) m^{2}_{i}\rho_{i,r-2}^{\beta \rangle \mu}\right] 
+
\frac{2}{15} \sigma^{\alpha \beta}  \left[ - (r-1)m^{4}_{i} \rho_{i,r-2} + (2r+3) m^{2}_{i}\rho_{i,r} - (r+4) \rho_{i,r+2}\right]\\
&
- \alpha^{(2)}_{i,r} 
 \sigma^{\alpha \beta}
=
\int dP E_{\mathbf{p}}^{r-1} p^{\langle \alpha} p^{\beta \rangle} C_{i}[f_{\mathbf{p}}]
\equiv
\mathcal{C}_{i,r-1}^{\alpha \beta}. 
\end{aligned}    
\end{equation}
In the above equations, we have defined  $\omega_{\mu \nu} = (1/2) (\nabla_{\mu}u_{\nu} - \nabla_{\nu}u_{\mu})$ and
\begin{equation}
\begin{aligned}
&
\alpha_{i,r}^{(0)} = 
-
c_{s}^{2} 
J_{r+1,0}^{(i)}   
+ J_{r+1,1}^{(i)}, 
\\
&
\alpha^{(2)}_{i,r} = 2 \beta J_{r+3,2}^{(i)},
\\
&
c_{s}^{2}
=
\frac{\partial P_{0}}{\partial \varepsilon_{0}}
=
\frac{(\varepsilon_{0} + P_{0})}{\beta \sum_{j}J_{3,0}^{(j)}} ,
\\
&
J_{n,q}^{(i)} = \frac{1}{(2q+1)!!}\int dP_{i} \left(- \Delta_{\mu \nu} p^{\mu}_{i} p^{\nu}_{i}\right)^{q} E_{{\bf p},i}^{n-2q} f_{0 {\bf p},i} \Tilde{f}_{0 {\bf p},i},
\end{aligned}    
\end{equation}
where $c_{s}^{2}$ is the speed of sound of the hadron-resonance gas model. In the high-temperature limit, in which we consider $z_{j} \equiv m_{j}/T \to 0$, $\forall j = 1, \cdots, N_{\mathrm{spec}}$ and assuming classical-statistics for all particle species, we have  
\begin{equation}
\begin{aligned}
 &
\label{eq:bulk-coeffs-high-T-HRG-cs2} 
c_{s}^{2} \simeq
 \frac{1}{3} - \frac{1}{36} \frac{\sum_{j} g_{j} z_{j}^{2}}{\sum_{j} g_{j}},   
\end{aligned}    
\end{equation}
which is a generalization of the $c_{s}^{2} \simeq 1/3 - (1/36) (m/T)^{2}$ for single particle species derived in Ref.~\cite{Denicol:2014vaa}.
Equations \eqref{eq:transient-l=0}--\eqref{eq:transient-l=2} are completely analogous to the moment equations of Refs.~\cite{Denicol:2012cn,Denicol:2021}, with vanishing chemical potential, for each particle species.

\subsection{Relaxation time approximation}

The computation of collisional moments on the right-hand side of Eqs.~\eqref{eq:transient-l=0}--\eqref{eq:transient-l=2} is the most non-trivial step in deriving transient hydrodynamic theories. Thus, phenomenological models for the collision term are often employed to simplify the derivation of a fluid-dynamical theory. In the relaxation time approximation proposed by Anderson and Witting \cite{andersonRTA:74}, the collision term is substituted by a phenomenological \textit{ansatz} which imposes that the single-particle distribution function relaxes exponentially to local equilibrium within a time scale $\tau_{R}$,  
\begin{equation}
\label{eq:chap-ensk0-no-ch}
\begin{aligned}
C_{i}\left[ f_{\bf p}\right] \simeq - \frac{E_{{\bf p},i}}{\tau_{R}} f_{0 {\bf p},i} \Tilde{f}_{0 {\bf p},i} \phi_{{\bf p},i},
\end{aligned}
\end{equation}
where we consider that the relaxation time, $\tau_{R}$, is the same for all particle species. This approximation is widely used to obtain solutions of the Boltzmann equation \cite{Kamata:2020mka,Noronha:2015jia,Denicol:2014xca,Ochsenfeld:2023wxz} and exact expressions for transport coefficients \cite{Romatschke:2011qp,Albright:2015fpa,Chakraborty:2010fr}. In particular, this approximation was employed in Ref.~\cite{Denicol:2014vaa}, where the second order transport coefficients currently implement MUSIC were derived.

It is further noted that the above approximation can only be employed if the relaxation time, $\tau_{R}$, does not depend on particle momenta and if one employs Landau matching conditions \cite{Rocha:2021zcw}. This is the case in the present section. Otherwise, the Anderson-Witting RTA is inconsistent with the local conservation laws \eqref{eq:consv-laws}. Indeed, considering for a moment that $\tau_{R}$ depends on momentum, and integrating both sides of the Boltzmann equation with $p^{\nu}$, we have $\partial_{\mu}T^{\mu \nu} = -\sum_{i}\int dP  p^{\nu} (E_{{\bf p},i}/\tau_{R{\bf p},i}) \delta f_{{\bf p},i}$, whose right-hand side does not necessarily vanish. Nevertheless, for a momentum- and species-independent $\tau_{R}$, the right-hand side will vanish as long as one adopts Landau's matching condition. 

With Ansatz \eqref{eq:chap-ensk0-no-ch}, all irreducible moments of the collision term become
\begin{equation}
\label{eq:coll-moment-HRG}
\begin{aligned}
&
\mathcal{C}_{i,r-1}^{\mu_{1} \cdots \mu_{\ell}} = - \frac{1}{\tau_{R}} \rho_{i,r}^{\mu_{1} \cdots \mu_{\ell}}.
\end{aligned}    
\end{equation}

\subsection{Transient hydrodynamic equations of motion}

The reduction of dynamical degrees of freedom from the full moment equations to that of the hydrodynamic currents ($\varepsilon_{0}$, $\Pi$, $u^{\mu}$, $\pi^{\mu \nu}$, in the present case) requires a truncation procedure. In the present case, we make use of the order of magnitude procedure \cite{struchtrup2005macroscopic,Fotakis:2022usk}. In this case, approximate relations between different moments of $\delta f_{\pp,i}$ are established by considering that the order of magnitude of the moments can be well estimated by the corresponding asymptotic constitutive relations, given in terms of space-like derivatives of temperature and 4-velocity. At leading order, we approximate scalar and rank-2 moments by their respective asymptotic Navier-Stokes values,

\begin{equation}
\begin{aligned}
&
\rho_{i,r} = - \zeta_{i,r} \theta + \mathcal{O}(2), 
\\
&
\rho_{i,r}^{\mu \nu} = 2 \eta_{i,r} \sigma^{\mu \nu} +
\mathcal{O}(2), 
\end{aligned}    
\end{equation}
where $\mathcal{O}(2)$ denotes terms that are of second-order or higher in powers of gradients or in powers of the dissipative currents. Then, these relations are rearranged so that one finds relations between generic moments solely in terms of $\Pi$ and $\pi^{\mu \nu}$. Hence, we derive 
\begin{subequations}
\label{eq:OoM-trun-HRG}
\begin{align}
&
\label{eq:OoM-scal-trun}
\rho_{i,r} \equiv \frac{\zeta_{i,r}}{\zeta} \Pi \equiv \mathcal{A}_{r}^{(i)} \Pi + \mathcal{O}(2), 
\\
&
\label{eq:OoM-tens-trun-HRG}
\rho_{i,r}^{\mu \nu} = \frac{\eta_{i,r}}{\eta} \pi^{\mu \nu} \equiv \mathcal{C}_{r}^{(i)} \pi^{\mu \nu} +
\mathcal{O}(2). 
\end{align}    
\end{subequations}
Microscopic expressions for the coefficients $\zeta_{r,i}$ and $\eta_{r,i}$, and thus $\mathcal{A}_{r}^{(i)}$ and $\mathcal{C}_{r}^{(i)}$ can be derived from Eqs.~\eqref{eq:transient-l=0} and \eqref{eq:transient-l=2} together with Eq.~\eqref{eq:coll-moment-HRG} for the collisional moments. For instance, we have
\begin{equation}
\label{eq:asymp-coeffs-HRG}
\begin{aligned}
&
\zeta_{i,r} = - \tau_{R} \alpha_{i,r}^{(0)},\\
&
\eta_{i,r} = \tau_{R} \alpha_{i,r}^{(2)},
\end{aligned}    
\end{equation}
where we identify $\zeta = - (1/3) \sum_{i} m_{i}^{2} \zeta_{i,0}$ and $\eta = \sum_{i} \eta_{i,0}$. We also have that 
\begin{equation}
\label{eq:OoM-relation-coeffs}
\begin{aligned}
&
\mathcal{A}_{r}^{(i)} 
=
 \frac{\left( - c_{s}^{2} J_{r+1,0}^{(i)} + J_{r+1,1}^{(i)}\right)}{\frac{1}{3} \sum_{j}
m_{j}^{2}
\left(J_{1,0}^{(j)} c_{s}^{2}  
- 
J_{1,1}^{(j)}\right)},
\\
&
\mathcal{C}_{r}^{(i)} 
=
\frac{J_{r+3,2}^{(i)}}{\sum_{j} J_{3,2}^{(j)}}.
\end{aligned}    
\end{equation}

\subsubsection{Bulk viscous pressure}
\label{sec:bulk-pres-HRG}

Now, we proceed to compute the equation of motion for the bulk viscous pressure. To that end, we take $r=0$ in Eq.~\eqref{eq:transient-l=2} multiply by $-m_{i}^{2}$, sum it over the particle-species index $i$ and use the order-of-magnitude relations \eqref{eq:OoM-trun-HRG} and \eqref{eq:OoM-relation-coeffs} to reexpress the non-hydrodynamic moment as dissipative currents. From this procedure, we derive 
\begin{equation}
\begin{aligned}
&
\tau_{\Pi} D\Pi + \Pi
=  
- \zeta \theta
- \delta_{\Pi \Pi} \Pi \theta  
+
\lambda_{\Pi \pi}
\pi^{\mu \nu} \sigma_{\mu \nu} + \mathcal{O}(3),
\end{aligned}
\end{equation}
where $\mathcal{O}(3)$ denote terms that are of third order or higher in the power-counting scheme explained in the previous section -- such terms are not considered in traditional theories of transient fluid dynamics.  The corresponding transport coefficients are determined microscopically as
\begin{subequations}
\label{eq:transp-coeffs-HRG-bulk}
\begin{align}
&
\label{eq:transp-coeffs-HRG-bulk-1}
\tau_{\Pi} = \tau_{R},
\\
&
\zeta =
\frac{\tau_{R}}{3} \sum_{i}
m_{i}^{2} 
\left(J_{1,0}^{(i)} c_{s}^{2}
- \beta
J_{1,1}^{(i)}
\right),
\\
&
\frac{\delta_{\Pi \Pi}}{\tau_{\Pi}} =  \frac{2}{3} - \frac{1}{9}\sum_{i} m_{i}^{4} \mathcal{A}_{-2}^{(i)}
- \frac{1}{3}\frac{\sum_{i} m_{i}^{2} J_{1,0}^{(i)}}{\sum_{j} J_{3,0}^{(j)}} ,
\\
&
\frac{\lambda_{\Pi \pi}}{\tau_{\Pi}} = 
\frac{1}{3} \sum_{i} m_{i}^{2}\mathcal{C}_{-2}^{(i)} + \frac{1}{3} \frac{\sum_{i} m_{i}^{2} J_{1,0}^{(i)}}{\sum_{j} J_{3,0}^{(j)}}.
\end{align}    
\end{subequations}

In connection with Ref.~\cite{Denicol:2014vaa}, we assess the behavior of the transport coefficients in the large temperature limit. Presently, we define such limit so that the ratio of the masses of all particles with respect to temperature are small, i.e.,  $z_{j} \equiv m_{j}/T \to 0$, $\forall j = 1, \cdots, N_{\mathrm{spec}}$, and classical statistics for all species ($a_{i} \to 0$) is assumed. In this limit, the transport coefficients behave asymptotically as 
\begin{subequations}
\label{eq:bulk-coeffs-high-T-HRG}
\begin{align}
&
 \label{eq:bulk-coeffs-high-T-HRG-zeta}
\frac{\zeta}{(\varepsilon_{0} + P_{0})\tau_{\Pi}} 
\simeq
\frac{1}{72 \sum_{j} g_{j}} \left[ \sum_{j} g_{j} z_{j}^{4} 
 - \frac{1}{6} \frac{\left(\sum_{j} g_{j} z_{j}^{2}\right)^{2}}{\sum_{j} g_{j}}\right]  ,
\\
&
\label{eq:bulk-coeffs-high-T-HRG-del-PIPI}
\frac{\delta_{\Pi \Pi}}{\tau_{\Pi}} \simeq
\frac{2}{3} - \frac{1}{6} \frac{\left( \sum_{j}g_{j} m_{j}^{4} \right)\left(\sum_{j} g_{j} z_{j}^{2}\right) }
{\left[ - \left(\sum_{j} g_{j} m_{j}^{2}\right)^{2} + 9 \left(\sum_{j} g_{j} m_{j}^{4}\right)\left(\sum_{j} g_{j}\right)\right]} 
-
\frac{1}{36} \frac{\sum_{j} g_{j} z_{j}^{2}}{\sum_{j} g_{j}},
\\
&
\frac{\lambda_{\Pi \pi}}{\tau_{\Pi}} \simeq 
\frac{1}{18} \frac{\sum_{j} g_{j} z_{j}^{2}}{\sum_{j} g_{j}} .
\end{align}    
\end{subequations}
In codes solving the hydrodynamic equations of motion, it is customary to employ expressions \eqref{eq:transp-coeffs-HRG-bulk} as a function of temperature as a tabulated input or to make use of expansions \eqref{eq:bulk-coeffs-high-T-HRG} above to derive simple expressions in terms of another thermodynamic quantity, such as the speed of sound, so that the dependence on the parameters of microscopic theory (the mass of particle species in the present case) is not manifest.  
Indeed from Eqs.~\eqref{eq:bulk-coeffs-high-T-HRG-zeta} we can provide relations between the bulk transport coefficients and the conformal violation of the speed of sound $\left(c_{s}^{2} - 1/3\right) \simeq \mathcal{O}(z^{2})$ (see Eq.~\eqref{eq:bulk-coeffs-high-T-HRG-cs2}). For the normalized bulk viscosity and the coefficient $\lambda_{\Pi \pi}$, we have, respectively,
\begin{equation}
\label{eq:pocket-form-HRG}
\begin{aligned}
&   
\frac{\zeta}{(\varepsilon_{0} + P_{0})\tau_{\Pi}} 
 \simeq
\left[18 \frac{\left(\sum_{j} g_{j}\right) \left(\sum_{j} g_{j} m_{j}^{4}\right)}{\left(\sum_{j} g_{j} m_{j}^{2} \right)^{2}} - 3 \right] \left(\frac{1}{3}-c_{s}^{2}\right)^{2},
\\
&
\frac{\lambda_{\Pi \pi}}{\tau_{\Pi}} 
\simeq 
2 \left(\frac{1}{3}-c_{s}^{2}\right).
\end{aligned}    
\end{equation}
whereas for $\delta_{\Pi \Pi}$ one would use the constant leading order result $\delta_{\Pi \Pi}/\tau_{\Pi} \simeq 2/3$. In the above expressions, we see that the result $\zeta/[(\varepsilon_{0} + P_{0})\tau_{\Pi}] \simeq 15 (1/3 - c_{s}^{2})^{2}$ calculated in Ref.~\cite{Denicol:2014vaa} is not valid for multiple species systems. Indeed, quadratic and quartic power averages of the mass ratios appear in a non-trivial way. For the particle content of UrQMD \cite{Bass:1998ca,Bleicher:1999xi}, $\zeta/[(\varepsilon_{0} + P_{0})\tau_{\Pi}] \simeq 16.91 (1/3 - c_{s}^{2})^{2}$ and $\zeta/[(\varepsilon_{0} + P_{0})\tau_{\Pi}] \simeq 19.36 (1/3 - c_{s}^{2})^{2}$ for SMASH particle content \cite{SMASH:2016zqf}\footnote{The PDG data used for the obtention of these results stems from the scikit HEP package Particle \cite{Rodrigues:2020syo}.}. Anyhow, the result $\zeta/[(\varepsilon_{0} + P_{0})\tau_{\Pi}] \simeq 15 (1/3 - c_{s}^{2})^{2}$ is recovered in the limiting case where there is only one particle species, i.e., $g_{j} = \delta_{j 0}$, for instance. We note that the expression $\lambda_{\Pi \pi}/\tau_{\Pi}$ is different from what was derived in Ref.~\cite{Denicol:2014vaa} and used in the MUSIC code \cite{Schenke:2010nt,Schenke:2010rr,Paquet:2015lta}. There, $\lambda_{\Pi \pi}/\tau_{\Pi} \simeq (8/5)\left(1/3-c_{s}^{2}\right)$.

\subsubsection{Shear-stress tensor}

In order to derive the equation of motion for the shear-stress tensor, we take Eq.~\eqref{eq:transient-l=2}, sum it over the particle-species index $i$ and employ the order-of-magnitude relations \eqref{eq:OoM-tens-trun-HRG} and \eqref{eq:OoM-relation-coeffs}. The final result is
\begin{equation}
\begin{aligned}
& 
\tau_{\pi} D \pi^{\langle \alpha \beta \rangle}
+
\pi^{\alpha \beta}
=
2 \eta  \sigma^{\alpha \beta}
-
\delta_{\pi \pi} \pi^{\alpha \beta} \theta 
-
2 
 \omega_{\mu}^{ \ \langle \alpha} \pi^{\beta \rangle \mu} 
- 
\tau_{\pi \pi} 
\sigma_{\mu}^{\ \langle \alpha} \pi^{\beta \rangle \mu} +
\lambda_{\pi \Pi} \Pi \sigma^{\alpha \beta}  + \mathcal{O}(3),
\end{aligned}    
\end{equation}
where the transport coefficients read
\begin{equation}
\begin{aligned}
&
\tau_{\pi} = \tau_{R},
\\
&
\eta 
=
\tau_{R}
\sum_{i} \beta J_{3,2}^{(i)},
\\
&
\frac{\delta_{\pi \pi}}{\tau_{\pi}} = \frac{4}{3} + \sum_{i} m^{2}_{i} 
\mathcal{C}_{-2}^{(i)},
\\
&
\frac{\tau_{\pi \pi}}{\tau_{\pi}}
=
\frac{10}{7}
+ \frac{4}{7} \sum_{i} m^{2}_{i} \mathcal{C}_{-2}^{(i)},
\\
&
\frac{\lambda_{\pi \Pi}}{\tau_{\pi}} = \frac{6}{5}  - 
\frac{2}{15}
\sum_{i} 
 m^{4}_{i} \mathcal{A}_{-2}^{(i)}.
\end{aligned}    
\end{equation}
We note, once more, that the shear relaxation time $\tau_{\pi}$ coincides with the RTA characteristic time $\tau_{R}$. Besides, it is seen that the second-order transport coefficients $\delta_{\pi \pi}$, $\tau_{\pi \pi}$ and $\lambda_{\pi \Pi}$ coincide with the values calculated in Ref.~\cite{Denicol:2014vaa} in the massless limit. Indeed, in the high temperature limit, $z_{j} = m_{j}/T \to 0$, $\forall j = 1, \cdots, N_{\mathrm{spec}}$, and assuming classical statistics for all species ($a_{i} \to 0$), we have the following asymptotic expressions for the transport coefficients in 
\begin{equation}
\label{eq:high-T-shear-HRG}
\begin{aligned}
&
\frac{\eta}{\tau_{\pi}(\varepsilon_{0} + P_{0})}
\simeq \frac{1}{5} 
-
\frac{1}{60} \frac{\sum_{j} g_{j} z_{j}^{2}}{\sum_{j} g_{j}},
\\
&
\frac{\delta_{\pi \pi}}{\tau_{\pi}} \simeq
\frac{4}{3} 
+
\frac{1}{12} \frac{\sum_{j} g_{j} z_{j}^{2}}{\sum_{j} g_{j}},
\\
&
\frac{\tau_{\pi \pi}}{\tau_{\pi}}
\simeq
\frac{10}{7}
+
\frac{1}{21} \frac{\sum_{j} g_{j} z_{j}^{2}}{\sum_{j} g_{j}},
\\
&
\frac{\lambda_{\pi \Pi}}{\tau_{\pi}} \simeq 
\frac{6}{5}  - 
\frac{1}{5}
\frac{\left( \sum_{j}g_{j} m_{j}^{4} \right)\left(\sum_{j} g_{j} z_{j}^{2}\right) }
{\left[ - \left(\sum_{j} g_{j} m_{j}^{2}\right)^{2} + 9 \left(\sum_{j} g_{j} m_{j}^{4}\right)\left(\sum_{j} g_{j}\right)\right]} ,
\end{aligned}    
\end{equation}
where it seen that all the transport coefficients coincide with the values derived in Ref.~\cite{Denicol:2014vaa}
and used in the MUSIC code at leading order. At sub-leading order, almost all transport coefficients are given in terms of the average of the square of the mass-to-temperature ratio of all particle species with a coefficient that does not depend on averages of the particle species (e.g.~$1/12$ for $\delta_{\pi \pi}/\tau_{\pi}$). The exception being $\lambda_{\pi \Pi}/\tau_{\pi}$, where this coefficient is given in terms of averages of quartic and quadratic powers of masses. Expressions analogous to Eqs.~\eqref{eq:pocket-form-HRG} can also be deduced from Eqs.~\eqref{eq:high-T-shear-HRG}. In this case, they reduce to the leading order constant results, which coincide with the values employed in the MUSIC simulation code \cite{Denicol:2014vaa}.

\section{Thermal-mass quasi-particle model}
\label{sec:th-mass-QPM}

Effective kinetic models have also been employed as a technique to obtain properties of quantum field theories in the weakly coupled regime \cite{Calzetta:1986cq,Jeon:1995zm,Arnold:2002zm,Berges:2005md, Gagnon:2007qt,Arnold:2003zc,Baym:1990uj,Albright:2015fpa,Tinti:2016bav}. In such formalisms, certain aspects emerging from the underlying microscopic interaction can be captured by an effective relativistic Boltzmann equation for quasiparticles with a temperature-dependent mass, 
\begin{equation}
\label{eq:BoltzmannQuasi}
p^{\mu} \partial_{\mu} f_{\bf p} + \frac{1}{2}\partial_{\mu} M^2(T) \partial^{\mu}_{(p)} f_{\bf p} 
=
\int dQ \ dQ^{\prime} \ dP^{\prime} W_{pp' \leftrightarrow qq'} (f_{\mathbf{q}}f_{\mathbf{q}'}  -  f_{\pp}f_{\pp'}) \equiv C_{Q}\left[ f_{\bf p}\right].
\end{equation}
In the present case, the thermal mass $M(T)$ is defined so that QCD thermodynamics is recovered. We stress at this point that the quasiparticles of the present model should not be confused with any of the fundamental QCD degrees of freedom. Instead, they are effective degrees of freedom defined in such a way as to reproduce thermodynamic properties of QCD. In Eq.~\eqref{eq:BoltzmannQuasi}, $f_{\bf p}$ is the single particle distribution function, $C_{Q}\left[ f_{\bf p}\right]$ is the corresponding quasi-particle collision term, in which $W_{pp' \leftrightarrow qq'}$ denotes the collision rate, which enforces the conservation of momentum in the collision processes. It is seen that the addition of a thermal mass leads to the emergence of a Boltzmann-Vlasov term, where we define
\begin{equation}
\partial^{\mu}_{(p)} \equiv \frac{\partial}{\partial p_{\mu}}.
\end{equation}
Besides, we employ the integral measure 
\begin{equation} 
\int dP = 4 \pi \int \frac{d^4p}{(2\pi)^4} \delta(p \cdot p - M^{2}(T)) .
\end{equation}
In the present model, which shall be hereon referred to as quasi-particle model (QPM), we consider that the fluid is locally neutral and that there are no charge four-currents. By integrating both sides of the Boltzmann equation with four-momentum components $p^{\mu}$, we can identify the energy-momentum tensor as  
\begin{equation}
\label{eq:Tmunudef}
T^{\mu\nu} \equiv \int dP p^\mu p^\nu f_{\bf p}  + g^{\mu\nu} B(T).
\end{equation}
Thus, it is readily seen that the presence of the thermal mass leads to a redefinition of the conserved current which employs the field $B$, called the background field, which obeys the constrain
\begin{equation}
\label{eq:dB}
\partial_{\mu} B = - \frac{1}{2} \partial_{\mu} M^2 \int dP f_\pp ,
\end{equation}
that is valid for arbitrary non-equilibrium configurations. In Ref.~\cite{Alqahtani:2015qja}, which assumed a highly symmetric flow configuration (Bjorken flow), this was seen as a boundary condition, which allowed the dynamical determination of $B$. In Ref.~\cite{Tinti:2016bav}, a different prescription for the dynamics of the $B$-field is assumed, also considering non-equilibrium effects, but relating $B$ to the dissipative hydrodynamic fields. In general, the dynamic constrain \eqref{eq:dB} is not trivially incorporated in power counting procedures such as the Chapman-Enskog expansion \cite{Rocha:2022fqz} and the procedure that shall be employed in Sec.~\ref{sec:transient-QPM} to derive transient hydrodynamic equations.

As discussed above, a fundamental ingredient of the hydrodynamic model in the imposition of matching conditions, which provide a definition of the reference local equilibrium state, i.e. they define the temperature and four-velocity. In this context, we impose that 
\begin{equation}
\label{eq:match1}
\int dP f_\pp \equiv \int dP f_{0 \pp} ,   
\end{equation}
with $f_{0\pp}$, the local equilibrium state being given by
\begin{equation}
\label{eq:EQL-qp}
f_{0\pp} \equiv g \exp\left( - \beta E_{\bf p} \right),
\end{equation}
where $E_{\bf p} = u_{\mu}p^{\mu}$ and $g$ is the degeneracy factor which is set to
\begin{equation}
g = \frac{\pi^4}{180}\left[ 4(N_c^2 - 1) + 7 N_c N_f \right] \,,
\end{equation}
with $N_c = N_f = 3$, so that the high-temperature limit of QCD is recovered \cite{Romatschke:2011qp}. Condition \eqref{eq:match1} effectively defines the temperature of the local equilibrium state and we still need to provide one additional matching condition to define the fluid four-velocity, $u^{\mu}$. To that end, we employ the constrain,
\begin{equation}
\begin{aligned}
\label{eq:matching_kinetic1}
    \int dP \, E_{\textbf{p}}\, p^{\langle \mu \rangle} f_{\textbf{p}}  = \int dP \, E_{\textbf{p}}\, p^{\langle \mu \rangle} f_{0\textbf{p}},
\end{aligned}
\end{equation}
that is \textit{one} of the conditions implied by Landau \cite{landau:59fluid}, which defined $u^{\mu}$ as the time-like normalized eigenvector of the energy momentum tensor. This condition implies that the fluid-comoving observer measures no heat flux, $h^{\mu} \equiv 0$. 

With conditions \eqref{eq:match1} and \eqref{eq:EQL-qp}, the energy-momentum tensor reads
\begin{equation}
\begin{aligned}
&
T^{\mu \nu}_{Q} = (\varepsilon_{0,Q} + \delta \varepsilon_{Q}) u^{\mu} u^{\nu} - (P_{0,Q} + \Pi_{Q}) \Delta^{\mu \nu} + \pi^{\mu \nu}_{Q}.
\end{aligned}    
\end{equation}
The equilibrium energy density, $\varepsilon_0$, and pressure, $P_0$, are given by   
\begin{subequations}
\begin{align}
\varepsilon_{0,Q} =& I_{2,0} + B = \frac{gT^4Z^2}{2\pi^2}[3K_2(Z) + Z K_1(Z) ] + B, \label{eq:eps} \\ 
P_{0,Q} =&I_{2,1} - B = \frac{gT^4Z^2}{2\pi^2}K_2(Z) - B, \label{eq:P} 
\\
I_{n,q} \equiv& \frac{1}{(2q+1)!!}\int dP \left(- \Delta_{\mu\nu} p^\mu p^\nu \right)^{q} E_{\bf p}^{n-2q} f_{0 {\bf p}},   
\end{align}    
\end{subequations}
where $Z = M(T)/T$ and $K_n(Z)$ denotes the $n$--th modified Bessel function of the second kind \cite{gradshteyn2014table}. In their turn, the viscous correction to the energy density, the bulk viscous pressure and the shear-stress tensor are obtained as 
\begin{equation}
\label{eq:diss-curr-QPM}
\begin{aligned}
\delta \varepsilon_{Q} & = \int dP E_{\bf p}^{2} \delta f_{\bf p},
\\
\Pi_{Q} & = - \frac{1}{3}\int dP \left( \Delta_{\mu\nu} p^\mu p^\nu \right) \delta f_{\bf p},
\\
\pi^{\mu \nu}_{Q} & = \int dP p^{\langle \mu} p^{\nu \rangle} f_{\bf p}.
\end{aligned}    
\end{equation}
We note that constrain \eqref{eq:match1} implies that  $\delta \varepsilon_{Q} = 3 \Pi_{Q}$. Thus, matching condition \eqref{eq:match1} implies that the dissipative corrections to the trace anomaly are automatically set to zero, i.e., $T^{\mu \nu}g_{\mu \nu} \equiv \varepsilon_{0,Q} - 3 P_{0,Q}$. As it occurred in Sec.~\ref{sec:HRG-model}, for the sake of simplicity, we shall omit the $Q$ subscript, until section \ref{sec:matching-redef-QPM}, where we shall discuss the mapping from the trace-anomaly matching to the Landau matching condition employed in the last section.

\subsection{Temperature-dependence of the quasiparticle mass}

\begin{figure}[!h]
    \centering
\begin{subfigure}{0.5\textwidth}
    \includegraphics[scale=0.36]{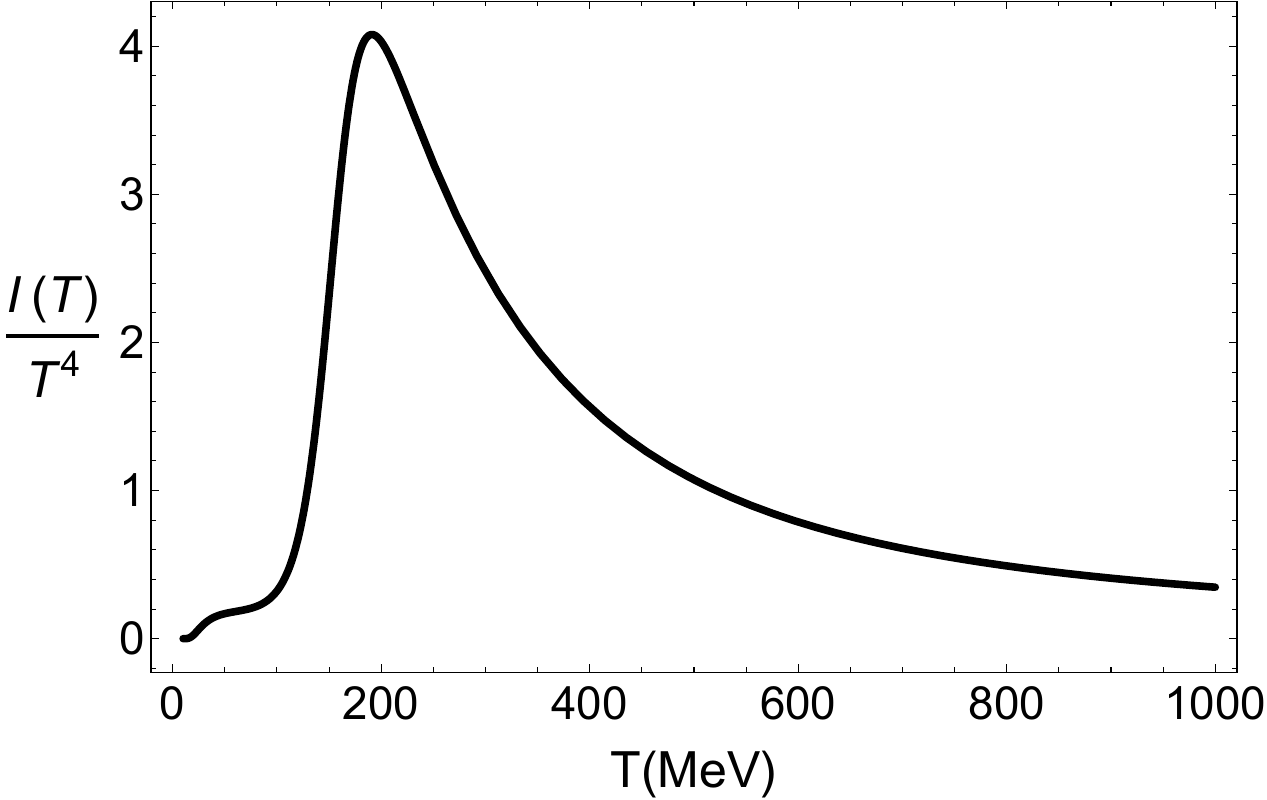}
    \caption{Normalized trace anomaly}
    \label{fig:tr-anom}    
\end{subfigure}\hfil
\begin{subfigure}{0.5\textwidth}
    \includegraphics[scale=0.36]{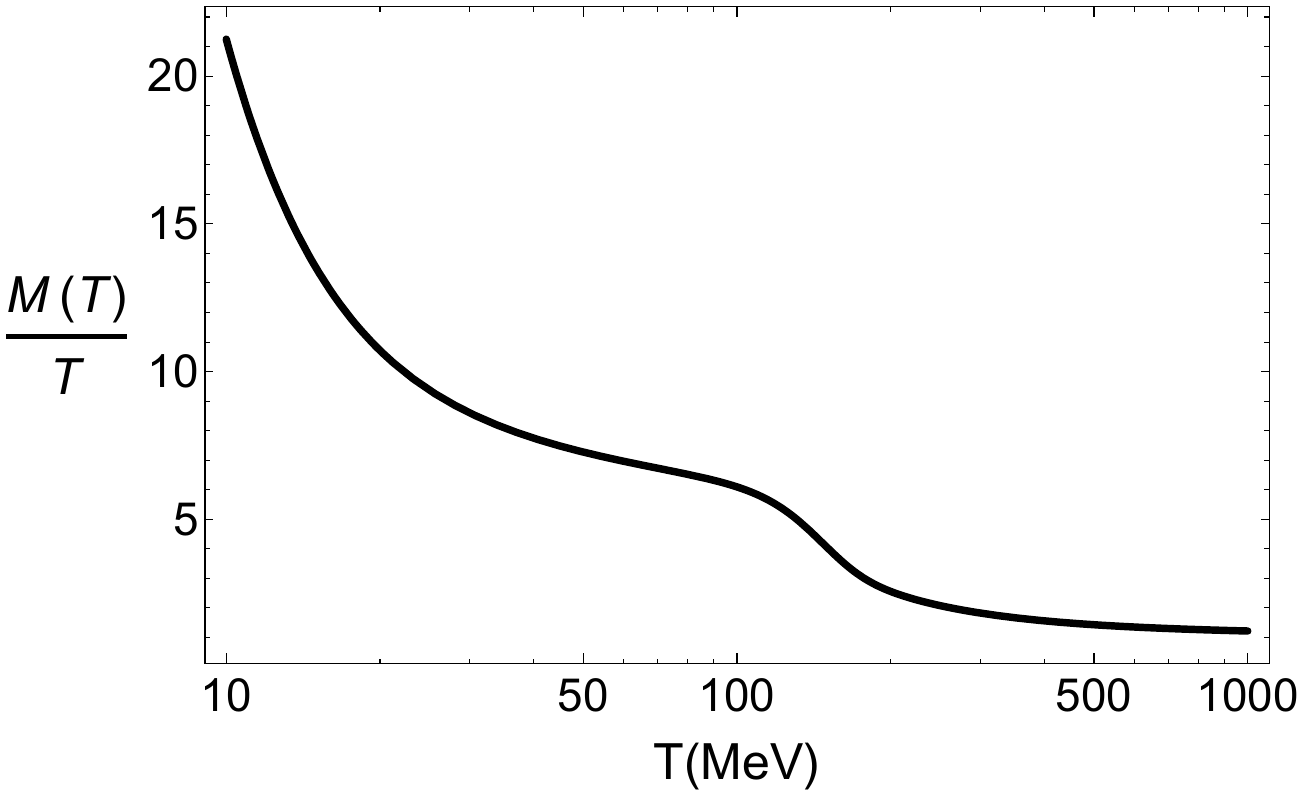}
    \caption{Normalized thermal mass}
    \label{fig:th-mass-1}    
\end{subfigure}\hfil
\caption{Normalized trace-anomaly obtained from Lattice QCD data \cite{Borsanyi:2010cj} and thermal mass as a function of temperature for the QPM.}
\label{fig:th-mass}
\end{figure}

The procedure for determining the temperature dependence of the quasiparticle mass is such that the basic thermodynamic relation is guaranteed,
\begin{equation}
\label{eq:thermo_consistency}
s_0 = \frac{\partial P_0}{\partial T}  = \beta (\varepsilon_{0} + P_{0}) = 
\frac{g T^3 Z^3 }{2\pi^2}K_3(Z).
\end{equation}
Thus, once the function $s_{0}(T)$ is especified (for the equation of state under consideration), the transcendental equation \eqref{eq:thermo_consistency} can be solved for $Z = M(T)/T$. Since $Z^{3}K_3(Z)$ is a monotonic function of $Z$, the solution of $s_{0}(T)$, at a given $T$, is unique. In the present case, we employ the Wuppertal-Budapest Collaboration~\cite{Borsanyi:2010cj} lattice QCD equation of state for $2 + 1$ quark flavors. The corresponding trace anomaly, $I(T) \equiv \varepsilon_0(T) - 3 P_0(T)$, has the analytic parametrization
\begin{equation}
\label{eq:WP_tr_anom}
\frac{I(T)}{T^4} = \left\lbrace \frac{h_0}{1 + h_3 t^2} + \frac{f_0\left[ \tanh(f_1 t + f_2 ) + 1 \right] }{1 + g_1 t + g_2 t^2} \right\rbrace\exp\left( - \frac{h_1}{t} - \frac{h_2}{t^2} \right) ,
\end{equation}
with parameters given by $h_0 = 0.1396$, $h_1 = - 0.1800$, $h_2 = 0.0350$, $f_0 = 2.76$, $f_1 = 6.79$, $f_2 = -5.29$, $g_1 = - 0.47$, $g_2 = 1.04$ and $h_3 = 0.01$, where $t \equiv T/( 0.2 \text{ GeV})$. In Fig.~\ref{fig:th-mass}, we display the Wuppertal-Budapest trace anomaly and the corresponding result for the temperature dependence of the quasiparticles mass. Similar plots can be seen in Refs.~\cite{Romatschke:2011qp,Alqahtani:2015qja,Rocha:2022fqz}. We note that, in the high-temperature limit, the behavior of $I(T)$ leads to a thermal mass that behaves as, $M(T)/T \to 1.1$ when $T \to \infty$. Besides, in the region near the crossover, the peak in the trace-anomaly leads to a non-monotonic behavior of the derivative of $M(T)/T$ with respect to $T$ in that region.

Moreover, from the trace anomaly, the pressure can be obtained by solving the following ODE, stemming from thermodynamic relations,
\begin{equation}
T\frac{\partial P_0(T)}{\partial T} - 4P_0(T) = I(T).
\end{equation}
Imposing the boundary condition $P(0) = 0$, we have the unique solution
\begin{equation}
\frac{P_0(T)}{T^4} = \int_0^T \frac{dX}{X} \frac{I(X)}{X^4} , \label{eq:P_from_I}
\end{equation}
from that, the energy and entropy densities can be obtained, respectively, by the relations $\varepsilon_0 = I + 3P_0$ and $s_{0} = \beta (\varepsilon_0 + P_0 )$.

The matching condition \eqref{eq:match1} simplifies the dynamical constrain \eqref{eq:dB}, since it implies that we can consider solutions of $B$ that are dependent only on temperature and, thus, can be determined solely by the quasiparticle mass \cite{Rocha:2022fqz}. Hence, the background field $B$ can be determined as if the system were in equilibrium,
\begin{equation}
\label{eq:dBeq}
\frac{\partial B}{\partial T} = - \frac{1}{2} \frac{\partial M^2}{\partial T} \int dP f_{0 \pp} =  - \frac{g T M^2}{2\pi^2}K_1(Z)\frac{\partial M(T)}{\partial T},
\end{equation}
which can be uniquely solved for $B(T)$ by 
\begin{equation}
\label{eq:B_fin}
B(T) = - \frac{g}{2\pi^2}\int_0^T dX\, M^2(X) X\, K_1\left(\frac{M(X)}{X}\right) \frac{\partial M(X)}{\partial X} , 
\end{equation}
where the initial condition $B(0) = 0$ \cite{Romatschke:2011qp,Alqahtani:2015qja} has been used. 

\begin{figure}[!h]
    \centering
\begin{subfigure}{0.5\textwidth}
    \includegraphics[scale=0.36]{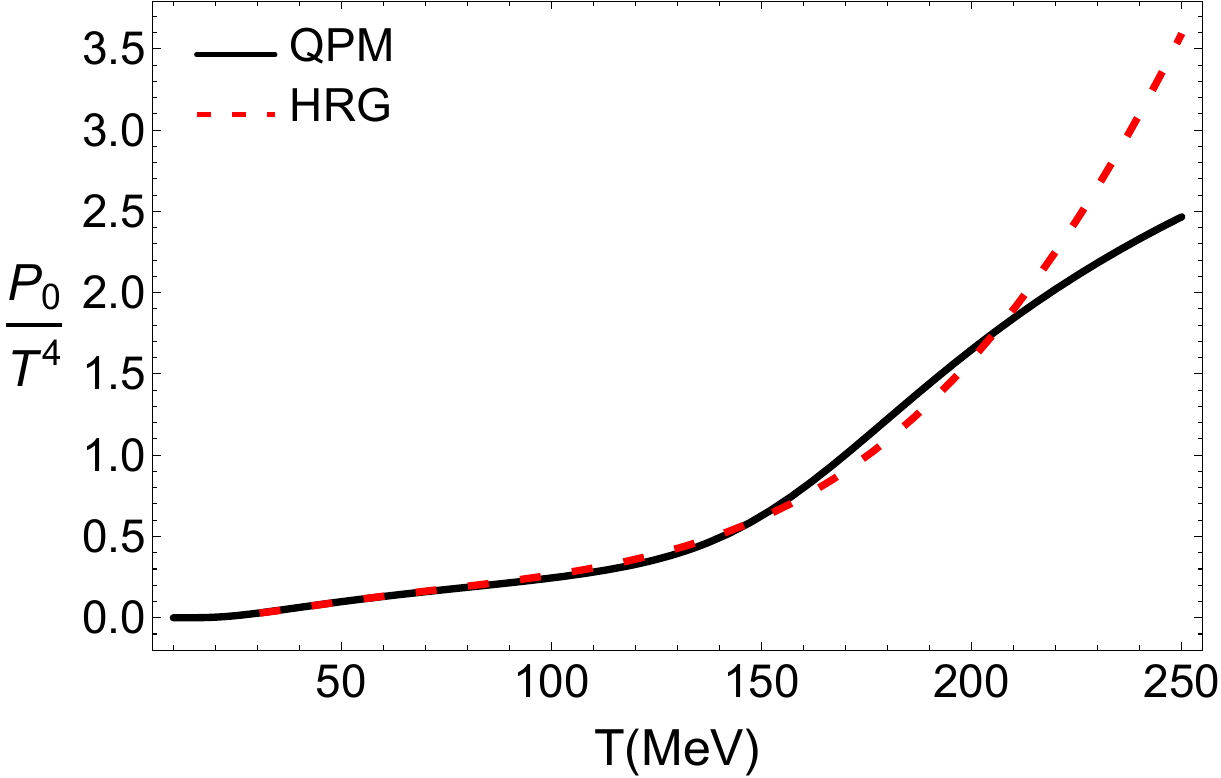}
    \caption{Normalized pressure}
    \label{fig:pres-1}    
\end{subfigure}\hfil
\begin{subfigure}{0.5\textwidth}
    \includegraphics[scale=0.35]{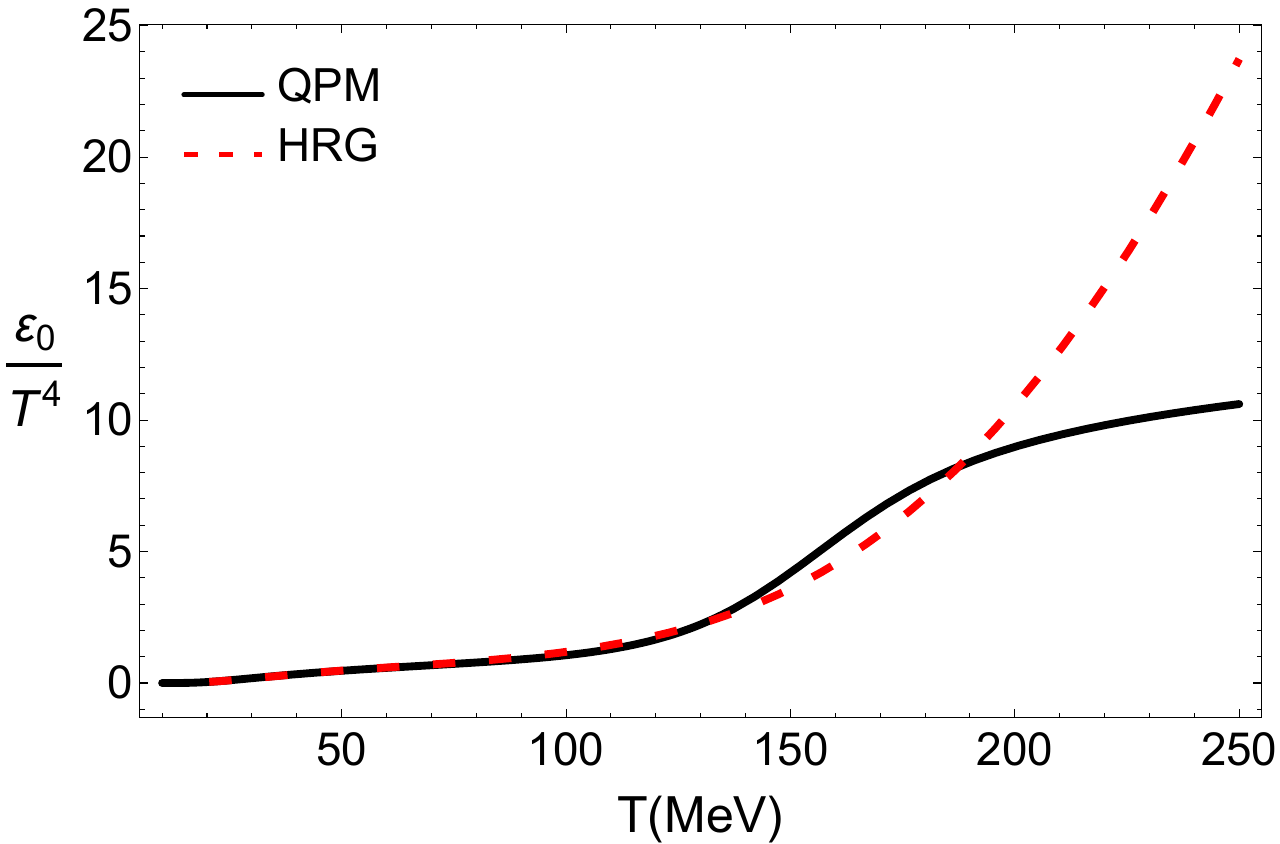}
    \caption{Normalized energy density}
    \label{fig:ener-2}    
\end{subfigure}\hfil
\\
\begin{subfigure}{0.5\textwidth}    \includegraphics[scale=0.30]{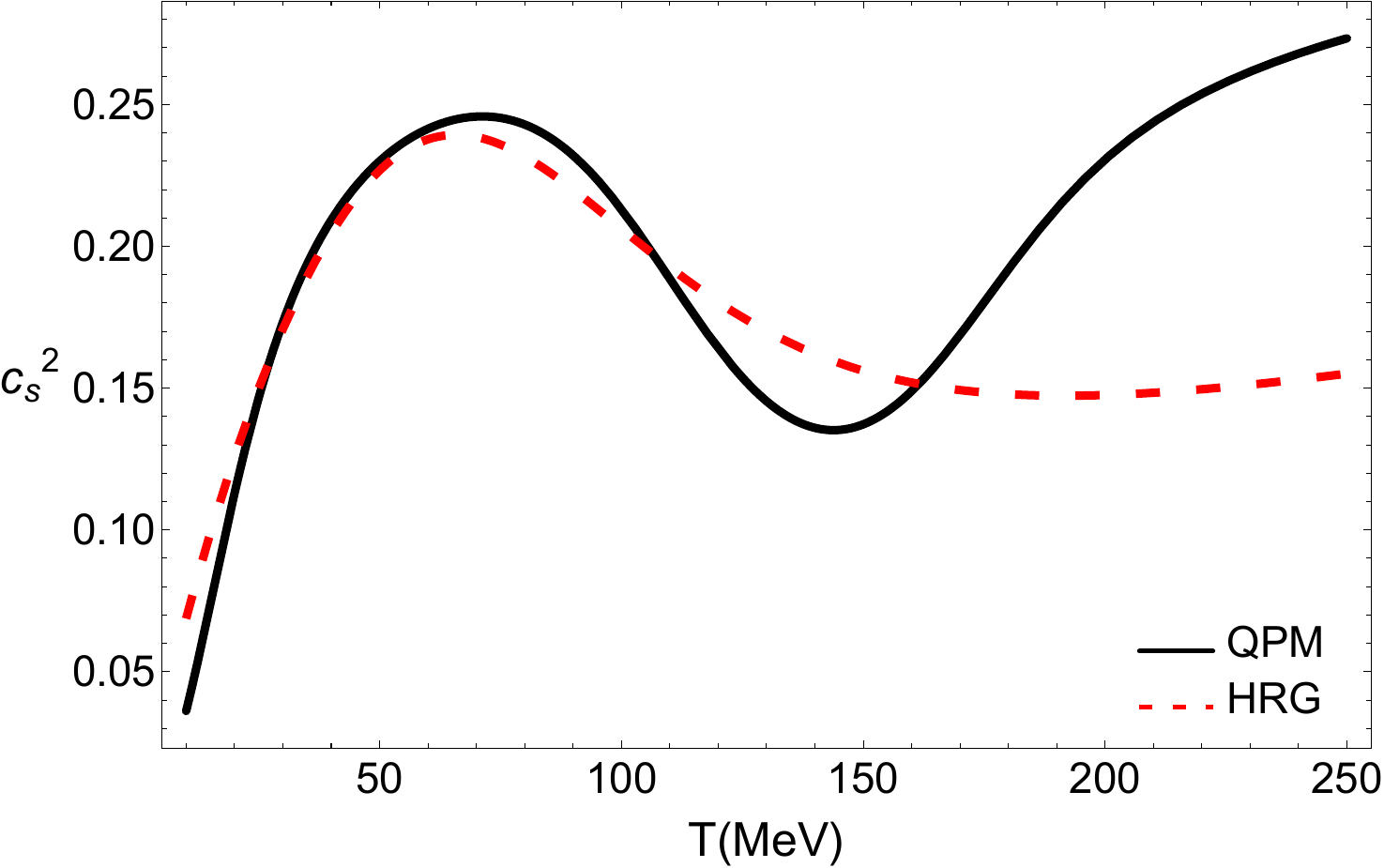}
    \caption{Speed of sound squared}    
    \label{fig:cs2}
\end{subfigure}\hfil
\begin{subfigure}{0.5\textwidth}    \includegraphics[scale=0.30]{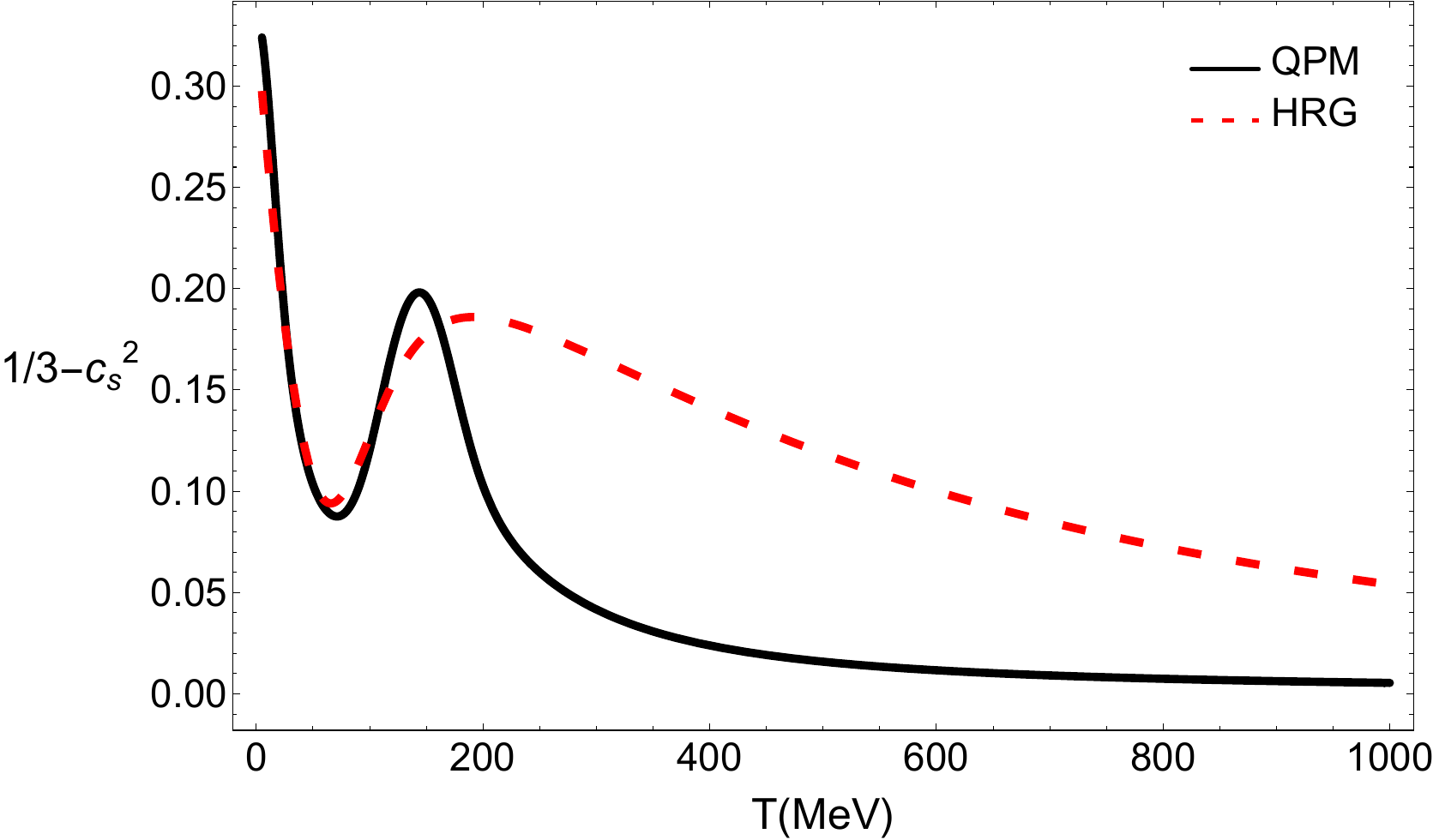}
    \caption{Conformality violation of the speed of sound squared}    
    \label{fig:cs2}
\end{subfigure}\hfil
\caption{Comparison between thermodynamic quantities for the QPM and the HRG as a function of temperature.}
\label{fig:eql-qtts}
\end{figure}

In Fig.~\ref{fig:eql-qtts}, the results obtained for equilibrium quantities within the QPM are compared with the ones for the HRG, with UrQMD particle content. We see that both models agree for  temperatures $T \lesssim 100$. As expected, as one increases the temperature the models start to differ significantly since the HRG model is not capable of describing the deconfined phase of nuclear matter. 

\subsection{Moment equations for the quasiparticle model}

Analogously to what occured in Sec.~\ref{sec:moments-HRG}, we shall employ the method of moments to derive trasient hydrodynamic equations. In the present case, we shall derive exact equations of motion for
\begin{equation}
\label{eq:irreducible_moments}
\begin{aligned}
\rho^{\mu_{1} \cdots \mu_{\ell}}_{r} = \int dP  E_{\mathbf{p}}^{r} p^{\langle \mu_{1}} \cdots p^{\mu_{\ell} \rangle} \delta f_{\mathbf{p}}.
\end{aligned}    
\end{equation}
Similarly to what occurred in the last section, we restrict our analyses to the equations of motion for the scalar and rank-2 moments, since they are the most relevant for the derivation of hydrodynamic equations of motion (at vanishing chemical potential). Hence, we have, for scalar moments 
\begin{equation}
\label{eq:moment-eoms-rank0}
\begin{aligned}
& D\rho_{r}
-
r Du_{\mu} \rho^{\mu}_{r-1}
+ 
\nabla_{\mu}\rho^{\mu}_{r-1} + \rho_{r} \theta
-(r-1) \sigma_{\mu \nu} \rho_{r-2}^{\mu \nu}
- (r-1) \frac{\theta}{3} \left(M^{2} \rho_{r-2} - \rho_{r}\right) 
-
\frac{1}{2} D M^{2} \  (r-1)\rho_{r-2}  
\\
&
+
\frac{-
\frac{\beta}{2} (\partial M^{2}/\partial \beta) I_{r-1,0}
-
I_{r+1,0}}{\left(I_{3,0} + \frac{\beta}{2} (\partial M^{2}/\partial \beta) I_{1,0} \right)} \left[ D\delta \varepsilon + (\delta \varepsilon + \Pi) \theta - \pi^{\mu \nu} \sigma_{\mu \nu} + \partial_{\mu}h^{\mu} + u_{\mu} D h^{\mu}\right]
\\
 &
-
\alpha_{r}^{(0)} \theta
 =
\int dP E_{\mathbf{p}}^{r-1}  C_{Q}[f_{\mathbf{p}}] ,
\end{aligned}    
\end{equation}
and for rank-2 moments we have
\begin{equation}
\label{eq:moment-eoms-rank2}
\begin{aligned}
&
D\rho^{\langle \alpha \beta \rangle}_{r}
-
r Du_{\mu} \rho^{\mu \alpha \beta}_{r-1} 
+
\frac{2}{5}  Du^{ \langle \alpha} \left[ \left( 2 r + 5\right) \rho_{r+1}^{\beta\rangle} - r M^{2} \rho_{r-1}^{\beta\rangle} \right]
\\
& 
+\Delta^{\alpha \beta}_{ \ \ \alpha' \beta'}
 \nabla_{\mu} \rho_{r-1}^{\alpha' \beta' \mu} 
-
(r-1) \sigma_{\mu \nu} \rho^{\mu \alpha \beta \nu}_{r-2}
+
\frac{2}{5} 
 \nabla^{\langle \alpha}\left( M^{2} \rho_{r-1}^{\beta \rangle} - \rho_{r+1}^{\beta \rangle} \right) 
+
\frac{\theta}{3} \left[ (r+4) \rho_{r}^{\alpha \beta} - (r-1) M^{2} \rho_{r-2}^{\alpha \beta}\right]
+
2 
 \omega_{\mu}^{ \ \langle \alpha} \rho^{\beta \rangle \mu}_{r} 
\\
& 
+ \frac{2}{7} 
\sigma_{\mu}^{\ \langle \alpha}  \left[ (2r+5) \rho_{r}^{\beta \rangle \mu} - 2(r-1) M^{2}\rho_{r-2}^{\beta \rangle \mu}\right] 
+
\frac{2}{15} \sigma^{\alpha \beta}  \left[ - (r-1)M^{4} \rho_{r-2} + (2r+3) M^{2}\rho_{r} - (r+4) \rho_{r+2}\right]
\\
&
-
\frac{1}{2} \beta D M^{2} (r-1) \rho_{r-2}^{\alpha \beta}
-
\Delta^{\alpha \beta}_{\ \ \ \alpha' \mu} 
\nabla^{\mu} M^{2} \rho_{r-1}^{\alpha'} 
-
\alpha_{r}^{(2)} 
 \sigma^{\alpha \beta}
 =
 \int dP E_{p}^{r-1} p^{\langle \alpha} p^{\beta \rangle}
C_{Q}[f_{p}],
\end{aligned}    
\end{equation}
where we made use of the definitions
\begin{equation}
\begin{aligned}
&
\alpha_{r}^{(0)} =  c_{s}^{2}\left(
\frac{\beta}{2} \frac{\partial M^{2}}{\partial \beta} I_{r-1,0}
+
I_{r+1,0} \right)  - \beta I_{r+1,1}, 
\\
&
\alpha_{r}^{(2)} = 2 \beta I_{r+3,2},
\\
&
c_{s}^{2} = \frac{\partial P_{0}}{\partial \varepsilon_{0}} =  \frac{I_{3,1}}{I_{3,0} + \frac{\beta}{2} (\partial M^{2}/\partial \beta)  I_{1,0}}.
\end{aligned}    
\end{equation}
Note that the expression for $c_{s}^{2}$ displays a contribution of the temperature-dependence of the mass, which remains relevant even in the high-temperature limit, i.e. when we consider that $M(T)/T \to 0$  
\begin{equation}
\label{eq:cs2-QPM-high-T}
\begin{aligned}
&
c_{s}^{2} \simeq \frac{1}{3} + \frac{M(T)}{36} \frac{\partial}{\partial T}\left(\frac{M(T)}{T}\right).
\end{aligned}    
\end{equation}
This differs from the $c_{s}^{2} - 1/3 \simeq - (1/36)(m/T)^{2}$ behavior, found when the mass $m$ does not depend on the temperature \cite{Denicol:2014vaa}. In Eqs.~\eqref{eq:moment-eoms-rank0}--\eqref{eq:moment-eoms-rank2}, we note the appearance of new terms with respect to Refs.~\cite{Denicol:2012cn,Denicol:2021,deBrito:2024vhm}, proportional to $DM^{2}$ or $\nabla^{\mu}M^{2}$ stemming from the fact that the mass is a function of temperature, that varies locally.

\subsection{Relaxation time approximation}

Now we shall address the collisional moments appearing on the right-hand side of Eqs.~\eqref{eq:moment-eoms-rank0}--\eqref{eq:moment-eoms-rank2}. In contrast to the last section, the matching conditions \eqref{eq:match1} and \eqref{eq:EQL-qp} do not correspond to Landau matching conditions, and the traditional Anderson-Witting RTA  \cite{andersonRTA:74} cannot be employed, since this approximation is inconsistent with the local conservation laws. Consequently, we adopt Relaxation Time approximation developed in Ref.~\cite{Rocha:2021zcw} to approximate the collision term in the quasi-particle model and render the computation of the equations of motion simpler. In practice, we employ 
\begin{equation}
\label{eq:nRTA-QPM}
\begin{aligned}
C_{Q}[f_{\bf p}] \simeq - \frac{E_{\mathbf{p}}}{\tau_{R}} f_{0\textbf{p}} \left[ \phi_{\mathbf{p}} - 
\frac{\lan \phi_{\mathbf{p}}  E_{\mathbf{p}}^{2} \ran_{0}}{I_{3,0}} E_{\bf p} 
+ 
\frac{\lan \phi_{\mathbf{p}}  E_{\mathbf{p}} p^{\langle \mu \rangle} \ran_{0}}{ I_{3,1}} p_{\langle \mu \rangle}   \right],
\end{aligned}
\end{equation}
where we use the notation
\begin{equation}
\begin{aligned}
&
\langle \cdots \rangle_{0} \equiv \int dP (\cdots) f_{0 {\bf p}},
\\
&
\phi_{\mathbf{p}} \equiv \frac{f_{\bf p} - f_{0 {\bf p}}}{f_{0 {\bf p}}},
\end{aligned}    
\end{equation}
and we consider, for simplicity, that the relaxation time, $\tau_{R}$, does not depend on momentum. In this case, the computation of the scalar and rank-2 tensor collisional moments present in Eqs.~\eqref{eq:moment-eoms-rank0}--\eqref{eq:moment-eoms-rank2} simplifies considerably and we have  
\begin{equation}
\label{eq:coll-mom-QPM}
\begin{aligned}
&
\int dP E_{\mathbf{p}}^{r-1}  C[f_{\mathbf{p}}]
=
- \frac{1}{\tau_{R}} \left( \rho_{r} -  \frac{I_{r+1,0}}{I_{3,0}} \rho_{2} \right),
\\
&
 \int dP E_{p}^{r-1} p^{\langle \alpha} p^{\beta \rangle}
C[f_{\mathbf{p}}]
=
- \frac{1}{\tau_{R}}  \rho_{r}^{\alpha \beta}.
\end{aligned}    
\end{equation}

\subsection{Transient hydrodynamic equations of motion}
\label{sec:transient-QPM}

The reduction of the dynamical degrees of freedom appearing in the full moment equations to that of the hydrodynamic currents ($\varepsilon_{0}$, $\Pi$, $u^{\mu}$, $\pi^{\mu \nu}$, in the present case) requires a truncation procedure. As before, we make use of the order of magnitude method \cite{struchtrup2004stable,struchtrup2005macroscopic,Fotakis:2022usk}. First, we derive the asymptotic expression for the scalar and rank-2 irreducible moments in a gradient expansion,
\begin{equation}
\begin{aligned}
&
\rho_{r} = - \zeta_{r} \theta + \mathcal{O}(2), 
\\
&
\rho_{r}^{\mu \nu} = 2 \eta_{r} \sigma^{\mu \nu} +
\mathcal{O}(2), 
\label{dontcare}
\end{aligned}    
\end{equation}
As in the last section, $\mathcal{O}(2)$ denotes terms that are of second-order or higher in powers of gradients or in powers of the dissipative currents. The expressions for $\zeta_{r}$ and $\eta_{r}$ can be obtained by combining Eqs.~\eqref{eq:moment-eoms-rank0}, \eqref{eq:moment-eoms-rank2} and \eqref{eq:coll-mom-QPM},
\begin{equation}
\label{eq:asymp-coeffs-QPM}
\begin{aligned}
&
\zeta_{r} =  - \alpha_{r}^{(0)}
+
\frac{I_{r+1,0}}{I_{1,0}} \alpha_{0}^{(0)}, 
\\
&
\eta_{r} = \frac{1}{2} \alpha_{r}^{(2)},
\end{aligned}    
\end{equation}
where we further identify the bulk viscosity $\zeta =(1/3)\zeta_{2}$ and the shear viscosity $\eta = \eta_{0}$.  Expressions \eqref{eq:asymp-coeffs-QPM} can be alternatively derived by computing momentum integrals of the first order solution of the Chapman-Enskog expansion, obtained in Ref.~\cite{Rocha:2022fqz} (Eq.~(53) of that reference, considering a momentum-independent $\tau_{R}$), using the relaxation time approximation \eqref{eq:nRTA-QPM} .  

Then, relations \eqref{dontcare} can be rearranged in order to express generic moments solely in terms of $\Pi$ and $\pi^{\mu \nu}$. Hence, we derive 
\begin{subequations}
\label{eq:OoM-trun}
\begin{align}
&
\label{eq:OoM-scal-trun}
\rho_{r} = \frac{\zeta_{r}}{\zeta} \Pi \equiv \mathcal{A}_{r} \Pi + \mathcal{O}(2), 
\\
&
\label{eq:OoM-tens-trun}
\rho_{r}^{\mu \nu} = \frac{\eta_{r}}{\eta} \pi^{\mu \nu} \equiv \mathcal{C}_{r} \pi^{\mu \nu} +
\mathcal{O}(2). 
\end{align}    
\end{subequations}
where we defined the following coefficients,
\begin{equation}
\begin{aligned}
&
\mathcal{A}_{r} 
=
3\frac{\alpha_{r}^{(0)}
-
(I_{r+1,0}/I_{1,0}) \alpha_{0}^{(0)}}{ \alpha_{2}^{(0)}
-
(I_{3,0}/I_{1,0}) \alpha_{0}^{(0)}},
\\
&
\mathcal{C}_{r} = \frac{I_{r+3,2}}{I_{3,2}}.
\end{aligned}    
\end{equation}
These relations allow to express non-hydrodynamic moments in terms of dissipative currents and thus provide closure for the equation of motion, when substituted in the moment equations \eqref{eq:moment-eoms-rank0}--\eqref{eq:moment-eoms-rank2}. This procedure is accurate up to $\mathcal{O}(2)$.

\subsubsection{Bulk viscous pressure}
\label{sec:bulk-pres-QPM}

First, we derive the equations of motion for the bulk viscous pressure. For this purpose, we take Eq.~\eqref{eq:moment-eoms-rank0} with $r=0$ (having in mind that $\delta \varepsilon = 3 \Pi$, see text below Eq.~\eqref{eq:diss-curr-QPM}), and substitute expressions \eqref{eq:coll-mom-QPM} and Eqs.~\eqref{eq:OoM-trun} to approximate the collisional integrals and the non-hydrodynamic moments appearing in the equations. We finally obtain
\begin{equation}
\label{eq:hydro-EoM-bulk-QPM-0}
\begin{aligned}
&
\tau_{\Pi}^{\star} D\Pi + \Pi
=  
- \zeta^{\star} \theta
- \delta_{\Pi \Pi}^{\star} \Pi \theta  
+
\lambda_{\Pi \pi}^{\star}
\pi^{\mu \nu} \sigma_{\mu \nu} + \mathcal{O}(3).
\end{aligned}    
\end{equation}
In this derivation, time-like derivatives of the inverse temperature are replaced, up to second-order in gradients, by space-like derivatives of the four-velocity using the conservation law
\begin{equation}
\label{eq:Db-Kn2}
\begin{aligned}
&
D \beta  
=
\beta c_{s}^{2} \theta + \mathcal{O}(2).
\end{aligned}    
\end{equation}
In Eq.~\eqref{eq:hydro-EoM-bulk-QPM-0} we employ the star indices in the transport coefficients as a reminder that these transport coefficients are computed without assuming Landau matching conditions and, thus, are not equivalent to the transport coefficient derived in the last section for the HRG. In Subsec.~\ref{sec:matching-redef-QPM}, we shall convert these equations of motion and corresponding transport coefficient to the Landau frame. The transport coefficients read
\begin{equation}
\label{eq:coeffs-bulk-QPM}
\begin{aligned}
&
\zeta^{\star} = -\frac{\tau_{R}}{3} \frac{I_{3,0}}{I_{1,0}} \left[
 \beta I_{1,1} 
 - 
 \beta c_{s}^{2} 
\left( I_{1,0}
+
\frac{\beta}{2} \frac{\partial M^{2}}{\partial \beta} I_{-1,0}\right)\right],
\\
&
\tau_{\Pi}^{\star} = \tau_{R} \frac{I_{1,0} +
(\beta/2) (\partial M^{2}/\partial \beta) I_{-1,0}
}{I_{3,0} + (\beta/2) (\partial M^{2}/\partial \beta) I_{1,0} } \frac{I_{3,0}} {I_{1,0}},
\\
&
\frac{\delta_{\Pi \Pi}^{\star}}{\tau_{\Pi}^{\star}} = 
\frac{4}{3} - \mathcal{A}_{-2} \left[ \frac{M^{2}}{9}  \frac{I_{3,0} + (\beta/2) (\partial M^{2}/\partial \beta)  I_{1,0}}
{I_{1,0} + (\beta/2) (\partial M^{2}/\partial \beta) I_{-1,0}} 
+
\frac{1}{6} \frac{\partial M^{2}}{\partial \beta} \frac{(\varepsilon_{0}+ P_{0})}{
I_{1,0} + (\beta/2) (\partial M^{2}/\partial \beta) I_{-1,0}
}\right],
\\
&
\frac{\lambda_{\Pi \pi}^{\star}}{\tau_{\Pi}^{\star}}
=
\frac{1}{3} 
+
\frac{1}{3}\mathcal{C}_{-2} \frac{I_{3,0} + (\beta/2) (\partial M^{2}/\partial \beta) I_{1,0} }{
I_{1,0} + (\beta/2) (\partial M^{2}/\partial \beta) I_{-1,0}}.
\end{aligned}    
\end{equation}
We notice a surprising feature particular of the thermal mass model: $\tau_{\Pi}^{\star} \neq \tau_{R}$, i.e., the fact that the relaxation time depends non-trivially on the temperature even when $\tau_{R}$ does not depend on temperature. This is in contrast to the HRG model (see Eq.~\eqref{eq:transp-coeffs-HRG-bulk-1}). This is an effect of the non-ideal equation of state implemented. Indeed, if the mass does not depend on temperature, the bulk relaxation time reduces to the RTA characteristic time, i.e., $\partial M /\partial \beta = 0 \Rightarrow \tau_{\Pi} = \tau_{R}$. 

Now, we analyze the high-temperature/small-mass limit behavior of the transport coefficients in the bulk pressure equation of motion, characterized by the limit $M(T)/T \to 0$. Before displaying the results, we must notice that this limit is not realized for QCD-thermodynamics-fitted quasi-particles, since $M(T)/T \to 1.1$, as $T \to \infty$ (see also Fig.~\ref{fig:th-mass-1}). However, since this asymptotic constant mass is relatively small, the results below should provide reasonable estimates,    
\begin{equation}
\label{eq:bulk-coeffs-expn-m=0}
\begin{aligned}
&
\frac{\tau_{\Pi}^{\star}}{\tau_{R}} \simeq 1 - \frac{5}{12} M(T)  \frac{\partial}{\partial T}\left(\frac{M(T)}{T}\right),
\\
&
\frac{\zeta^{\star}}{\tau_{R}(\varepsilon_{0}+ P_{0})} \simeq - \frac{5}{36} M(T)  \frac{\partial}{\partial T}\left(\frac{M(T)}{T}\right)
+
\left[\frac{T}{6}\frac{\partial}{\partial T}\left(\frac{M(T)}{T}\right) - \frac{5}{432} \left[T\frac{\partial}{\partial T}\left(\frac{M(T)}{T}\right)\right]^{2}\right] \left(\frac{M(T)}{T}\right)^{2},
\\
&
\frac{\delta_{\Pi \Pi}^{\star}}{\tau_{\Pi}^{\star}} 
\simeq
\frac{4}{3} + \frac{6 \pi T}{5} \frac{\partial}{\partial T}\left(\frac{M(T)}{T}\right),
\\
&
\frac{\lambda_{\Pi \pi}^{\star}}{\tau_{\Pi}^{\star}}
\simeq
\frac{2}{3} 
+
\frac{5}{36} M(T)  \frac{\partial}{\partial T}\left(\frac{M(T)}{T}\right), 
\end{aligned}    
\end{equation}
It is noted that the expression for $\zeta^{\star}/[\tau_{R}(\varepsilon_{0}+ P_{0})]$ coincides with Eq.~(54) of Ref.~\cite{Rocha:2022fqz} (with the parameter $\gamma = 0$, in that reference). Regarding the bulk relaxation time $\tau_{\Pi}$, it approaches the RTA characteristic time, $\tau_{R}$. Besides, it is seen that the sub-leading term for $\delta_{\Pi \Pi}/\tau_{\Pi}$ cannot be expressed as something proportional to $(1/3 - c_{s}^{2})$, due to the fact that $\mathcal{A}_{-2} \propto (T/M(T))$ in this regime. 



\subsubsection{Shear-stress tensor}

Now we turn our attention to the equation of motion for the shear-stress tensor. For this purpose, we take Eq.~\eqref{eq:moment-eoms-rank0} with $r=0$, and Eq.~\eqref{eq:moment-eoms-rank2} for the rank-2 moments of the collision term. Employing also Eq.~\eqref{eq:Db-Kn2}, we derive 
\begin{equation}
\label{eq:hydro-EoM-shear-QPM-0}
\begin{aligned}
& 
\tau_{\pi}^{\star} D \pi^{\langle \alpha \beta \rangle}
+
\pi^{\alpha \beta}
=
2 \eta^{\star}  \sigma^{\alpha \beta}
-
\delta_{\pi \pi}^{\star} \pi^{\alpha \beta} \theta 
-
2 \tau_{\pi}^{\star}
 \omega_{\mu}^{ \ \langle \alpha} \pi^{\beta \rangle \mu} 
- 
\tau_{\pi \pi}^{\star} 
\sigma_{\mu}^{\ \langle \alpha} \pi^{\beta \rangle \mu} +
\lambda_{\pi \Pi}^{\star} \Pi \sigma^{\alpha \beta} + \mathcal{O}(3),  
\end{aligned}    
\end{equation}
where the various transport coefficients have the following expressions
\begin{equation}
\label{eq:shear-coeffs-tr-anom}
\begin{aligned}
&
\eta^{\star} = \tau_{R} \beta I_{3,2},
\\
&
\tau_{\pi}^{\star} = \tau_{R},
\\
&
\frac{\delta_{\pi \pi}^{\star}}{\tau_{\pi}^{\star}} = \frac{4}{3} 
+
\mathcal{C}_{-2} \left(\frac{M^{2}}{3}
+
 \frac{\beta}{2}  c_{s}^{2} \frac{\partial M^{2}}{\partial \beta} 
 \right),
\\
&
\frac{\tau_{\pi \pi}^{\star}}{\tau_{\pi}^{\star}} 
=
\frac{10}{7} + \frac{4}{7}M^{2} \mathcal{C}_{-2},
\\
&
\frac{\lambda_{\pi \Pi}^{\star}}{\tau_{\pi}^{\star}}  
=
\frac{8}{5}
-
\frac{2}{5} \mathcal{A}_{-2} M^{4}.
\end{aligned}    
\end{equation}
In contrast to the bulk equations of motion, we note that the shear relaxation time coincides with the RTA characteristic timescale.

Once again, we display the small-mass limit behavior for the transport coefficients,
\begin{equation}
\label{eq:shear-coeffs-QPM-trace-anom}
\begin{aligned}
&
\frac{\eta^{\star}}{\tau_{\pi}^{\star} (\varepsilon_{0} + P_{0})} \simeq \frac{1}{5} - \frac{1}{60} \left(\frac{M(T)}{T}\right)^{2},
\\
&
\frac{\delta_{\pi \pi}^{\star}}{\tau_{\pi}^{\star}}\simeq \frac{4}{3} 
-
\frac{1}{36} M(T)  \frac{\partial}{\partial T}\left(\frac{M(T)}{T}\right),
\\
&
\frac{\tau_{\pi \pi}^{\star}}{\tau_{\pi}^{\star}} 
\simeq
\frac{10}{7} + \frac{1}{21}\left(\frac{M(T)}{T}\right)^{2} ,
\\
&
\frac{\lambda_{\pi \Pi}^{\star}}{\tau_{\pi}^{\star}}  
\simeq
\frac{8}{5}
-
\frac{9 \pi}{25} \left(\frac{M(T)}{T}\right)^{3},
\end{aligned}    
\end{equation}
and it is readily seen that, with the exception of $\delta_{\pi \pi}^{\star}/\tau_{\pi}^{\star}$, all transport coefficients have sub-leading contributions that depend on powers of the mass-to-temperature ratio, instead of its derivative.

\subsection{Matching redefinition for the quasi-particle model} 
\label{sec:matching-redef-QPM}

At this point, we note that the equations of motion for the thermal-mass quasi-particle and the hadron-resonance gas models have been derived in different matching conditions. Hence, in order to have a consistent comparison of the results, we will perform a matching transformation connecting the matching conditions employed for the quasi-particle model to the Landau matching conditions employed for the hadron-resonance gas model. Thus, we must re-express the degrees of freedom of the energy-momentum tensor derived for the quasi-particle model in Sec.~\ref{sec:th-mass-QPM}. We perform that consistently with the hydrodynamic power-counting employed so far. For the sake of the present discussion, we shall denote the energy-momentum tensor in each matching condition as
\begin{equation}
\begin{aligned}
&
T^{\mu \nu}_{L} = \varepsilon_{0}(T_{L}) u^{\mu}_{L} u^{\nu}_{L} - [P_{0}(T_{L}) + \Pi_{L}] \Delta^{\mu \nu}_{L} + \pi^{\mu \nu}_{L}, \\
&
T^{\mu \nu}_{Q} = [\varepsilon_{0}(T_{Q}) +  \delta \varepsilon_{Q}] u^{\mu}_{Q} u^{\nu}_{Q} - [P_{0}(T_{Q}) + \Pi_{Q}] \Delta^{\mu \nu}_{Q} + \pi^{\mu \nu}_{Q},
\end{aligned}    
\end{equation}
where one is reminded that $\delta \varepsilon_{Q} = 3 \Pi_{Q}$ in the matching conditions \eqref{eq:match1}. In both cases, $u^{\mu}$ is defined by the Landau frame condition $T^{\mu}_{L,Q\ \nu} u^{\nu}_{L,Q} = \varepsilon_{L,Q} u^{\mu}_{L,Q}$, where we emphasize that $\varepsilon_{L,Q}$ is the \textit{total} energy density, which reduces to the equilibrium energy density for Landau matching, $\varepsilon_{L} = \varepsilon_{0,L}$, and to $\varepsilon_{Q} = \varepsilon_{0,Q} + \delta \varepsilon_{Q}$,  for conditions \eqref{eq:match1}. 

The redistribution of the degrees of freedom is established from the fact that the energy momentum tensors, $T_{L}^{\mu \nu}$ and $T_{Q}^{\mu \nu}$ are the same. We can then relate the degrees of freedom appearing in each matching procedure, leading to,
\begin{subequations}
\label{eq:redistr-trace-anom-Landau-mtch}
\begin{align}
u^{\mu}_{L} &= u^{\mu}_{Q},
\\
\varepsilon_{0}(T_{L}) & = \varepsilon_{0}(T_{Q}) + \delta \varepsilon_{Q}, 
\label{dontcare2}
\\
P_{0}(T_{L}) + \Pi_{L} &= P_{0}(T_{Q}) + \Pi_{Q},
\label{dontcare3}
\\
\label{eq:match-con-shear}
\pi^{\mu \nu}_{L} &= \pi^{\mu \nu}_{Q},
\end{align}    
\end{subequations}
where we note that the shear-stress tensor is unaffected by the matching re-definitions. Note that the thermodynamic energy density and pressure differ due to the different definitions of temperature. We define the temperature difference, $\delta T \equiv T_L - T_Q$, and relate it to $\delta \varepsilon_{Q}$ using Eq.~\eqref{dontcare2} 
\begin{equation}
\label{eq:matching-corrections0}
\delta T = \frac{\delta \varepsilon_{Q}}{\partial \varepsilon_{0,L}/\partial T_{L}}  
+
\frac{1}{2}
\frac{\partial^{2} \varepsilon_{0}}{\partial T_{L}^{2}} \frac{\delta \varepsilon_{Q}^{2}}{(\partial \varepsilon_{0}/\partial T_{L})^{3}} 
+
\mathcal{O}(3) 
=
\frac{3 \Pi_{Q}}{\partial \varepsilon_{0,L}/\partial T_{L}} 
+
\frac{9}{2}
\frac{\partial^{2} \varepsilon_{0}}{\partial T_{L}^{2}} \frac{\Pi_{Q}^{2}}{(\partial \varepsilon_{0}/\partial T_{L})^{3}} 
+ \mathcal{O}(3),
\end{equation}
where $\mathcal{O}(3)$ denotes terms that are cubic in the dissipative currents and, thus, are of third order in our power-counting scheme.
The bulk viscous pressure in each matching condition are related using Eq.~\eqref{dontcare3}
\begin{equation}
\label{eq:matching-corrections1}
\begin{aligned}    
&
\Pi_{Q} =  \Pi_{L} + c_{s}^{2}(T_{L}) \delta \varepsilon_{Q} 
-
\frac{1}{2}
\frac{\partial c_{s}^{2}}{\partial \varepsilon_{0}}
 \delta \varepsilon_{Q}^{2}
+  
\mathcal{O}(3) 
=
\Pi_{L} + 3 c_{s}^{2}(T_{L}) \Pi_{Q} 
-
\frac{9}{2}
\frac{\partial c_{s}^{2}}{\partial \varepsilon_{0}}
 \Pi_{Q}^{2}
+  \mathcal{O}(3).
\end{aligned}
\end{equation}
Combining these expressions, we obtain the following relation between the temperature and the bulk viscous in both equilibrium frames,
\begin{equation}
\label{eq:matching-corrections1}
\begin{aligned}    
T_Q &= T_L - \frac{3 \Pi_{L}}{(\partial \varepsilon_{0}/\partial T_{L})(1 - 3 c_{s}^{2})}  + \mathcal{O}(2),
\\
\Pi_{Q} & = \frac{\Pi_{L}}{1 - 3 c_{s}^{2}}  -
\frac{9}{2} 
\frac{\partial c_{s}^{2}}{\partial \varepsilon_{0}}
\frac{\Pi_{L}^{2}}{(1- 3 c_{s}^{2})^{3}}
+
\mathcal{O}(3).
\end{aligned}
\end{equation}
The equations of motion for the bulk viscous pressure and shear-stress tensor presented in Eqs.~\eqref{eq:hydro-EoM-bulk-QPM-0} and \eqref{eq:hydro-EoM-shear-QPM-0}, respectively, can be converted to the Landau frame using the expressions above and only retaining terms that are of second order or smaller.

\subsection{The redefined quasi-particle transport coefficients}

Now we employ the matching connection equations \eqref{eq:matching-corrections1} in the quasi-particle model transient equations of motion \eqref{eq:hydro-EoM-bulk-QPM-0} and \eqref{eq:hydro-EoM-shear-QPM-0}, respectively, for the bulk viscous pressure and the shear-stress tensor. We then derive the following equations of motion,
\begin{equation}
\label{eq:starred-transient-EoMs}
\begin{aligned}
&
\tau_{\Pi} D\Pi + \Pi
=  
- \zeta \theta
- \delta_{\Pi \Pi} \Pi \theta  
+
\lambda_{\Pi \pi}
\pi^{\mu \nu} \sigma_{\mu \nu}
+
\mathcal{O}(3),
\\
&
\tau_{\pi} D \pi^{\langle \alpha \beta \rangle}
+
\pi^{\alpha \beta}
=
2 \eta  \sigma^{\alpha \beta}
-
\delta_{\pi \pi} \pi^{\alpha \beta} \theta 
-
2 \tau_{\pi}
 \omega_{\mu}^{ \ \langle \alpha} \pi^{\beta \rangle \mu} 
- 
\tau_{\pi \pi} 
\sigma_{\mu}^{\ \langle \alpha} \pi^{\beta \rangle \mu} +
\lambda_{\pi \Pi} \Pi \sigma^{\alpha \beta}
+
\mathcal{O}(3),  
\end{aligned}    
\end{equation}
where, for the sake of simplicity, we dropped the $Q$ subscripts used in the beginning of the present section. We note that the term $\propto \Pi^{2}$, which emerges as a consequence of the matching redefinition, has been incorporated into $\propto \Pi \theta$. With this modification, Eqs.~\eqref{eq:starred-transient-EoMs} are of the same form of Eqs.~\eqref{eq:hydro-EoM-bulk-QPM-0} and \eqref{eq:hydro-EoM-shear-QPM-0}, with the important difference that the transport coefficients are modified. Most transport coefficients maintain their functional form (with just their temperature being modified),  
\begin{subequations}
\label{eq:eoms-after-mtch-con}
\begin{align}
&
\label{eq:eoms-after-mtch-con-bulk}
\frac{\zeta}{\tau_{\Pi}}
=
\frac{\zeta^{\star}}{\tau_{\Pi}^{\star}}, 
\quad
\frac{\lambda_{\Pi \pi}}{\tau_{\Pi}}
=
\frac{\lambda_{\Pi \pi}^{\star}}{\tau_{\Pi}^{\star}}, 
\\
&
\label{eq:eoms-after-mtch-con-shear}
\tau_{\pi} = \tau_{\pi}^{\star}, 
\quad
\frac{\eta}{\tau_{\pi}}
=
\frac{\eta^{\star}}{\tau_{\pi}^{\star}},
\quad
\frac{\delta_{\pi \pi}}{\tau_{\pi}}
=
\frac{\delta_{\pi \pi}^{\star}}{\tau_{\pi}^{\star}},
\quad
\frac{\tau_{\pi \pi}}{\tau_{\pi}}
= 
\frac{\tau_{\pi \pi}^{\star}}{\tau_{\pi}^{\star}},
\end{align}    
\end{subequations}
whereas some coefficients change non-trivially, so that
\begin{subequations}
\label{eq:eoms-after-mtch-con-hard}
\begin{align}
&
\tau_{\Pi} = \tau_{\Pi}^{\star} (1 - 3 c_{s}^{2}), \\
&
\label{eq:del-Pi-Pi-after-con}
\frac{\delta_{\Pi \Pi}}{\tau_{\Pi}} = 
\frac{\delta_{\Pi \Pi}^{\star}}{\tau_{\Pi}^{\star}} - \frac{3}{\tau_{\Pi}^{\star}(1- 3 c_{s}^{2})} \frac{\partial \zeta^{\star}}{\partial \varepsilon_{0}}
-   
\frac{9}{2} 
\frac{1}{\tau_{\Pi}^{\star}}
\frac{\partial c_{s}^{2}}{\partial \varepsilon_{0}}
\frac{\zeta^{\star} }{(1- 3 c_{s}^{2})^{2}},
\\
&
\label{eq:lam-pi-Pi-after-con}
\frac{\lambda_{\pi \Pi}}{\tau_{\pi}}
=
\frac{\lambda_{\pi \Pi}^{\star}}{\tau_{\pi}^{\star}}
-
\frac{6}{\tau_{\pi}^{\star}(1- 3 c_{s}^{2})} \frac{\partial \eta^{\star}}{\partial \varepsilon_{0}}. 
\end{align}    
\end{subequations}
We remark that the thermodynamic quantities on the right-hand side of Eqs.\ \eqref{eq:eoms-after-mtch-con} and \eqref{eq:eoms-after-mtch-con-hard} are calculated using the temperature evaluated in the Landau frame. We note that, since $\zeta$ and $\tau_\Pi$ are modified by the same factor in the matching transformation, the ratio $\zeta/\tau_{\Pi}$ is not modified by the frame transformation. We also note that $\delta_{\Pi \Pi}^{\star}$ is the only transport coefficient that depends on the second derivative of the thermal mass in the temperature.

With the corrections implied by Eqs.~\eqref{eq:eoms-after-mtch-con}, the high temperature ($M(T)/T \to 0$) behavior of the transport coefficients is also expected to change. Indeed, the expressions for the transport coefficients corresponding to the bulk viscous pressure equation of motion read,  
\begin{subequations}
\label{eq:bulk-coeffs-QPM-landau-m=0}
\begin{align}
&
\frac{\tau_{\Pi}}{\tau_{R}} \simeq  - \frac{M(T)}{12} \frac{\partial}{\partial T}\left(\frac{M(T)}{T}\right)
+
\frac{M(T)^{2}}{36} \left[\frac{\partial}{\partial T}\left(\frac{M(T)}{T}\right) \right]^{2}, 
\\
&
\frac{\zeta}{(\varepsilon_{0} + P_{0})\tau_{R}} 
\simeq 
\frac{5}{432} M(T)^{2}  \left[\frac{\partial}{\partial T}\left(\frac{M(T)}{T}\right) \right]^{2} 
+
\left\{\frac{5}{2592}\left[T \frac{\partial}{\partial T}\left(\frac{M(T)}{T}\right)\right]^{3} - \frac{5}{72} \left[T\frac{\partial}{\partial T}\left(\frac{M(T)}{T}\right)\right]^{2}\right\} \left(\frac{M(T)}{T}\right)^{3},
\\
&
\frac{\delta_{\Pi \Pi}}{\tau_{\Pi}}
\simeq
-\frac{5}{2} \frac{T^{2}}{M(T)} \frac{\partial}{\partial T}\left(\frac{M(T)}{T}\right)
- 
\frac{47}{6}
+
\left(3 + \frac{11 \pi}{10}\right)
T\frac{\partial}{\partial T}\left(\frac{M(T)}{T}\right)
-
\frac{35}{24}
\left[T\frac{\partial}{\partial T}\left(\frac{M(T)}{T}\right) \right]^{2}
-
\frac{5}{2} \frac{T\frac{\partial^{2}}{\partial T^{2}}\left(\frac{M(T)}{T}\right)}{\frac{\partial}{\partial T}\left(\frac{M(T)}{T}\right)}
,
\\
&
\frac{\lambda_{\Pi \pi}}{\tau_{\Pi}} \simeq 
\frac{2}{3} 
+
\frac{5}{36} M(T)  \frac{\partial}{\partial T}\left(\frac{M(T)}{T}\right),
\end{align}    
\end{subequations}
where we assumed that for the relaxation time $\tau_{R}$ does not depend on temperature for these expansions. The same shall be assumed when calculating the asymptotic expansion below. 
In the equation of motion for the shear-stress tensor, only $\lambda_{\pi \Pi}/\tau_{\pi}$ changes non-trivially with the redefinition of the hydrodynamic frame. Nevertheless, for the sake of convenience, we display the asymptotic behavior of all transport coefficients related to the shear-stress tensor equation of motion below, 
\begin{equation}
\label{eq:shear-coeffs-QPM-landau-m=0}
\begin{aligned}
&
\frac{\eta}{\tau_{\pi} (\varepsilon_{0} + P_{0})} \simeq \frac{1}{5} - \frac{1}{60} \left(\frac{M(T)}{T}\right)^{2},
\\
&
\frac{\delta_{\pi \pi}}{\tau_{\pi}}\simeq \frac{4}{3} 
-
\frac{1}{36} M(T)  \frac{\partial}{\partial T}\left(\frac{M(T)}{T}\right),
\\
&
\frac{\tau_{\pi \pi}}{\tau_{\pi}} 
\simeq
\frac{10}{7} + \frac{1}{21}\left(\frac{M(T)}{T}\right)^{2} ,
\\
&
\frac{\lambda_{\pi \Pi}}{\tau_{\pi}}
\simeq
\frac{96}{5}
\left[M(T)\frac{\partial}{\partial T}\left(\frac{M(T)}{T}\right) \right]^{-1}.
\end{aligned}    
\end{equation}

Keeping only the leading contribution of $\mathcal{O}(M(T)/T)$, we can derive simple thermodynamic expressions for most of the transport coefficients
\begin{equation}
\label{eq:coeffs-QPM-landau-pocket}
\begin{aligned}
&
\frac{\tau_{\Pi}}{\tau_{\pi}} \simeq  3 \left(\frac{1}{3}-c_{s}^{2}\right),
\quad
\frac{\zeta}{(\varepsilon_{0} + P_{0})\tau_{\Pi}} \simeq 
5\left(\frac{1}{3}-c_{s}^{2}\right),
\quad
\frac{\lambda_{\Pi \pi}}{\tau_{\Pi}} \simeq 
\frac{2}{3},
\\
&
\frac{\eta}{\tau_{\pi} (\varepsilon_{0} + P_{0})} \simeq \frac{1}{5},
\quad
\frac{\delta_{\pi \pi}}{\tau_{\pi}}\simeq \frac{4}{3},
\quad
\frac{\tau_{\pi \pi}}{\tau_{\pi}} 
\simeq
\frac{10}{7},
\quad
\frac{\lambda_{\pi \Pi}}{\tau_{\pi}}
\simeq
-\frac{3456}{5}\left(\frac{1}{3}-c_{s}^{2}\right)^{-1}.
\end{aligned}    
\end{equation}
Similar relations were derived for the HRG model, see Eqs.~\eqref{eq:bulk-coeffs-high-T-HRG}, \eqref{eq:pocket-form-HRG}, and \eqref{eq:high-T-shear-HRG}. We further note that similar asymptotic expressions, derived for a single component gas \cite{Denicol:2014vaa}, are often used in fluid-dynamical simulations of heavy-ion collisions \cite{Ryu:2015vwa,Paquet:2015lta,Ryu:2017qzn}. We note that the introduction of a temperature dependence mass to recover QCD thermodynamic modified qualitatively the asymptotic expressions for most of these transport coefficients. For instance, $\zeta/[\tau_\Pi(\varepsilon_0+P_0)]$ no longer behaves as $\sim 15 (1/3 - c_s^2)^2$, displaying a dependence usually associated to holography calculations, $\sim (1/3 - c_s^2)$ \cite{Finazzo:2014cna,Kanitscheider:2009as}. Finally, we remark that the transport coefficient $\delta_{\Pi \Pi}/\tau_{\Pi}$ could not be trivially expressed solely in terms of the conformality violation, even in the asymptotic regime. In order to so, we must assume a specific dependence for the thermal mass.

\section{Transport coefficient comparison: QPM, HRG and MUSIC}
\label{sec:compar-HRG-QPM}

\begin{figure}[!h]
    \centering
\begin{subfigure}{0.5\textwidth}
    \includegraphics[scale=0.30]{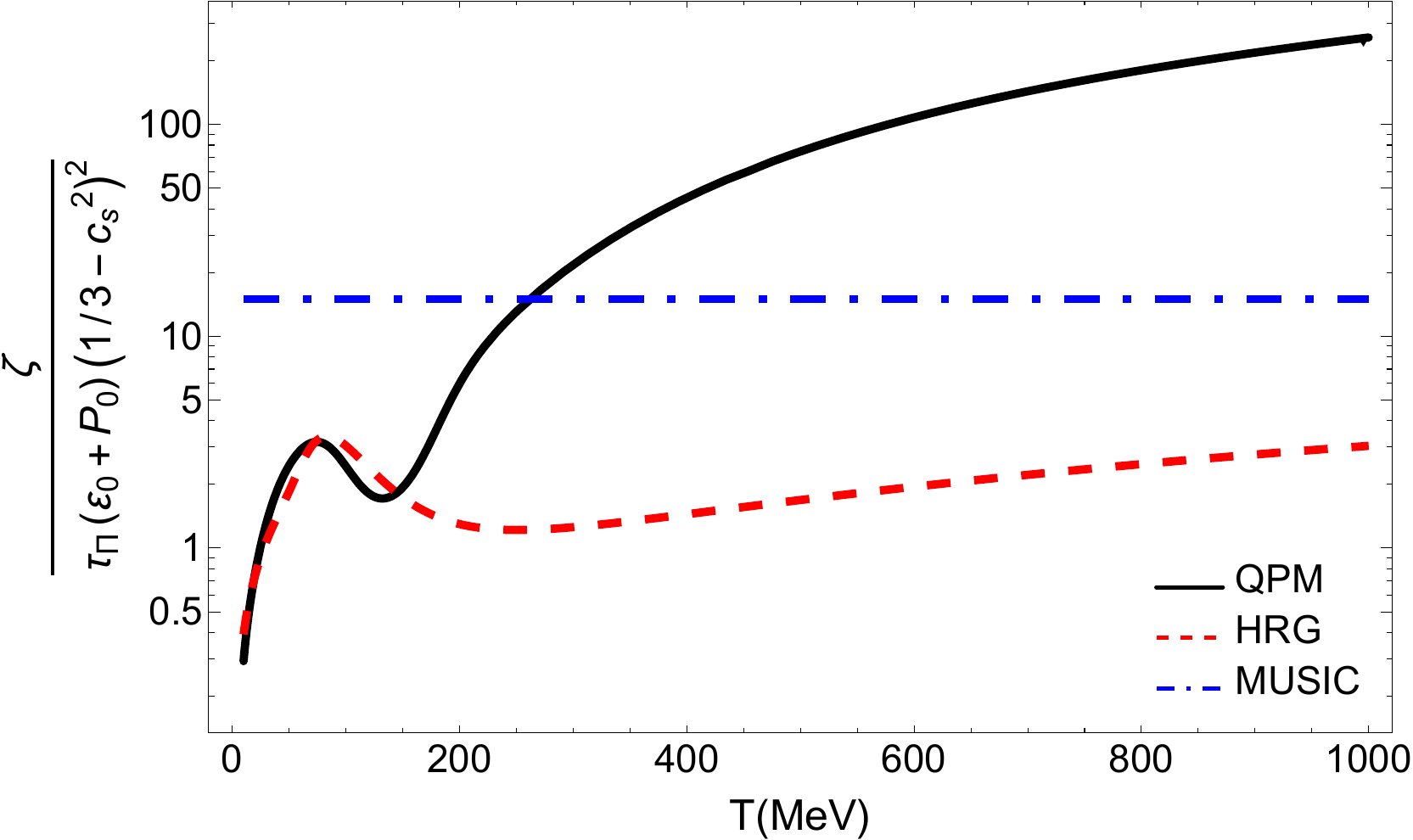}
    \caption{Bulk viscosity normalized with its relaxation time.}
    \label{fig:bulk-1}    
\end{subfigure}\hfil
\begin{subfigure}{0.5\textwidth}    \includegraphics[scale=0.30]{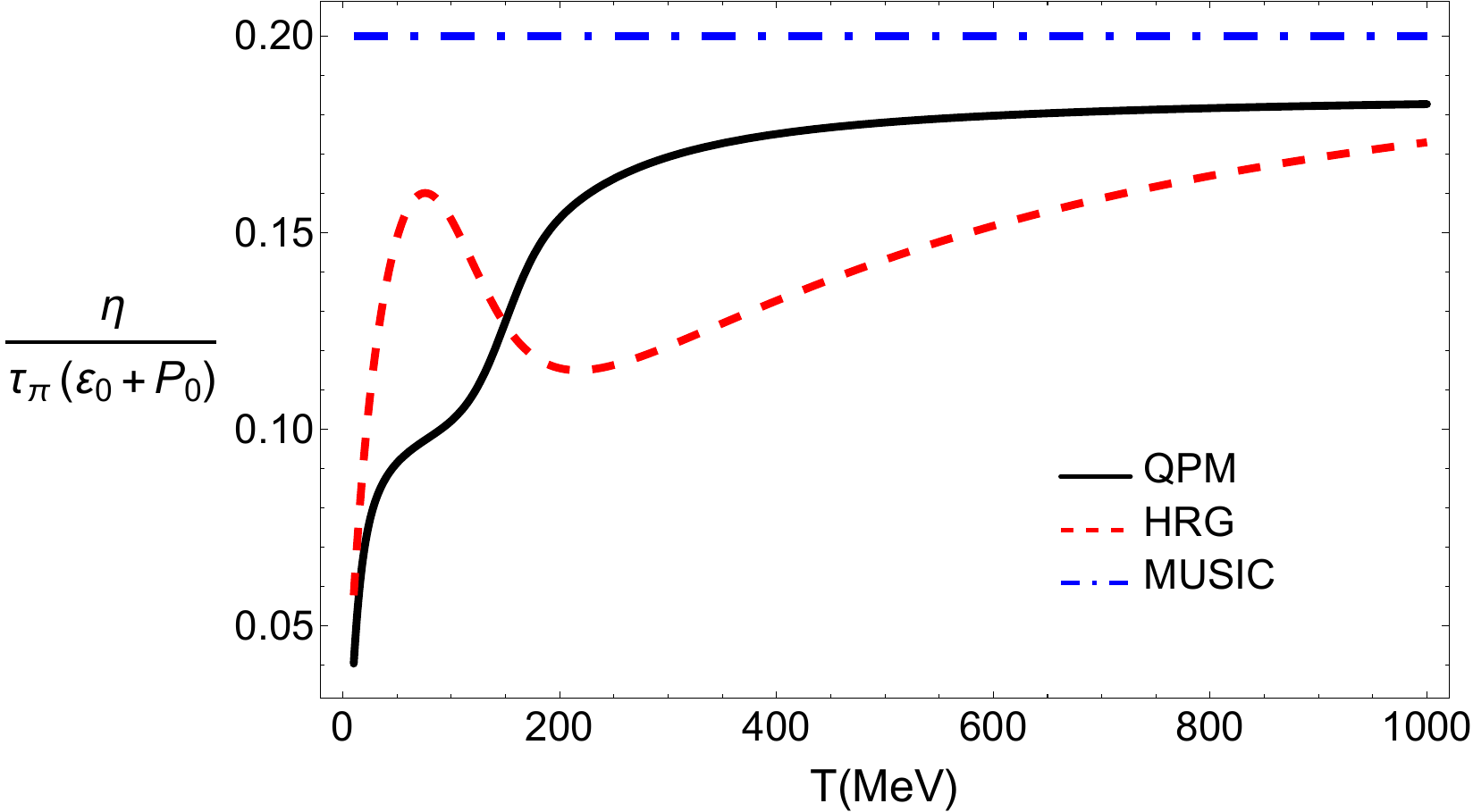}
    \caption{Normalized shear viscosity.}
    \label{fig:shear}
\end{subfigure}\hfil
\caption{Comparison between first order transport coefficients of the QPM, HRG and the ones employed in the MUSIC code as a function of temperature. (a) Bulk viscosity normalized with $\tau_{\Pi}$, enthapy density and conformal violation of the speed of sound. (b) Shear viscosity normalized with $\tau_{\pi}$ and enthapy density.}
\label{fig:1st-order-coeffs}
\end{figure}

\begin{figure}[!h]
    \centering
\begin{subfigure}{0.5\textwidth}
    \includegraphics[scale=0.25]{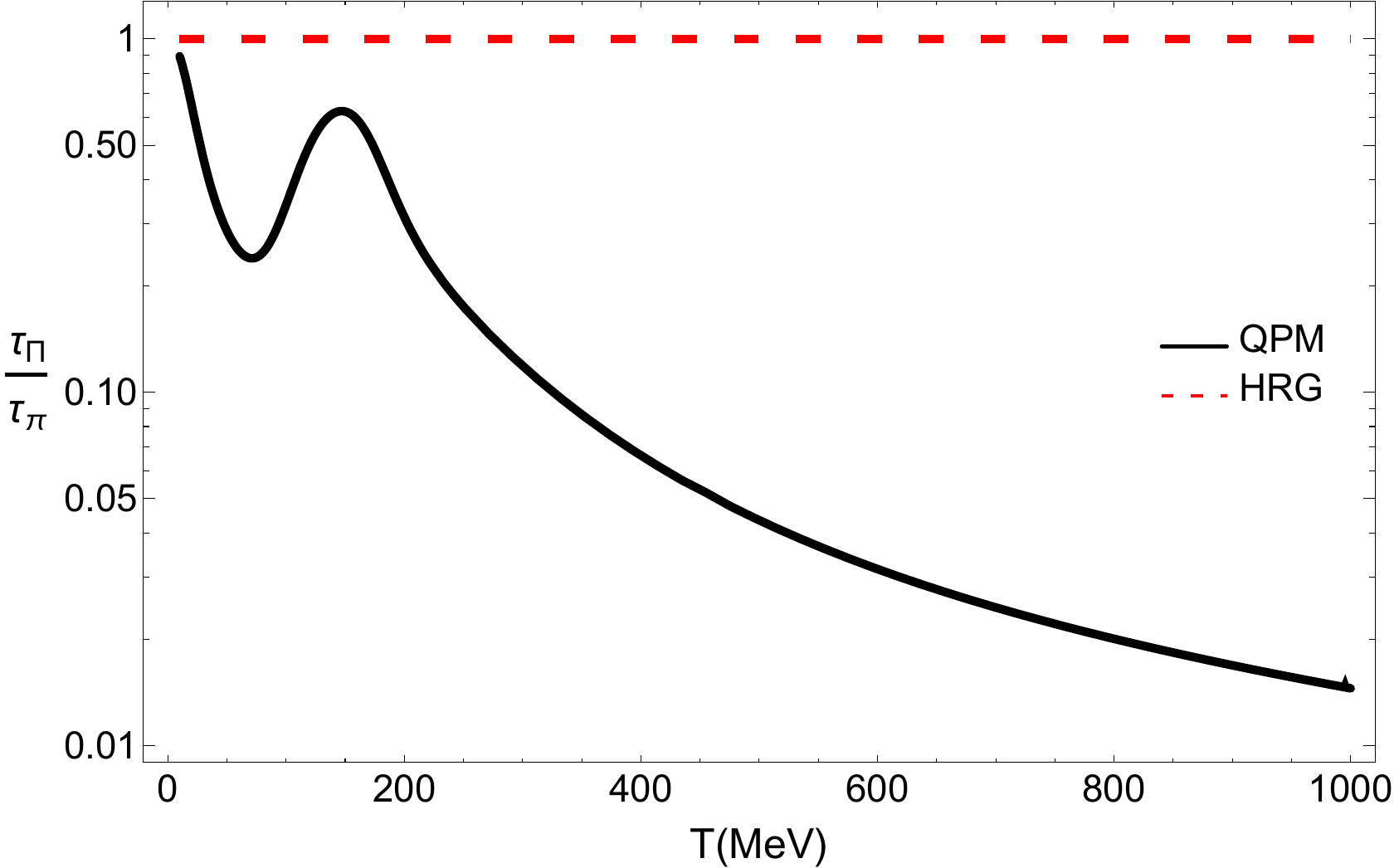}
    \caption{Bulk relaxation time.}
    \label{fig:t_PI}    
\end{subfigure}\hfil
\begin{subfigure}{0.5\textwidth}
    \includegraphics[scale=0.28]{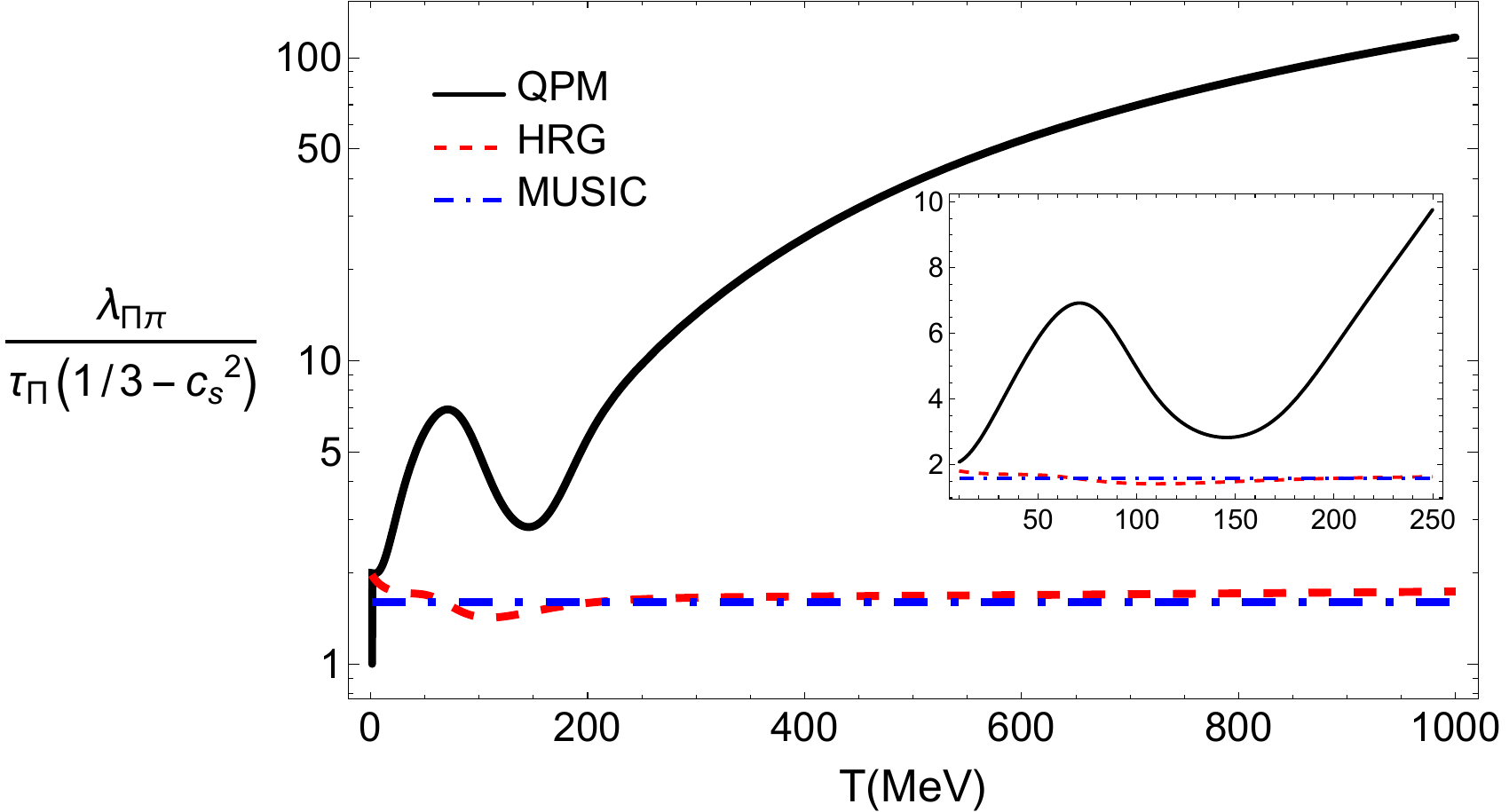}
    \caption{Bulk to shear coupling.}
    \label{fig:lam_PI-pi}    
\end{subfigure}\hfil
\begin{subfigure}{0.4\textwidth}
    \includegraphics[scale=0.27]{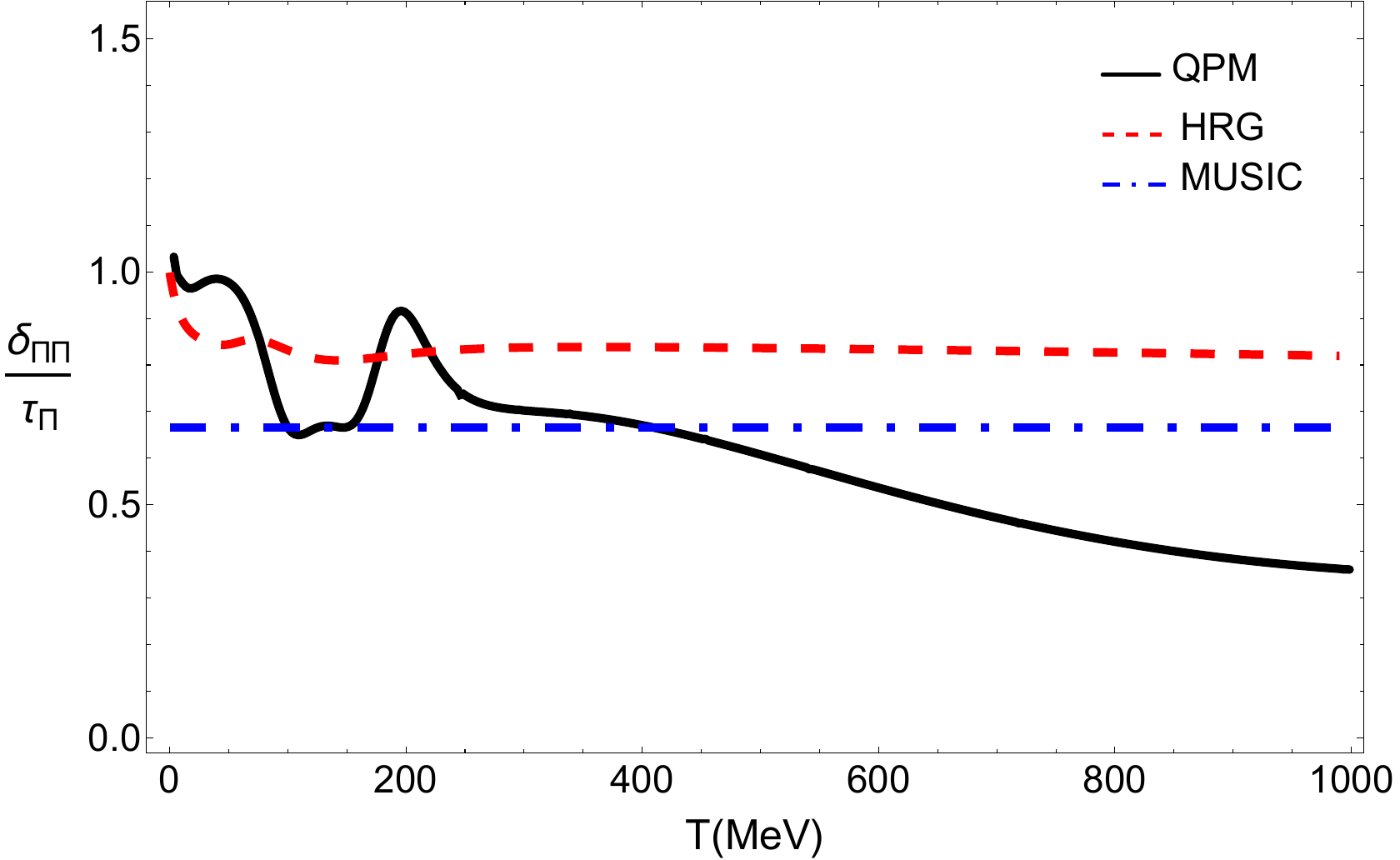}
    \caption{Bulk to bulk and expansion coefficient coupling.}
    \label{fig:del-PI-PI}    
\end{subfigure}\hfil
\caption{Comparison between second order transport coefficients in the bulk viscous pressure equation of motion for the QPM, HRG and the ones employed in the
MUSIC code as a function of temperature.}
\label{fig:2nd-order-coeffs-bulk}
\end{figure}

\begin{figure}[!h]
    \centering
\begin{subfigure}{0.5\textwidth}
    \includegraphics[scale=0.30]{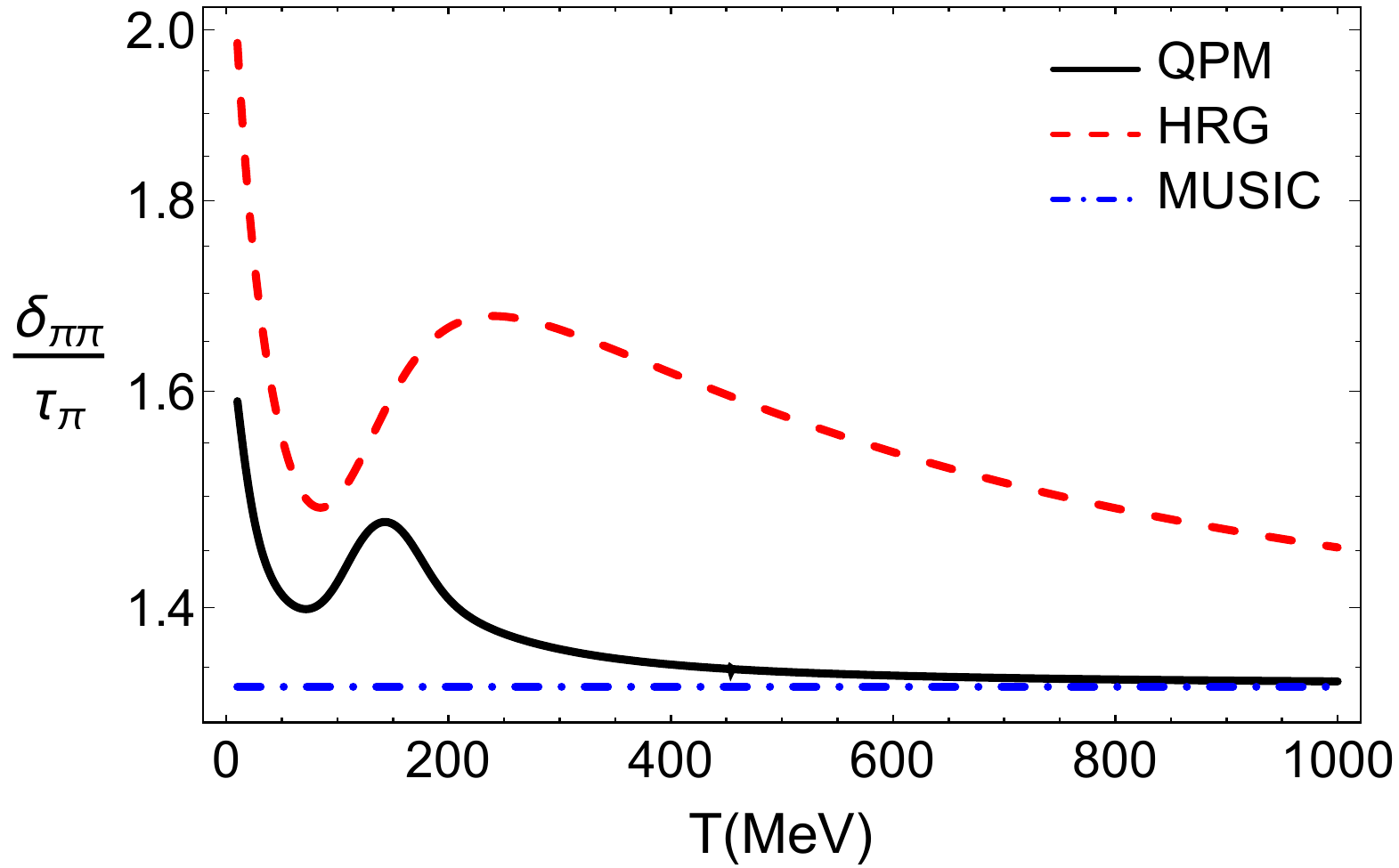}
    \caption{Shear-stress to expansion coefficient coupling.}
    \label{fig:del_pi-pi}    
\end{subfigure}\hfil
\begin{subfigure}{0.5\textwidth}
    \includegraphics[scale=0.30]{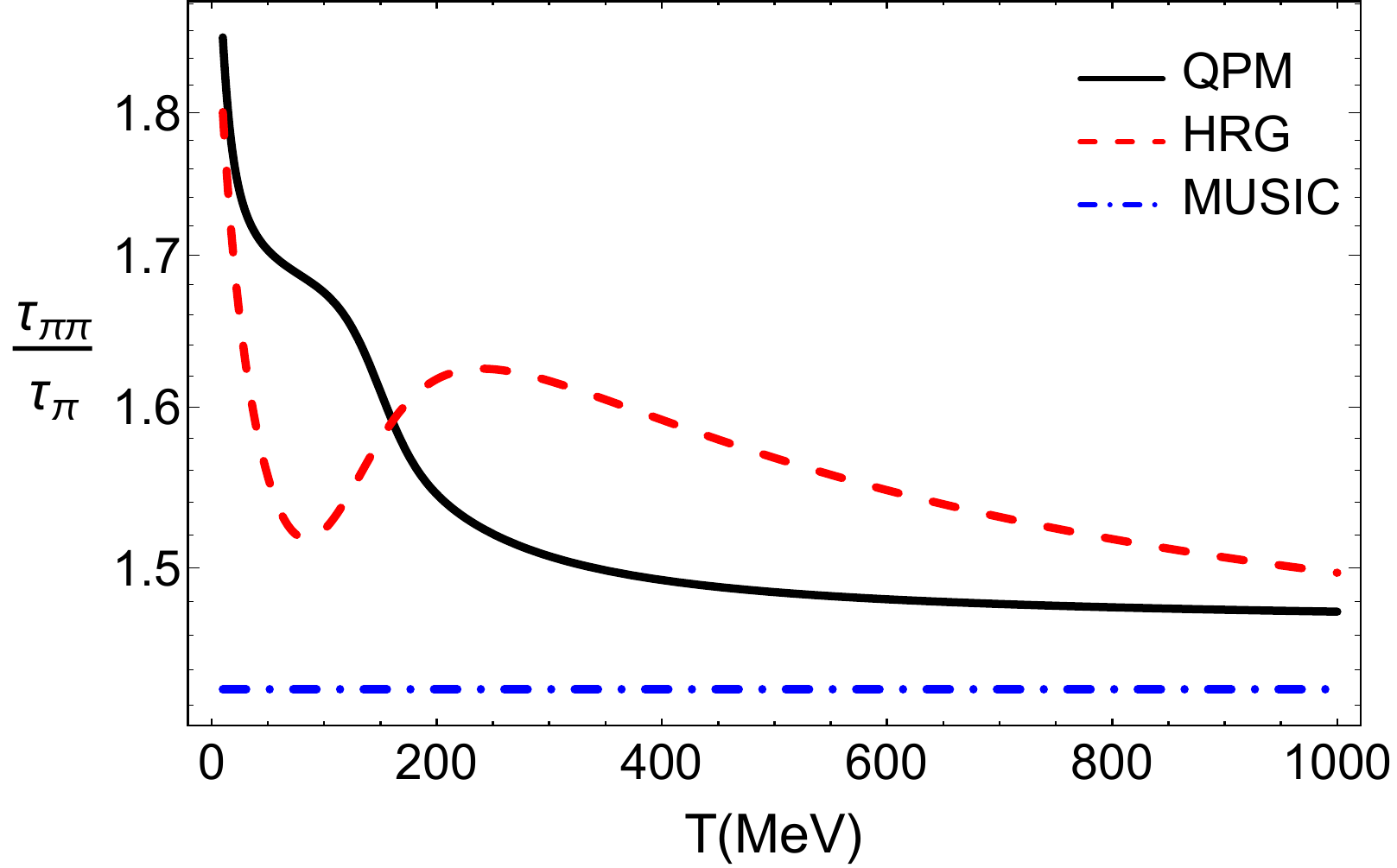}
    \caption{Shear-stress to shear-stress and shear tensor coupling}
    \label{fig:tau_pi-pi}    
\end{subfigure}\hfil
\\
\begin{subfigure}{0.5\textwidth}    \includegraphics[scale=0.30]{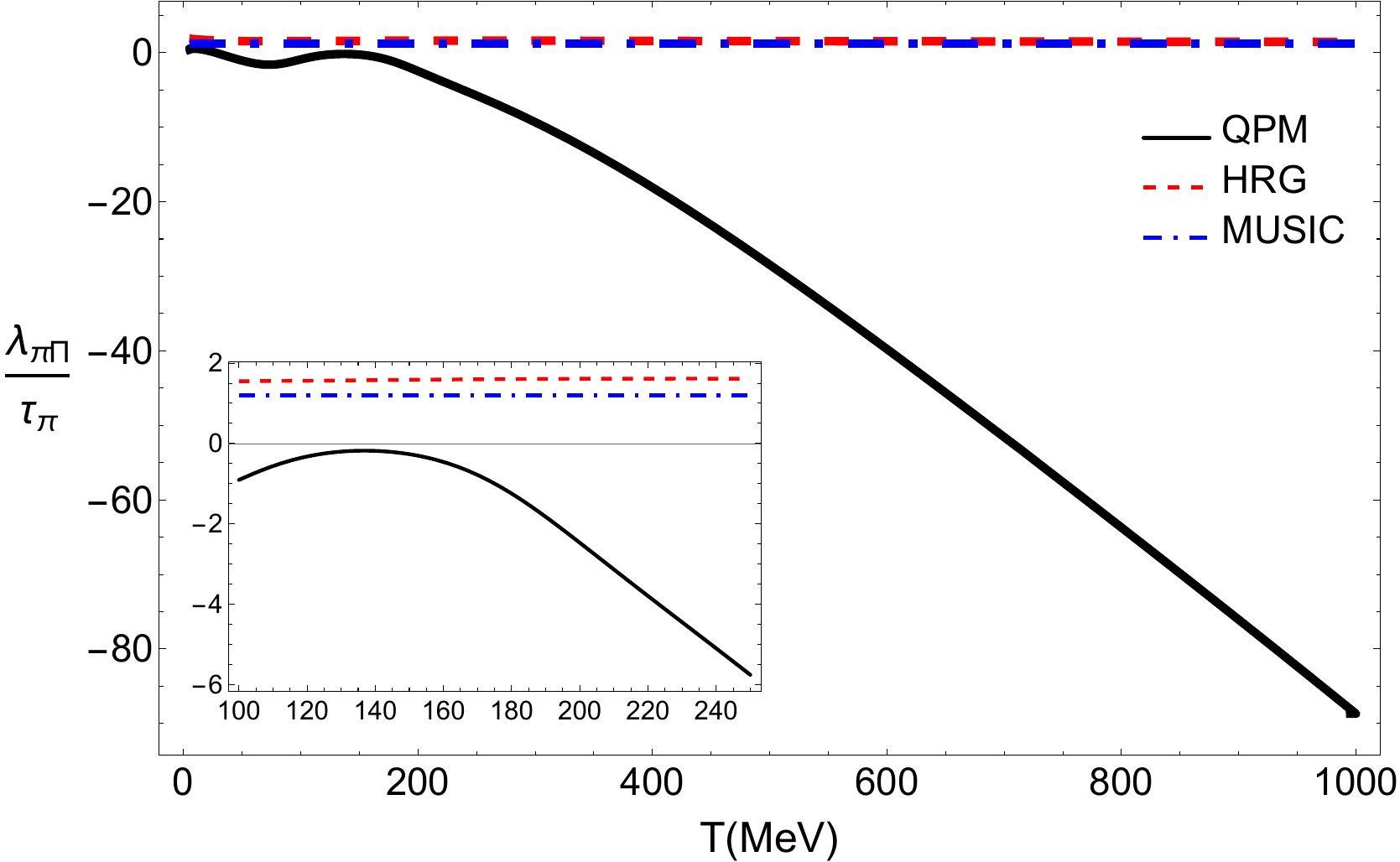}
    \caption{Shear to bulk and shear tensor coupling. }
    \label{fig:lam-pi-Pi}
\end{subfigure}\hfil  
\caption{Comparison between second order transport coefficients in the shear-stress tensor equation of motion for the QPM, HRG and the ones employed in the
MUSIC code as a function of temperature.}
\label{fig:2nd-order-coeffs-shear}
\end{figure}

Now that the transport coefficients of both the HRG and QPM models are expressed in the Landau frame, we can compare them. In Figs.~\ref{fig:1st-order-coeffs}--\ref{fig:2nd-order-coeffs-shear}, we display the quasi-particle model coefficients in comparison to hadron-resonance gas model (with UrQMD particle content) and also the values employed in the MUSIC simulation code \cite{Paquet:2015lta,Schenke:2010nt,Schenke:2010rr}. First, in Figs.~\ref{fig:1st-order-coeffs}, we display the ratios $\zeta/[\tau_\Pi(\varepsilon_0+P_0)(1/3-c_s^2)^2]$ (left panel) and $\eta/[\tau_\pi(\varepsilon_0+P_0)]$ (right panel). Figure \ref{fig:bulk-1}  shows that the QPM and the HRG models agree with each other until $T \simeq 150$ MeV, but both are significantly below the value employed in MUSIC. At larger temperatures, both the HRG and QPM models grow monotonically. The QPM grows over the MUSIC constant value, since this transport coefficient behaves asymptotically as\footnote{This is in contrast to the $\sim 16.91 (1/3-c_s^2)^2$ asymptotic behavior of the HRG model, as discussed in Subsec.~\ref{sec:bulk-pres-HRG}, and the $15 (1/3-c_s^2)^2$ parametrization adopted in the MUSIC simulation code} $\sim 5(1/3-c_s^2)$. Figure \ref{fig:shear} shows that the QPM and the HRG models do not agree with each other unless we are in the high temperature limit. In the latter case, all models, including the parametrization used in MUSIC, behave as $\eta/[\tau_\pi(\varepsilon_0+P_0)] \sim 0.2$. 

In Figs.~\ref{fig:2nd-order-coeffs-bulk} and \ref{fig:2nd-order-coeffs-shear} we plot the second-order transport coefficients. We note that the dependence on the free parameter $\tau_R$ of the transport coefficients $\delta_{\Pi\Pi}$ and $\lambda_{\pi\Pi}$ cannot be removed by simply normalizing these coefficients by their corresponding relaxation times, as is the case for all the other transport coefficients. This happens because these coefficients depend also on derivatives of the relaxation time on the temperature (see Eqs.~\eqref{eq:del-Pi-Pi-after-con} and \eqref{eq:lam-pi-Pi-after-con}). When plotting these quantities, we always assume a constant (temperature-independent) relaxation time. 

In Fig.~\ref{fig:2nd-order-coeffs-bulk}, we display the second-order transport coefficients related to the bulk viscous pressure equation of motion. In Fig.~\ref{fig:t_PI}, it is seen that the bulk relaxation time of the QPM 
is always smaller than the shear relaxation time, while in the HRG model they are identical \footnote{This happens because of our assumption of an momentum-independent relaxation time in the RTA model.}. In the QPM, the bulk relaxation time goes to zero at very large temperature -- this does not lead to issues related to acausality since the bulk viscosity (normalized by the enthalpy) decreases even faster with increasing temperature. We further note that in QP displays a peak around the crossover transition region, $T \simeq 200$ MeV. In Fig.~\ref{fig:lam_PI-pi} we show $\lambda_{\Pi \pi}/[\tau_{\Pi} (1/3-c_{s}^{2})]$ as a function of temperature. Our results show a good quantitative agreement between the HRG and MUSIC values for this transport coefficient. The QPM model, on the other hand, displays significantly larger values for this quantity for almost all temperatures. This emerges from the fact that $\lambda_{\Pi \pi}/\tau_{\Pi} \simeq 2/3$, as estimated in Eq.~\eqref{eq:coeffs-QPM-landau-pocket}, and thus all temperature dependence emerges due to the normalization factor, $1/3 - c_{s}^{2}$. For the coefficient $\delta_{\Pi \Pi}/\tau_{\Pi}$, displayed in Fig.~\eqref{fig:del-PI-PI}, the HRG model leads to an almost constant value $\sim 0.8$ for most temperatures, whereas the QPM calculations lead to values that vary significantly with the temperature.

In Fig.~\ref{fig:2nd-order-coeffs-shear}, we display second-order transport coefficients appearing in the equations of motion for the shear-stress tensor. In Fig.\ \ref{fig:del_pi-pi} we plot $\delta_{\pi \pi}/\tau_{\pi}$ as a function of temperature and see a qualitatively similar behavior for the QPM and HRG calculations, with the former converging faster to the values used in MUSIC. A similar behavior is observed for $\tau_{\pi \pi}/\tau_{\pi}$ in Fig.~\ref{fig:tau_pi-pi}, with the important differences that the HRG and QPM models cross around $T=200$ MeV and that the QPM results do not converge to the values used in MUSIC. Finally, for the coefficient $\lambda_{\pi \Pi}/\tau_{\pi}$ shown in Fig.\ \ref{fig:lam-pi-Pi}, the HRG model converges slowly to a value compatible to the one of MUSIC, while the QPM calculation saturates to zero.

\begin{figure}[!h]
    \centering
\begin{subfigure}{0.5\textwidth}
    \includegraphics[scale=0.30]{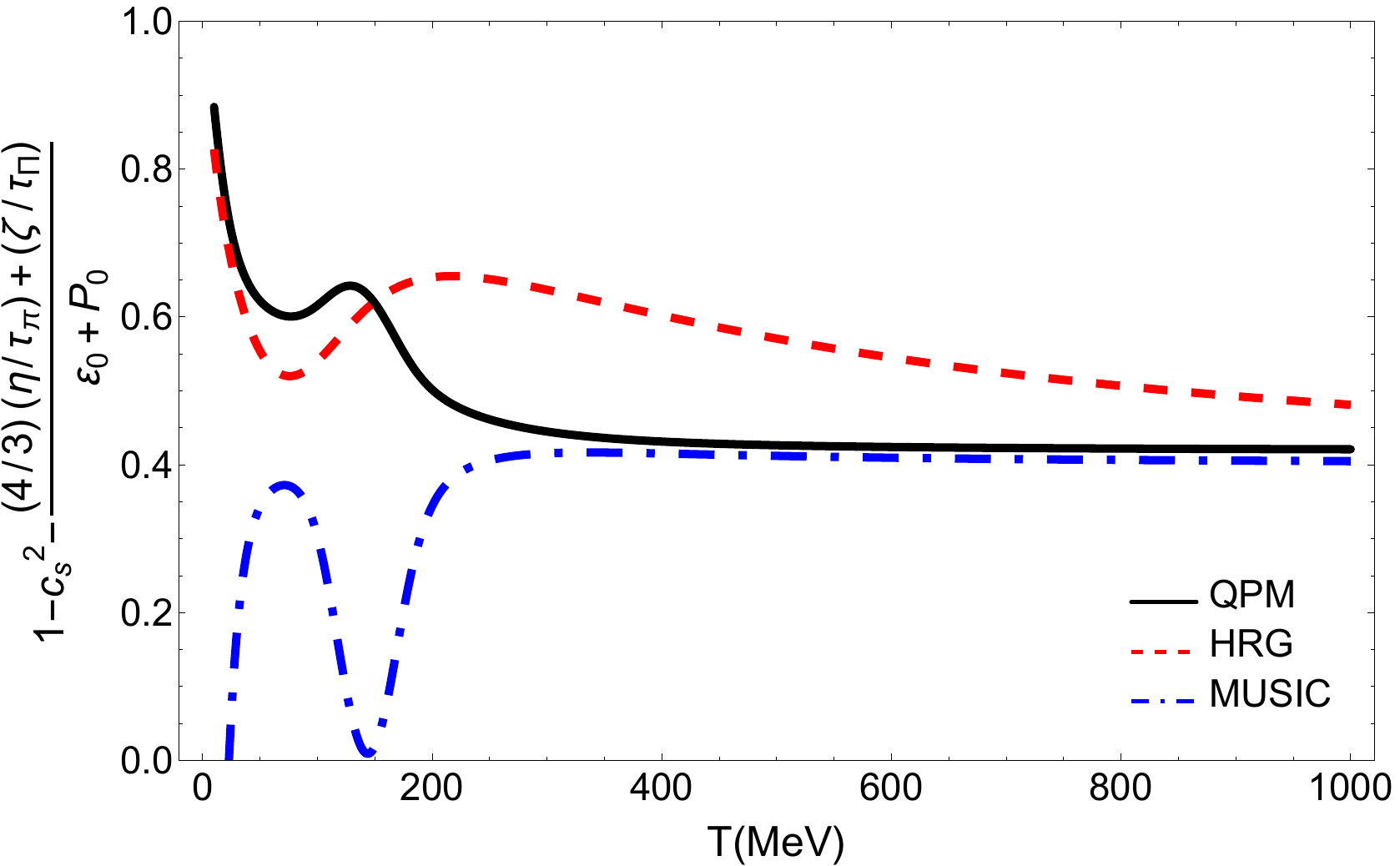}
\end{subfigure}\hfil
\caption{Comparison between linear causality constrain condition for the
QPM, HRG and the coefficients employed in the MUSIC code as a function of temperature.}
\label{fig:lin-caus}
\end{figure}

Finally, now we assess the linear causality and stability properties emerging from the fluid-dynamical theories derived. Indeed, in order for a transient hydrodynamic theory to be linearly causal, the first-order transport coefficients and their corresponding relaxation times must satisfy \cite{Olson:1990rzl,Hiscock:1983zz,Pu:2009fj}
\begin{equation}
\label{eq:lin-causality}
\begin{aligned}
&    
1- c_{s}^{2}
-
\frac{1}{\varepsilon_{0} + P_{0}} \left(\frac{4}{3} \frac{\eta}{\tau_{\pi}} +
\frac{\zeta}{\tau_{\Pi}}\right)
\geq
0
.
\end{aligned}    
\end{equation}
The above inequality arises from the requirement that the group velocity of the perturbations is smaller than the velocity of light \cite{Olson:1990rzl,Hiscock:1983zz,Pu:2009fj,Denicol:2021,Brito:2020nou}. In Fig.~\ref{fig:lin-caus}, we display the left-hand side of Eq.~\eqref{eq:lin-causality} as a function of temperature. For the curve corresponding to MUSIC, we use the same lattice equation of state employed to obtain the QPM results. At high temperatures, we see that all three models converge to the same constant asymptotic value that obeys the causality constraint. We note that the HRG model and the QPM approach this asymptotic value from above i.e., they are farther from violating the causality conditions in the cross over region. On the other hand, the transport coefficients employed in MUSIC approach the constant asymptotic value from below, i.e., they are closer to violating the causality constraint. As a matter of fact, around $T = 143$ MeV the left-hand side of \eqref{eq:lin-causality} yields 0.01, almost violating the linear causality condition.  As pointed out in Ref.\ \cite{Krupczak:2023jpa},
this almost violation of the \textit{linear} causality condition is what leads to the violation of causality\footnote{nonlinear causality conditions were derived in Ref.\ \cite{Bemfica:2020xym}.} in heavy-ion collision simulations. Given that a causal evolution is a central property of any relativistic dissipative hydrodynamic theory, we think that the QPM would be a good candidate to update the second-order transport coefficients in codes such as MUSIC.

\section{Conclusions}
\label{sec:concl}

In this work, we have derived the transport coefficients for transient hydrodynamics within kinetic theory for two systems: a hadron resonance gas model and a quasiparticle model with a thermal mass tuned to reproduce lattice QCD thermodynamics. In both cases, we derive transient hydrodynamics assuming the relaxation time approximation \cite{Rocha:2021zcw} to simplify the collision term. We find that both the QPM and HRG transport coefficients can be rather different from what is currently present in the MUSIC simulation code, which contains transport coefficients calculated assuming a single component gas in the high temperature limit \cite{Denicol:2014vaa}. 

In particular, we discussed the normalized bulk viscosity coefficient, $\zeta/[\tau_{\Pi}(\varepsilon_{0} + P_{0})(1/3-c_{s}^{2})^{2}]$. In HRG, we have found that this quantity can differ significantly from the result $\zeta/[\tau_{\Pi}(\varepsilon_{0} + P_{0})(1/3-c_{s}^{2})^{2}] \simeq 15$, computed in Ref.~\cite{Denicol:2014vaa} and used in MUSIC. In the asymptotically high temperature regime, we found similar values of $\zeta/[\tau_{\Pi}(\varepsilon_{0} + P_{0})(1/3-c_{s}^{2})^{2}] \simeq 16.91$ for UrQMD particle content, while for the temperature range usually appearing in heavy-ion collision simulations, it is seen that this quantity assumes values much smaller, $\zeta/[\tau_{\Pi}(\varepsilon_{0} + P_{0})(1/3-c_{s}^{2})^{2}] \simeq 3.04$. On the other hand, in the QPM, $\zeta/[\tau_{\Pi}(\varepsilon_{0} + P_{0})(1/3-c_{s}^{2})^{2}]$ does not saturate to a constant, increasing indefinitely with temperature. Indeed, this happens because this quantity is actually linearly proportional to the conformal violation of the speed of sound, $\zeta/[\tau_{\Pi}(\varepsilon_{0} + P_{0})] \simeq 5 (1/3-c_{s}^{2})$ -- a behavior qualitatively similar to the one found in strongly coupled plasma calculations within holography \cite{Huang:2010sa,Finazzo:2014cna,Kanitscheider:2009as}. Finally, we note that both the HRG model and the QPM provide values of $\zeta/[\tau_{\Pi}(\varepsilon_{0} + P_{0})(1/3-c_{s}^{2})^{2}]$, that are in reasonable agreement around the crossover region.  

Regarding the remaining transport coefficients, it is seen that the value of the normalized shear viscosity, $\eta/[\tau_{\pi}(\varepsilon_{0}+ P_{0})]$, is similar for the HRG and the QPM near the crossover region, but both are smaller than what is implemented in MUSIC. Besides, the bulk-to-shear coupling, $\lambda_{\Pi \pi}/[\tau_{\Pi} (1/3-c_{s}^{2})]$, possess similar values for HRG and MUSIC, but the QPM calculations lead to much larger values at high temperatures. In its turn, the shear-to-bulk coupling, $\lambda_{\pi \Pi}/\tau_{\pi}$, calculated in the QPM also behaves differently with respect to one calculated in the HRG model and the one used in MUSIC: the QPM calculation leads to a negative value for this transport coefficient, while the HRG model leads to a positive value. Finally, the transport coefficients, $\delta_{\Pi \Pi}/\tau_{\Pi}$, $\delta_{\pi \pi}/\tau_{\pi}$ and $\tau_{\pi \pi}/\tau_{\pi}$, are quantitatively similar in both models, even though they display a qualitatively different dependence with temperature. It was also demonstrated that the transport coefficients in both models satisfy with a greater margin the linear causality condition than the transport coefficients currently implemented in MUSIC. The latter almost violate the causality constrains in regions relevant to heavy-ion phenomenology.  In future studies, it would be relevant to assess the role of baryon chemical potential to the analysis and to assess the effect on heavy-ion observables and on the hydrodynamic evolution of the QGP of updating the transport coefficients in simulation codes such as MUSIC. 

\section*{Acknowledgements}

G.~S.~R. thanks M.~N.~Ferreira, M.~Singh, J.-F.~Paquet, J.~Noronha and J.~Noronha-Hostler for fruitful discussions. G.~S.~R. is supported by Vanderbilt University, in part by the U.S. Department of Energy, Office of Science under Award Number DE-SC-0024347, by Conselho Nacional de Desenvolvimento Científico e Tecnológico (CNPq), grant No. 142548/2019-7, and partly funded by Coordenação de Aperfeiçoamento de Pessoal de Nível Superior (CAPES) Finance code 001, award No. 88881.650299/2021-01. G.~S.~D.~also acknowledges CNPq as well as Fundação Carlos Chagas Filho de Amparo à Pesquisa do Estado do Rio de Janeiro (FAPERJ), grant No.~E-26/202.747/2018.

\appendix

\bibliographystyle{apsrev4-1}
\bibliography{liography}

\begin{thebibliography}{75}%
\makeatletter
\providecommand \@ifxundefined [1]{%
 \@ifx{#1\undefined}
}%
\providecommand \@ifnum [1]{%
 \ifnum #1\expandafter \@firstoftwo
 \else \expandafter \@secondoftwo
 \fi
}%
\providecommand \@ifx [1]{%
 \ifx #1\expandafter \@firstoftwo
 \else \expandafter \@secondoftwo
 \fi
}%
\providecommand \natexlab [1]{#1}%
\providecommand \enquote  [1]{``#1''}%
\providecommand \bibnamefont  [1]{#1}%
\providecommand \bibfnamefont [1]{#1}%
\providecommand \citenamefont [1]{#1}%
\providecommand \href@noop [0]{\@secondoftwo}%
\providecommand \href [0]{\begingroup \@sanitize@url \@href}%
\providecommand \@href[1]{\@@startlink{#1}\@@href}%
\providecommand \@@href[1]{\endgroup#1\@@endlink}%
\providecommand \@sanitize@url [0]{\catcode `\\12\catcode `\$12\catcode
  `\&12\catcode `\#12\catcode `\^12\catcode `\_12\catcode `\%12\relax}%
\providecommand \@@startlink[1]{}%
\providecommand \@@endlink[0]{}%
\providecommand \url  [0]{\begingroup\@sanitize@url \@url }%
\providecommand \@url [1]{\endgroup\@href {#1}{\urlprefix }}%
\providecommand \urlprefix  [0]{URL }%
\providecommand \Eprint [0]{\href }%
\providecommand \doibase [0]{http://dx.doi.org/}%
\providecommand \selectlanguage [0]{\@gobble}%
\providecommand \bibinfo  [0]{\@secondoftwo}%
\providecommand \bibfield  [0]{\@secondoftwo}%
\providecommand \translation [1]{[#1]}%
\providecommand \BibitemOpen [0]{}%
\providecommand \bibitemStop [0]{}%
\providecommand \bibitemNoStop [0]{.\EOS\space}%
\providecommand \EOS [0]{\spacefactor3000\relax}%
\providecommand \BibitemShut  [1]{\csname bibitem#1\endcsname}%
\let\auto@bib@innerbib\@empty
\bibitem [{\citenamefont {Gale}\ \emph {et~al.}(2013)\citenamefont {Gale},
  \citenamefont {Jeon},\ and\ \citenamefont {Schenke}}]{Gale:2013da}%
  \BibitemOpen
  \bibfield  {author} {\bibinfo {author} {\bibfnamefont {C.}~\bibnamefont
  {Gale}}, \bibinfo {author} {\bibfnamefont {S.}~\bibnamefont {Jeon}}, \ and\
  \bibinfo {author} {\bibfnamefont {B.}~\bibnamefont {Schenke}},\ }\href
  {\doibase 10.1142/S0217751X13400113} {\bibfield  {journal} {\bibinfo
  {journal} {Int. J. Mod. Phys. A}\ }\textbf {\bibinfo {volume} {28}},\
  \bibinfo {pages} {1340011} (\bibinfo {year} {2013})},\ \Eprint
  {http://arxiv.org/abs/1301.5893} {arXiv:1301.5893 [nucl-th]} \BibitemShut
  {NoStop}%
\bibitem [{\citenamefont {Heinz}\ and\ \citenamefont
  {Snellings}(2013)}]{Heinz:2013th}%
  \BibitemOpen
  \bibfield  {author} {\bibinfo {author} {\bibfnamefont {U.}~\bibnamefont
  {Heinz}}\ and\ \bibinfo {author} {\bibfnamefont {R.}~\bibnamefont
  {Snellings}},\ }\href {\doibase 10.1146/annurev-nucl-102212-170540}
  {\bibfield  {journal} {\bibinfo  {journal} {Ann. Rev. Nucl. Part. Sci.}\
  }\textbf {\bibinfo {volume} {63}},\ \bibinfo {pages} {123} (\bibinfo {year}
  {2013})},\ \Eprint {http://arxiv.org/abs/1301.2826} {arXiv:1301.2826
  [nucl-th]} \BibitemShut {NoStop}%
\bibitem [{\citenamefont {D.~de Souza}\ \emph {et~al.}(2016)\citenamefont
  {D.~de Souza}, \citenamefont {Koide},\ and\ \citenamefont
  {Kodama}}]{DerradideSouza:2015kpt}%
  \BibitemOpen
  \bibfield  {author} {\bibinfo {author} {\bibfnamefont {R.}~\bibnamefont
  {D.~de Souza}}, \bibinfo {author} {\bibfnamefont {T.}~\bibnamefont {Koide}},
  \ and\ \bibinfo {author} {\bibfnamefont {T.}~\bibnamefont {Kodama}},\ }\href
  {\doibase 10.1016/j.ppnp.2015.09.002} {\bibfield  {journal} {\bibinfo
  {journal} {Prog. Part. Nucl. Phys.}\ }\textbf {\bibinfo {volume} {86}},\
  \bibinfo {pages} {35} (\bibinfo {year} {2016})},\ \Eprint
  {http://arxiv.org/abs/1506.03863} {arXiv:1506.03863 [nucl-th]} \BibitemShut
  {NoStop}%
\bibitem [{\citenamefont {Florkowski}\ \emph {et~al.}(2018)\citenamefont
  {Florkowski}, \citenamefont {Heller},\ and\ \citenamefont
  {Spalinski}}]{Florkowski:2017olj}%
  \BibitemOpen
  \bibfield  {author} {\bibinfo {author} {\bibfnamefont {W.}~\bibnamefont
  {Florkowski}}, \bibinfo {author} {\bibfnamefont {M.~P.}\ \bibnamefont
  {Heller}}, \ and\ \bibinfo {author} {\bibfnamefont {M.}~\bibnamefont
  {Spalinski}},\ }\href {\doibase 10.1088/1361-6633/aaa091} {\bibfield
  {journal} {\bibinfo  {journal} {Rept. Prog. Phys.}\ }\textbf {\bibinfo
  {volume} {81}},\ \bibinfo {pages} {046001} (\bibinfo {year} {2018})},\
  \Eprint {http://arxiv.org/abs/1707.02282} {arXiv:1707.02282 [hep-ph]}
  \BibitemShut {NoStop}%
\bibitem [{\citenamefont {Rocha}\ \emph {et~al.}(2023)\citenamefont {Rocha},
  \citenamefont {Wagner}, \citenamefont {Denicol}, \citenamefont {Noronha},\
  and\ \citenamefont {Rischke}}]{Rocha:2023ilf}%
  \BibitemOpen
  \bibfield  {author} {\bibinfo {author} {\bibfnamefont {G.~S.}\ \bibnamefont
  {Rocha}}, \bibinfo {author} {\bibfnamefont {D.}~\bibnamefont {Wagner}},
  \bibinfo {author} {\bibfnamefont {G.~S.}\ \bibnamefont {Denicol}}, \bibinfo
  {author} {\bibfnamefont {J.}~\bibnamefont {Noronha}}, \ and\ \bibinfo
  {author} {\bibfnamefont {D.~H.}\ \bibnamefont {Rischke}},\ }\href {\doibase
  10.3390/e26030189} {\  (\bibinfo {year} {2023}),\ 10.3390/e26030189},\
  \Eprint {http://arxiv.org/abs/2311.15063} {arXiv:2311.15063 [nucl-th]}
  \BibitemShut {NoStop}%
\bibitem [{\citenamefont {Struchtrup}(2005)}]{struchtrup2005macroscopic}%
  \BibitemOpen
  \bibfield  {author} {\bibinfo {author} {\bibfnamefont {H.}~\bibnamefont
  {Struchtrup}},\ }in\ \href@noop {} {\emph {\bibinfo {booktitle} {Macroscopic
  transport equations for rarefied gas flows}}}\ (\bibinfo  {publisher}
  {Springer},\ \bibinfo {year} {2005})\ pp.\ \bibinfo {pages}
  {145--160}\BibitemShut {NoStop}%
\bibitem [{\citenamefont {Batchelor}\ and\ \citenamefont
  {Batchelor}(1967)}]{batchelor1967introduction}%
  \BibitemOpen
  \bibfield  {author} {\bibinfo {author} {\bibfnamefont {C.~K.}\ \bibnamefont
  {Batchelor}}\ and\ \bibinfo {author} {\bibfnamefont {G.~K.}\ \bibnamefont
  {Batchelor}},\ }\href@noop {} {\emph {\bibinfo {title} {An introduction to
  fluid dynamics}}}\ (\bibinfo  {publisher} {Cambridge university press},\
  \bibinfo {year} {1967})\BibitemShut {NoStop}%
\bibitem [{\citenamefont {Pichon}(1965)}]{pichon:65etude}%
  \BibitemOpen
  \bibfield  {author} {\bibinfo {author} {\bibfnamefont {G.}~\bibnamefont
  {Pichon}},\ }in\ \href@noop {} {\emph {\bibinfo {booktitle} {Annales de l'IHP
  Physique th{\'e}orique}}},\ Vol.~\bibinfo {volume} {2}\ (\bibinfo {year}
  {1965})\ pp.\ \bibinfo {pages} {21--85}\BibitemShut {NoStop}%
\bibitem [{\citenamefont {Hiscock}\ and\ \citenamefont
  {Lindblom}(1985)}]{hiscock:85generic}%
  \BibitemOpen
  \bibfield  {author} {\bibinfo {author} {\bibfnamefont {W.~A.}\ \bibnamefont
  {Hiscock}}\ and\ \bibinfo {author} {\bibfnamefont {L.}~\bibnamefont
  {Lindblom}},\ }\href@noop {} {\bibfield  {journal} {\bibinfo  {journal}
  {Physical Review D}\ }\textbf {\bibinfo {volume} {31}},\ \bibinfo {pages}
  {725} (\bibinfo {year} {1985})}\BibitemShut {NoStop}%
\bibitem [{\citenamefont {Hiscock}\ and\ \citenamefont
  {Lindblom}(1983)}]{Hiscock:1983zz}%
  \BibitemOpen
  \bibfield  {author} {\bibinfo {author} {\bibfnamefont {W.~A.}\ \bibnamefont
  {Hiscock}}\ and\ \bibinfo {author} {\bibfnamefont {L.}~\bibnamefont
  {Lindblom}},\ }\href {\doibase 10.1016/0003-4916(83)90288-9} {\bibfield
  {journal} {\bibinfo  {journal} {Annals Phys.}\ }\textbf {\bibinfo {volume}
  {151}},\ \bibinfo {pages} {466} (\bibinfo {year} {1983})}\BibitemShut
  {NoStop}%
\bibitem [{\citenamefont {Paquet}(2023)}]{Paquet:2023rfd}%
  \BibitemOpen
  \bibfield  {author} {\bibinfo {author} {\bibfnamefont {J.-F.}\ \bibnamefont
  {Paquet}},\ }\href@noop {} {\  (\bibinfo {year} {2023})},\ \Eprint
  {http://arxiv.org/abs/2310.17618} {arXiv:2310.17618 [nucl-th]} \BibitemShut
  {NoStop}%
\bibitem [{\citenamefont {Israel}(1979)}]{israel1979jm}%
  \BibitemOpen
  \bibfield  {author} {\bibinfo {author} {\bibfnamefont {W.}~\bibnamefont
  {Israel}},\ }in\ \href@noop {} {\emph {\bibinfo {booktitle} {Roy. Soc. Lond.
  A}}},\ Vol.\ \bibinfo {volume} {365}\ (\bibinfo {year} {1979})\ p.~\bibinfo
  {pages} {43}\BibitemShut {NoStop}%
\bibitem [{\citenamefont {Israel}\ and\ \citenamefont
  {Stewart}(1979)}]{Israel:1979wp}%
  \BibitemOpen
  \bibfield  {author} {\bibinfo {author} {\bibfnamefont {W.}~\bibnamefont
  {Israel}}\ and\ \bibinfo {author} {\bibfnamefont {J.~M.}\ \bibnamefont
  {Stewart}},\ }\href {\doibase 10.1016/0003-4916(79)90130-1} {\bibfield
  {journal} {\bibinfo  {journal} {Annals Phys.}\ }\textbf {\bibinfo {volume}
  {118}},\ \bibinfo {pages} {341} (\bibinfo {year} {1979})}\BibitemShut
  {NoStop}%
\bibitem [{\citenamefont {Wagner}\ and\ \citenamefont
  {Gavassino}(2023)}]{Wagner:2023jgq}%
  \BibitemOpen
  \bibfield  {author} {\bibinfo {author} {\bibfnamefont {D.}~\bibnamefont
  {Wagner}}\ and\ \bibinfo {author} {\bibfnamefont {L.}~\bibnamefont
  {Gavassino}},\ }\href@noop {} {\  (\bibinfo {year} {2023})},\ \Eprint
  {http://arxiv.org/abs/2309.14828} {arXiv:2309.14828 [nucl-th]} \BibitemShut
  {NoStop}%
\bibitem [{\citenamefont {Denicol}\ and\ \citenamefont
  {Rischke}(2021)}]{Denicol:2021}%
  \BibitemOpen
  \bibfield  {author} {\bibinfo {author} {\bibfnamefont {G.~S.}\ \bibnamefont
  {Denicol}}\ and\ \bibinfo {author} {\bibfnamefont {D.~H.}\ \bibnamefont
  {Rischke}},\ }\href@noop {} {\emph {\bibinfo {title} {Microscopic Foundations
  of Relativistic Fluid Dynamics}}}\ (\bibinfo  {publisher} {Springer},\
  \bibinfo {year} {2021})\BibitemShut {NoStop}%
\bibitem [{\citenamefont {Pu}\ \emph {et~al.}(2010)\citenamefont {Pu},
  \citenamefont {Koide},\ and\ \citenamefont {Rischke}}]{Pu:2009fj}%
  \BibitemOpen
  \bibfield  {author} {\bibinfo {author} {\bibfnamefont {S.}~\bibnamefont
  {Pu}}, \bibinfo {author} {\bibfnamefont {T.}~\bibnamefont {Koide}}, \ and\
  \bibinfo {author} {\bibfnamefont {D.~H.}\ \bibnamefont {Rischke}},\ }\href
  {\doibase 10.1103/PhysRevD.81.114039} {\bibfield  {journal} {\bibinfo
  {journal} {Phys. Rev. D}\ }\textbf {\bibinfo {volume} {81}},\ \bibinfo
  {pages} {114039} (\bibinfo {year} {2010})},\ \Eprint
  {http://arxiv.org/abs/0907.3906} {arXiv:0907.3906 [hep-ph]} \BibitemShut
  {NoStop}%
\bibitem [{\citenamefont {Brito}\ and\ \citenamefont
  {Denicol}(2020)}]{Brito:2020nou}%
  \BibitemOpen
  \bibfield  {author} {\bibinfo {author} {\bibfnamefont {C.~V.}\ \bibnamefont
  {Brito}}\ and\ \bibinfo {author} {\bibfnamefont {G.~S.}\ \bibnamefont
  {Denicol}},\ }\href {\doibase 10.1103/PhysRevD.102.116009} {\bibfield
  {journal} {\bibinfo  {journal} {Phys. Rev. D}\ }\textbf {\bibinfo {volume}
  {102}},\ \bibinfo {pages} {116009} (\bibinfo {year} {2020})},\ \Eprint
  {http://arxiv.org/abs/2007.16141} {arXiv:2007.16141 [nucl-th]} \BibitemShut
  {NoStop}%
\bibitem [{\citenamefont {Pereira~de Brito}(2021)}]{PereiradeBrito:2021jul}%
  \BibitemOpen
  \bibfield  {author} {\bibinfo {author} {\bibfnamefont {C.~V.}\ \bibnamefont
  {Pereira~de Brito}},\ }\emph {\bibinfo {title} {{On the linear stability and
  causality of transient fluid dynamics}}},\ \href@noop {} {Master's thesis},\
  \bibinfo  {school} {Federal Fluminense University} (\bibinfo {year}
  {2021})\BibitemShut {NoStop}%
\bibitem [{\citenamefont {Sammet}\ \emph {et~al.}(2023)\citenamefont {Sammet},
  \citenamefont {Mayer},\ and\ \citenamefont {Rischke}}]{Sammet:2023bfo}%
  \BibitemOpen
  \bibfield  {author} {\bibinfo {author} {\bibfnamefont {J.}~\bibnamefont
  {Sammet}}, \bibinfo {author} {\bibfnamefont {M.}~\bibnamefont {Mayer}}, \
  and\ \bibinfo {author} {\bibfnamefont {D.~H.}\ \bibnamefont {Rischke}},\
  }\href@noop {} {\  (\bibinfo {year} {2023})},\ \Eprint
  {http://arxiv.org/abs/2302.01070} {arXiv:2302.01070 [hep-th]} \BibitemShut
  {NoStop}%
\bibitem [{\citenamefont {Gavassino}(2022)}]{Gavassino:2021owo}%
  \BibitemOpen
  \bibfield  {author} {\bibinfo {author} {\bibfnamefont {L.}~\bibnamefont
  {Gavassino}},\ }\href {\doibase 10.1103/PhysRevX.12.041001} {\bibfield
  {journal} {\bibinfo  {journal} {Phys. Rev. X}\ }\textbf {\bibinfo {volume}
  {12}},\ \bibinfo {pages} {041001} (\bibinfo {year} {2022})},\ \Eprint
  {http://arxiv.org/abs/2111.05254} {arXiv:2111.05254 [gr-qc]} \BibitemShut
  {NoStop}%
\bibitem [{\citenamefont {Gavassino}\ \emph {et~al.}(2023)\citenamefont
  {Gavassino}, \citenamefont {Disconzi},\ and\ \citenamefont
  {Noronha}}]{Gavassino:2023mad}%
  \BibitemOpen
  \bibfield  {author} {\bibinfo {author} {\bibfnamefont {L.}~\bibnamefont
  {Gavassino}}, \bibinfo {author} {\bibfnamefont {M.~M.}\ \bibnamefont
  {Disconzi}}, \ and\ \bibinfo {author} {\bibfnamefont {J.}~\bibnamefont
  {Noronha}},\ }\href@noop {} {\  (\bibinfo {year} {2023})},\ \Eprint
  {http://arxiv.org/abs/2307.05987} {arXiv:2307.05987 [hep-th]} \BibitemShut
  {NoStop}%
\bibitem [{\citenamefont {Gavassino}\ \emph {et~al.}(2022)\citenamefont
  {Gavassino}, \citenamefont {Antonelli},\ and\ \citenamefont
  {Haskell}}]{Gavassino:2021kjm}%
  \BibitemOpen
  \bibfield  {author} {\bibinfo {author} {\bibfnamefont {L.}~\bibnamefont
  {Gavassino}}, \bibinfo {author} {\bibfnamefont {M.}~\bibnamefont
  {Antonelli}}, \ and\ \bibinfo {author} {\bibfnamefont {B.}~\bibnamefont
  {Haskell}},\ }\href {\doibase 10.1103/PhysRevLett.128.010606} {\bibfield
  {journal} {\bibinfo  {journal} {Phys. Rev. Lett.}\ }\textbf {\bibinfo
  {volume} {128}},\ \bibinfo {pages} {010606} (\bibinfo {year} {2022})},\
  \Eprint {http://arxiv.org/abs/2105.14621} {arXiv:2105.14621 [gr-qc]}
  \BibitemShut {NoStop}%
\bibitem [{\citenamefont {Bemfica}\ \emph {et~al.}(2019)\citenamefont
  {Bemfica}, \citenamefont {Disconzi},\ and\ \citenamefont
  {Noronha}}]{Bemfica:2019cop}%
  \BibitemOpen
  \bibfield  {author} {\bibinfo {author} {\bibfnamefont {F.~S.}\ \bibnamefont
  {Bemfica}}, \bibinfo {author} {\bibfnamefont {M.~M.}\ \bibnamefont
  {Disconzi}}, \ and\ \bibinfo {author} {\bibfnamefont {J.}~\bibnamefont
  {Noronha}},\ }\href {\doibase 10.1103/PhysRevLett.122.221602} {\bibfield
  {journal} {\bibinfo  {journal} {Phys. Rev. Lett.}\ }\textbf {\bibinfo
  {volume} {122}},\ \bibinfo {pages} {221602} (\bibinfo {year} {2019})},\
  \Eprint {http://arxiv.org/abs/1901.06701} {arXiv:1901.06701 [gr-qc]}
  \BibitemShut {NoStop}%
\bibitem [{\citenamefont {Bemfica}\ \emph {et~al.}(2021)\citenamefont
  {Bemfica}, \citenamefont {Disconzi}, \citenamefont {Hoang}, \citenamefont
  {Noronha},\ and\ \citenamefont {Radosz}}]{Bemfica:2020xym}%
  \BibitemOpen
  \bibfield  {author} {\bibinfo {author} {\bibfnamefont {F.~S.}\ \bibnamefont
  {Bemfica}}, \bibinfo {author} {\bibfnamefont {M.~M.}\ \bibnamefont
  {Disconzi}}, \bibinfo {author} {\bibfnamefont {V.}~\bibnamefont {Hoang}},
  \bibinfo {author} {\bibfnamefont {J.}~\bibnamefont {Noronha}}, \ and\
  \bibinfo {author} {\bibfnamefont {M.}~\bibnamefont {Radosz}},\ }\href
  {\doibase 10.1103/PhysRevLett.126.222301} {\bibfield  {journal} {\bibinfo
  {journal} {Phys. Rev. Lett.}\ }\textbf {\bibinfo {volume} {126}},\ \bibinfo
  {pages} {222301} (\bibinfo {year} {2021})},\ \Eprint
  {http://arxiv.org/abs/2005.11632} {arXiv:2005.11632 [hep-th]} \BibitemShut
  {NoStop}%
\bibitem [{\citenamefont {Disconzi}(2023)}]{Disconzi:2023rtt}%
  \BibitemOpen
  \bibfield  {author} {\bibinfo {author} {\bibfnamefont {M.~M.}\ \bibnamefont
  {Disconzi}},\ }\href@noop {} {\  (\bibinfo {year} {2023})},\ \Eprint
  {http://arxiv.org/abs/2308.09844} {arXiv:2308.09844 [math.AP]} \BibitemShut
  {NoStop}%
\bibitem [{\citenamefont {Plumberg}\ \emph {et~al.}(2022)\citenamefont
  {Plumberg}, \citenamefont {Almaalol}, \citenamefont {Dore}, \citenamefont
  {Noronha},\ and\ \citenamefont {Noronha-Hostler}}]{Plumberg:2021bme}%
  \BibitemOpen
  \bibfield  {author} {\bibinfo {author} {\bibfnamefont {C.}~\bibnamefont
  {Plumberg}}, \bibinfo {author} {\bibfnamefont {D.}~\bibnamefont {Almaalol}},
  \bibinfo {author} {\bibfnamefont {T.}~\bibnamefont {Dore}}, \bibinfo {author}
  {\bibfnamefont {J.}~\bibnamefont {Noronha}}, \ and\ \bibinfo {author}
  {\bibfnamefont {J.}~\bibnamefont {Noronha-Hostler}},\ }\href {\doibase
  10.1103/PhysRevC.105.L061901} {\bibfield  {journal} {\bibinfo  {journal}
  {Phys. Rev. C}\ }\textbf {\bibinfo {volume} {105}},\ \bibinfo {pages}
  {L061901} (\bibinfo {year} {2022})},\ \Eprint
  {http://arxiv.org/abs/2103.15889} {arXiv:2103.15889 [nucl-th]} \BibitemShut
  {NoStop}%
\bibitem [{\citenamefont {Krupczak}\ \emph {et~al.}(2023)\citenamefont
  {Krupczak} \emph {et~al.}}]{Krupczak:2023jpa}%
  \BibitemOpen
  \bibfield  {author} {\bibinfo {author} {\bibfnamefont {R.}~\bibnamefont
  {Krupczak}} \emph {et~al.},\ }\href@noop {} {\  (\bibinfo {year} {2023})},\
  \Eprint {http://arxiv.org/abs/2311.02210} {arXiv:2311.02210 [nucl-th]}
  \BibitemShut {NoStop}%
\bibitem [{\citenamefont {Schenke}\ \emph {et~al.}(2010)\citenamefont
  {Schenke}, \citenamefont {Jeon},\ and\ \citenamefont
  {Gale}}]{Schenke:2010nt}%
  \BibitemOpen
  \bibfield  {author} {\bibinfo {author} {\bibfnamefont {B.}~\bibnamefont
  {Schenke}}, \bibinfo {author} {\bibfnamefont {S.}~\bibnamefont {Jeon}}, \
  and\ \bibinfo {author} {\bibfnamefont {C.}~\bibnamefont {Gale}},\ }\href
  {\doibase 10.1103/PhysRevC.82.014903} {\bibfield  {journal} {\bibinfo
  {journal} {Phys. Rev. C}\ }\textbf {\bibinfo {volume} {82}},\ \bibinfo
  {pages} {014903} (\bibinfo {year} {2010})},\ \Eprint
  {http://arxiv.org/abs/1004.1408} {arXiv:1004.1408 [hep-ph]} \BibitemShut
  {NoStop}%
\bibitem [{\citenamefont {Ryu}\ \emph {et~al.}(2015)\citenamefont {Ryu},
  \citenamefont {Paquet}, \citenamefont {Shen}, \citenamefont {Denicol},
  \citenamefont {Schenke}, \citenamefont {Jeon},\ and\ \citenamefont
  {Gale}}]{Ryu:2015vwa}%
  \BibitemOpen
  \bibfield  {author} {\bibinfo {author} {\bibfnamefont {S.}~\bibnamefont
  {Ryu}}, \bibinfo {author} {\bibfnamefont {J.~F.}\ \bibnamefont {Paquet}},
  \bibinfo {author} {\bibfnamefont {C.}~\bibnamefont {Shen}}, \bibinfo {author}
  {\bibfnamefont {G.~S.}\ \bibnamefont {Denicol}}, \bibinfo {author}
  {\bibfnamefont {B.}~\bibnamefont {Schenke}}, \bibinfo {author} {\bibfnamefont
  {S.}~\bibnamefont {Jeon}}, \ and\ \bibinfo {author} {\bibfnamefont
  {C.}~\bibnamefont {Gale}},\ }\href {\doibase 10.1103/PhysRevLett.115.132301}
  {\bibfield  {journal} {\bibinfo  {journal} {Phys. Rev. Lett.}\ }\textbf
  {\bibinfo {volume} {115}},\ \bibinfo {pages} {132301} (\bibinfo {year}
  {2015})},\ \Eprint {http://arxiv.org/abs/1502.01675} {arXiv:1502.01675
  [nucl-th]} \BibitemShut {NoStop}%
\bibitem [{\citenamefont {Paquet}\ \emph {et~al.}(2016)\citenamefont {Paquet},
  \citenamefont {Shen}, \citenamefont {Denicol}, \citenamefont {Luzum},
  \citenamefont {Schenke}, \citenamefont {Jeon},\ and\ \citenamefont
  {Gale}}]{Paquet:2015lta}%
  \BibitemOpen
  \bibfield  {author} {\bibinfo {author} {\bibfnamefont {J.-F.}\ \bibnamefont
  {Paquet}}, \bibinfo {author} {\bibfnamefont {C.}~\bibnamefont {Shen}},
  \bibinfo {author} {\bibfnamefont {G.~S.}\ \bibnamefont {Denicol}}, \bibinfo
  {author} {\bibfnamefont {M.}~\bibnamefont {Luzum}}, \bibinfo {author}
  {\bibfnamefont {B.}~\bibnamefont {Schenke}}, \bibinfo {author} {\bibfnamefont
  {S.}~\bibnamefont {Jeon}}, \ and\ \bibinfo {author} {\bibfnamefont
  {C.}~\bibnamefont {Gale}},\ }\href {\doibase 10.1103/PhysRevC.93.044906}
  {\bibfield  {journal} {\bibinfo  {journal} {Phys. Rev. C}\ }\textbf {\bibinfo
  {volume} {93}},\ \bibinfo {pages} {044906} (\bibinfo {year} {2016})},\
  \Eprint {http://arxiv.org/abs/1509.06738} {arXiv:1509.06738 [hep-ph]}
  \BibitemShut {NoStop}%
\bibitem [{\citenamefont {Ryu}\ \emph {et~al.}(2018)\citenamefont {Ryu},
  \citenamefont {Paquet}, \citenamefont {Shen}, \citenamefont {Denicol},
  \citenamefont {Schenke}, \citenamefont {Jeon},\ and\ \citenamefont
  {Gale}}]{Ryu:2017qzn}%
  \BibitemOpen
  \bibfield  {author} {\bibinfo {author} {\bibfnamefont {S.}~\bibnamefont
  {Ryu}}, \bibinfo {author} {\bibfnamefont {J.-F.}\ \bibnamefont {Paquet}},
  \bibinfo {author} {\bibfnamefont {C.}~\bibnamefont {Shen}}, \bibinfo {author}
  {\bibfnamefont {G.}~\bibnamefont {Denicol}}, \bibinfo {author} {\bibfnamefont
  {B.}~\bibnamefont {Schenke}}, \bibinfo {author} {\bibfnamefont
  {S.}~\bibnamefont {Jeon}}, \ and\ \bibinfo {author} {\bibfnamefont
  {C.}~\bibnamefont {Gale}},\ }\href {\doibase 10.1103/PhysRevC.97.034910}
  {\bibfield  {journal} {\bibinfo  {journal} {Phys. Rev. C}\ }\textbf {\bibinfo
  {volume} {97}},\ \bibinfo {pages} {034910} (\bibinfo {year} {2018})},\
  \Eprint {http://arxiv.org/abs/1704.04216} {arXiv:1704.04216 [nucl-th]}
  \BibitemShut {NoStop}%
\bibitem [{\citenamefont {Denicol}\ \emph
  {et~al.}(2014{\natexlab{a}})\citenamefont {Denicol}, \citenamefont {Jeon},\
  and\ \citenamefont {Gale}}]{Denicol:2014vaa}%
  \BibitemOpen
  \bibfield  {author} {\bibinfo {author} {\bibfnamefont {G.~S.}\ \bibnamefont
  {Denicol}}, \bibinfo {author} {\bibfnamefont {S.}~\bibnamefont {Jeon}}, \
  and\ \bibinfo {author} {\bibfnamefont {C.}~\bibnamefont {Gale}},\ }\href
  {\doibase 10.1103/PhysRevC.90.024912} {\bibfield  {journal} {\bibinfo
  {journal} {Phys. Rev. C}\ }\textbf {\bibinfo {volume} {90}},\ \bibinfo
  {pages} {024912} (\bibinfo {year} {2014}{\natexlab{a}})},\ \Eprint
  {http://arxiv.org/abs/1403.0962} {arXiv:1403.0962 [nucl-th]} \BibitemShut
  {NoStop}%
\bibitem [{\citenamefont {Hagedorn}(1965)}]{Hagedorn:1965st}%
  \BibitemOpen
  \bibfield  {author} {\bibinfo {author} {\bibfnamefont {R.}~\bibnamefont
  {Hagedorn}},\ }\href@noop {} {\bibfield  {journal} {\bibinfo  {journal}
  {Nuovo Cim. Suppl.}\ }\textbf {\bibinfo {volume} {3}},\ \bibinfo {pages}
  {147} (\bibinfo {year} {1965})}\BibitemShut {NoStop}%
\bibitem [{\citenamefont {Venugopalan}\ and\ \citenamefont
  {Prakash}(1992)}]{Venugopalan:1992hy}%
  \BibitemOpen
  \bibfield  {author} {\bibinfo {author} {\bibfnamefont {R.}~\bibnamefont
  {Venugopalan}}\ and\ \bibinfo {author} {\bibfnamefont {M.}~\bibnamefont
  {Prakash}},\ }\href {\doibase 10.1016/0375-9474(92)90005-5} {\bibfield
  {journal} {\bibinfo  {journal} {Nucl. Phys. A}\ }\textbf {\bibinfo {volume}
  {546}},\ \bibinfo {pages} {718} (\bibinfo {year} {1992})}\BibitemShut
  {NoStop}%
\bibitem [{\citenamefont {Karsch}\ \emph {et~al.}(2003)\citenamefont {Karsch},
  \citenamefont {Redlich},\ and\ \citenamefont {Tawfik}}]{Karsch:2003vd}%
  \BibitemOpen
  \bibfield  {author} {\bibinfo {author} {\bibfnamefont {F.}~\bibnamefont
  {Karsch}}, \bibinfo {author} {\bibfnamefont {K.}~\bibnamefont {Redlich}}, \
  and\ \bibinfo {author} {\bibfnamefont {A.}~\bibnamefont {Tawfik}},\ }\href
  {\doibase 10.1140/epjc/s2003-01228-y} {\bibfield  {journal} {\bibinfo
  {journal} {Eur. Phys. J. C}\ }\textbf {\bibinfo {volume} {29}},\ \bibinfo
  {pages} {549} (\bibinfo {year} {2003})},\ \Eprint
  {http://arxiv.org/abs/hep-ph/0303108} {arXiv:hep-ph/0303108} \BibitemShut
  {NoStop}%
\bibitem [{\citenamefont {Huovinen}\ and\ \citenamefont
  {Petreczky}(2010)}]{Huovinen:2009yb}%
  \BibitemOpen
  \bibfield  {author} {\bibinfo {author} {\bibfnamefont {P.}~\bibnamefont
  {Huovinen}}\ and\ \bibinfo {author} {\bibfnamefont {P.}~\bibnamefont
  {Petreczky}},\ }\href {\doibase 10.1016/j.nuclphysa.2010.02.015} {\bibfield
  {journal} {\bibinfo  {journal} {Nucl. Phys. A}\ }\textbf {\bibinfo {volume}
  {837}},\ \bibinfo {pages} {26} (\bibinfo {year} {2010})},\ \Eprint
  {http://arxiv.org/abs/0912.2541} {arXiv:0912.2541 [hep-ph]} \BibitemShut
  {NoStop}%
\bibitem [{\citenamefont {Bazavov}\ \emph {et~al.}(2009)\citenamefont {Bazavov}
  \emph {et~al.}}]{Bazavov:2009zn}%
  \BibitemOpen
  \bibfield  {author} {\bibinfo {author} {\bibfnamefont {A.}~\bibnamefont
  {Bazavov}} \emph {et~al.},\ }\href {\doibase 10.1103/PhysRevD.80.014504}
  {\bibfield  {journal} {\bibinfo  {journal} {Phys. Rev. D}\ }\textbf {\bibinfo
  {volume} {80}},\ \bibinfo {pages} {014504} (\bibinfo {year} {2009})},\
  \Eprint {http://arxiv.org/abs/0903.4379} {arXiv:0903.4379 [hep-lat]}
  \BibitemShut {NoStop}%
\bibitem [{\citenamefont {Borsanyi}\ \emph {et~al.}(2010)\citenamefont
  {Borsanyi}, \citenamefont {Endrodi}, \citenamefont {Fodor}, \citenamefont
  {Jakovac}, \citenamefont {Katz}, \citenamefont {Krieg}, \citenamefont
  {Ratti},\ and\ \citenamefont {Szabo}}]{Borsanyi:2010cj}%
  \BibitemOpen
  \bibfield  {author} {\bibinfo {author} {\bibfnamefont {S.}~\bibnamefont
  {Borsanyi}}, \bibinfo {author} {\bibfnamefont {G.}~\bibnamefont {Endrodi}},
  \bibinfo {author} {\bibfnamefont {Z.}~\bibnamefont {Fodor}}, \bibinfo
  {author} {\bibfnamefont {A.}~\bibnamefont {Jakovac}}, \bibinfo {author}
  {\bibfnamefont {S.~D.}\ \bibnamefont {Katz}}, \bibinfo {author}
  {\bibfnamefont {S.}~\bibnamefont {Krieg}}, \bibinfo {author} {\bibfnamefont
  {C.}~\bibnamefont {Ratti}}, \ and\ \bibinfo {author} {\bibfnamefont {K.~K.}\
  \bibnamefont {Szabo}},\ }\href {\doibase 10.1007/JHEP11(2010)077} {\bibfield
  {journal} {\bibinfo  {journal} {JHEP}\ }\textbf {\bibinfo {volume} {11}},\
  \bibinfo {pages} {077} (\bibinfo {year} {2010})},\ \Eprint
  {http://arxiv.org/abs/1007.2580} {arXiv:1007.2580 [hep-lat]} \BibitemShut
  {NoStop}%
\bibitem [{\citenamefont {Ratti}\ and\ \citenamefont
  {Bellwied}(2021)}]{Ratti:2021ubw}%
  \BibitemOpen
  \bibfield  {author} {\bibinfo {author} {\bibfnamefont {C.}~\bibnamefont
  {Ratti}}\ and\ \bibinfo {author} {\bibfnamefont {R.}~\bibnamefont
  {Bellwied}},\ }\href {\doibase 10.1007/978-3-030-67235-5} {\emph {\bibinfo
  {title} {{The Deconfinement Transition of QCD: Theory Meets Experiment}}}},\
  \bibinfo {series} {Lecture Notes in Physics}, Vol.\ \bibinfo {volume} {981}\
  (\bibinfo {year} {2021})\BibitemShut {NoStop}%
\bibitem [{\citenamefont {Romatschke}(2012)}]{Romatschke:2011qp}%
  \BibitemOpen
  \bibfield  {author} {\bibinfo {author} {\bibfnamefont {P.}~\bibnamefont
  {Romatschke}},\ }\href {\doibase 10.1103/PhysRevD.85.065012} {\bibfield
  {journal} {\bibinfo  {journal} {Phys. Rev. D}\ }\textbf {\bibinfo {volume}
  {85}},\ \bibinfo {pages} {065012} (\bibinfo {year} {2012})},\ \Eprint
  {http://arxiv.org/abs/1108.5561} {arXiv:1108.5561 [gr-qc]} \BibitemShut
  {NoStop}%
\bibitem [{\citenamefont {Albright}\ and\ \citenamefont
  {Kapusta}(2016)}]{Albright:2015fpa}%
  \BibitemOpen
  \bibfield  {author} {\bibinfo {author} {\bibfnamefont {M.}~\bibnamefont
  {Albright}}\ and\ \bibinfo {author} {\bibfnamefont {J.~I.}\ \bibnamefont
  {Kapusta}},\ }\href {\doibase 10.1103/PhysRevC.93.014903} {\bibfield
  {journal} {\bibinfo  {journal} {Phys. Rev. C}\ }\textbf {\bibinfo {volume}
  {93}},\ \bibinfo {pages} {014903} (\bibinfo {year} {2016})},\ \Eprint
  {http://arxiv.org/abs/1508.02696} {arXiv:1508.02696 [nucl-th]} \BibitemShut
  {NoStop}%
\bibitem [{\citenamefont {Alqahtani}\ \emph {et~al.}(2015)\citenamefont
  {Alqahtani}, \citenamefont {Nopoush},\ and\ \citenamefont
  {Strickland}}]{Alqahtani:2015qja}%
  \BibitemOpen
  \bibfield  {author} {\bibinfo {author} {\bibfnamefont {M.}~\bibnamefont
  {Alqahtani}}, \bibinfo {author} {\bibfnamefont {M.}~\bibnamefont {Nopoush}},
  \ and\ \bibinfo {author} {\bibfnamefont {M.}~\bibnamefont {Strickland}},\
  }\href {\doibase 10.1103/PhysRevC.92.054910} {\bibfield  {journal} {\bibinfo
  {journal} {Phys. Rev. C}\ }\textbf {\bibinfo {volume} {92}},\ \bibinfo
  {pages} {054910} (\bibinfo {year} {2015})},\ \Eprint
  {http://arxiv.org/abs/1509.02913} {arXiv:1509.02913 [hep-ph]} \BibitemShut
  {NoStop}%
\bibitem [{\citenamefont {Chakraborty}\ and\ \citenamefont
  {Kapusta}(2011)}]{Chakraborty:2010fr}%
  \BibitemOpen
  \bibfield  {author} {\bibinfo {author} {\bibfnamefont {P.}~\bibnamefont
  {Chakraborty}}\ and\ \bibinfo {author} {\bibfnamefont {J.~I.}\ \bibnamefont
  {Kapusta}},\ }\href {\doibase 10.1103/PhysRevC.83.014906} {\bibfield
  {journal} {\bibinfo  {journal} {Phys. Rev. C}\ }\textbf {\bibinfo {volume}
  {83}},\ \bibinfo {pages} {014906} (\bibinfo {year} {2011})},\ \Eprint
  {http://arxiv.org/abs/1006.0257} {arXiv:1006.0257 [nucl-th]} \BibitemShut
  {NoStop}%
\bibitem [{\citenamefont {Tinti}\ \emph {et~al.}(2017)\citenamefont {Tinti},
  \citenamefont {Jaiswal},\ and\ \citenamefont {Ryblewski}}]{Tinti:2016bav}%
  \BibitemOpen
  \bibfield  {author} {\bibinfo {author} {\bibfnamefont {L.}~\bibnamefont
  {Tinti}}, \bibinfo {author} {\bibfnamefont {A.}~\bibnamefont {Jaiswal}}, \
  and\ \bibinfo {author} {\bibfnamefont {R.}~\bibnamefont {Ryblewski}},\ }\href
  {\doibase 10.1103/PhysRevD.95.054007} {\bibfield  {journal} {\bibinfo
  {journal} {Phys. Rev. D}\ }\textbf {\bibinfo {volume} {95}},\ \bibinfo
  {pages} {054007} (\bibinfo {year} {2017})},\ \Eprint
  {http://arxiv.org/abs/1612.07329} {arXiv:1612.07329 [nucl-th]} \BibitemShut
  {NoStop}%
\bibitem [{\citenamefont {Rocha}\ \emph {et~al.}(2022)\citenamefont {Rocha},
  \citenamefont {Ferreira}, \citenamefont {Denicol},\ and\ \citenamefont
  {Noronha}}]{Rocha:2022fqz}%
  \BibitemOpen
  \bibfield  {author} {\bibinfo {author} {\bibfnamefont {G.~S.}\ \bibnamefont
  {Rocha}}, \bibinfo {author} {\bibfnamefont {M.~N.}\ \bibnamefont {Ferreira}},
  \bibinfo {author} {\bibfnamefont {G.~S.}\ \bibnamefont {Denicol}}, \ and\
  \bibinfo {author} {\bibfnamefont {J.}~\bibnamefont {Noronha}},\ }\href
  {\doibase 10.1103/PhysRevD.106.036022} {\bibfield  {journal} {\bibinfo
  {journal} {Phys. Rev. D}\ }\textbf {\bibinfo {volume} {106}},\ \bibinfo
  {pages} {036022} (\bibinfo {year} {2022})},\ \Eprint
  {http://arxiv.org/abs/2203.15571} {arXiv:2203.15571 [nucl-th]} \BibitemShut
  {NoStop}%
\bibitem [{\citenamefont {Workman}\ \emph {et~al.}(2022)\citenamefont {Workman}
  \emph {et~al.}}]{ParticleDataGroup:2022pth}%
  \BibitemOpen
  \bibfield  {author} {\bibinfo {author} {\bibfnamefont {R.~L.}\ \bibnamefont
  {Workman}} \emph {et~al.} (\bibinfo {collaboration} {Particle Data Group}),\
  }\href {\doibase 10.1093/ptep/ptac097} {\bibfield  {journal} {\bibinfo
  {journal} {PTEP}\ }\textbf {\bibinfo {volume} {2022}},\ \bibinfo {pages}
  {083C01} (\bibinfo {year} {2022})}\BibitemShut {NoStop}%
\bibitem [{\citenamefont {Aoki}\ \emph {et~al.}(2006)\citenamefont {Aoki},
  \citenamefont {Endrodi}, \citenamefont {Fodor}, \citenamefont {Katz},\ and\
  \citenamefont {Szabo}}]{Aoki:2006we}%
  \BibitemOpen
  \bibfield  {author} {\bibinfo {author} {\bibfnamefont {Y.}~\bibnamefont
  {Aoki}}, \bibinfo {author} {\bibfnamefont {G.}~\bibnamefont {Endrodi}},
  \bibinfo {author} {\bibfnamefont {Z.}~\bibnamefont {Fodor}}, \bibinfo
  {author} {\bibfnamefont {S.~D.}\ \bibnamefont {Katz}}, \ and\ \bibinfo
  {author} {\bibfnamefont {K.~K.}\ \bibnamefont {Szabo}},\ }\href {\doibase
  10.1038/nature05120} {\bibfield  {journal} {\bibinfo  {journal} {Nature}\
  }\textbf {\bibinfo {volume} {443}},\ \bibinfo {pages} {675} (\bibinfo {year}
  {2006})},\ \Eprint {http://arxiv.org/abs/hep-lat/0611014}
  {arXiv:hep-lat/0611014} \BibitemShut {NoStop}%
\bibitem [{\citenamefont {Anderson}\ and\ \citenamefont
  {Witting}(1974)}]{andersonRTA:74}%
  \BibitemOpen
  \bibfield  {author} {\bibinfo {author} {\bibfnamefont {J.~L.}\ \bibnamefont
  {Anderson}}\ and\ \bibinfo {author} {\bibfnamefont {H.}~\bibnamefont
  {Witting}},\ }\href@noop {} {\bibfield  {journal} {\bibinfo  {journal}
  {Physica}\ }\textbf {\bibinfo {volume} {74}},\ \bibinfo {pages} {466}
  (\bibinfo {year} {1974})}\BibitemShut {NoStop}%
\bibitem [{\citenamefont {Rocha}\ \emph {et~al.}(2021)\citenamefont {Rocha},
  \citenamefont {Denicol},\ and\ \citenamefont {Noronha}}]{Rocha:2021zcw}%
  \BibitemOpen
  \bibfield  {author} {\bibinfo {author} {\bibfnamefont {G.~S.}\ \bibnamefont
  {Rocha}}, \bibinfo {author} {\bibfnamefont {G.~S.}\ \bibnamefont {Denicol}},
  \ and\ \bibinfo {author} {\bibfnamefont {J.}~\bibnamefont {Noronha}},\ }\href
  {\doibase 10.1103/PhysRevLett.127.042301} {\bibfield  {journal} {\bibinfo
  {journal} {Phys. Rev. Lett.}\ }\textbf {\bibinfo {volume} {127}},\ \bibinfo
  {pages} {042301} (\bibinfo {year} {2021})},\ \Eprint
  {http://arxiv.org/abs/2103.07489} {arXiv:2103.07489 [nucl-th]} \BibitemShut
  {NoStop}%
\bibitem [{\citenamefont {Landau}\ and\ \citenamefont
  {Lifshitz}(1959)}]{landau:59fluid}%
  \BibitemOpen
  \bibfield  {author} {\bibinfo {author} {\bibfnamefont {L.}~\bibnamefont
  {Landau}}\ and\ \bibinfo {author} {\bibfnamefont {E.}~\bibnamefont
  {Lifshitz}},\ }\href@noop {} {\bibfield  {journal} {\bibinfo  {journal}
  {Course of Theoretical Physics, Pergamon Press, London}\ }\textbf {\bibinfo
  {volume} {6}} (\bibinfo {year} {1959})}\BibitemShut {NoStop}%
\bibitem [{\citenamefont {Denicol}\ \emph {et~al.}(2012)\citenamefont
  {Denicol}, \citenamefont {Niemi}, \citenamefont {Molnar},\ and\ \citenamefont
  {Rischke}}]{Denicol:2012cn}%
  \BibitemOpen
  \bibfield  {author} {\bibinfo {author} {\bibfnamefont {G.~S.}\ \bibnamefont
  {Denicol}}, \bibinfo {author} {\bibfnamefont {H.}~\bibnamefont {Niemi}},
  \bibinfo {author} {\bibfnamefont {E.}~\bibnamefont {Molnar}}, \ and\ \bibinfo
  {author} {\bibfnamefont {D.~H.}\ \bibnamefont {Rischke}},\ }\href {\doibase
  10.1103/PhysRevD.85.114047} {\bibfield  {journal} {\bibinfo  {journal} {Phys.
  Rev. D}\ }\textbf {\bibinfo {volume} {85}},\ \bibinfo {pages} {114047}
  (\bibinfo {year} {2012})},\ \bibinfo {note} {[Erratum: Phys.Rev.D 91, 039902
  (2015)]},\ \Eprint {http://arxiv.org/abs/1202.4551} {arXiv:1202.4551
  [nucl-th]} \BibitemShut {NoStop}%
\bibitem [{\citenamefont {de~Brito}\ and\ \citenamefont
  {Denicol}(2024)}]{deBrito:2024vhm}%
  \BibitemOpen
  \bibfield  {author} {\bibinfo {author} {\bibfnamefont {C.~V.~P.}\
  \bibnamefont {de~Brito}}\ and\ \bibinfo {author} {\bibfnamefont {G.~S.}\
  \bibnamefont {Denicol}},\ }\href@noop {} {\  (\bibinfo {year} {2024})},\
  \Eprint {http://arxiv.org/abs/2401.10098} {arXiv:2401.10098 [nucl-th]}
  \BibitemShut {NoStop}%
\bibitem [{\citenamefont {Kamata}\ \emph {et~al.}(2020)\citenamefont {Kamata},
  \citenamefont {Martinez}, \citenamefont {Plaschke}, \citenamefont
  {Ochsenfeld},\ and\ \citenamefont {Schlichting}}]{Kamata:2020mka}%
  \BibitemOpen
  \bibfield  {author} {\bibinfo {author} {\bibfnamefont {S.}~\bibnamefont
  {Kamata}}, \bibinfo {author} {\bibfnamefont {M.}~\bibnamefont {Martinez}},
  \bibinfo {author} {\bibfnamefont {P.}~\bibnamefont {Plaschke}}, \bibinfo
  {author} {\bibfnamefont {S.}~\bibnamefont {Ochsenfeld}}, \ and\ \bibinfo
  {author} {\bibfnamefont {S.}~\bibnamefont {Schlichting}},\ }\href {\doibase
  10.1103/PhysRevD.102.056003} {\bibfield  {journal} {\bibinfo  {journal}
  {Phys. Rev. D}\ }\textbf {\bibinfo {volume} {102}},\ \bibinfo {pages}
  {056003} (\bibinfo {year} {2020})},\ \Eprint
  {http://arxiv.org/abs/2004.06751} {arXiv:2004.06751 [hep-ph]} \BibitemShut
  {NoStop}%
\bibitem [{\citenamefont {Noronha}\ and\ \citenamefont
  {Denicol}(2015)}]{Noronha:2015jia}%
  \BibitemOpen
  \bibfield  {author} {\bibinfo {author} {\bibfnamefont {J.}~\bibnamefont
  {Noronha}}\ and\ \bibinfo {author} {\bibfnamefont {G.~S.}\ \bibnamefont
  {Denicol}},\ }\href {\doibase 10.1103/PhysRevD.92.114032} {\bibfield
  {journal} {\bibinfo  {journal} {Phys. Rev. D}\ }\textbf {\bibinfo {volume}
  {92}},\ \bibinfo {pages} {114032} (\bibinfo {year} {2015})},\ \Eprint
  {http://arxiv.org/abs/1502.05892} {arXiv:1502.05892 [hep-ph]} \BibitemShut
  {NoStop}%
\bibitem [{\citenamefont {Denicol}\ \emph
  {et~al.}(2014{\natexlab{b}})\citenamefont {Denicol}, \citenamefont {Heinz},
  \citenamefont {Martinez}, \citenamefont {Noronha},\ and\ \citenamefont
  {Strickland}}]{Denicol:2014xca}%
  \BibitemOpen
  \bibfield  {author} {\bibinfo {author} {\bibfnamefont {G.~S.}\ \bibnamefont
  {Denicol}}, \bibinfo {author} {\bibfnamefont {U.~W.}\ \bibnamefont {Heinz}},
  \bibinfo {author} {\bibfnamefont {M.}~\bibnamefont {Martinez}}, \bibinfo
  {author} {\bibfnamefont {J.}~\bibnamefont {Noronha}}, \ and\ \bibinfo
  {author} {\bibfnamefont {M.}~\bibnamefont {Strickland}},\ }\href {\doibase
  10.1103/PhysRevLett.113.202301} {\bibfield  {journal} {\bibinfo  {journal}
  {Phys. Rev. Lett.}\ }\textbf {\bibinfo {volume} {113}},\ \bibinfo {pages}
  {202301} (\bibinfo {year} {2014}{\natexlab{b}})},\ \Eprint
  {http://arxiv.org/abs/1408.5646} {arXiv:1408.5646 [hep-ph]} \BibitemShut
  {NoStop}%
\bibitem [{\citenamefont {Ochsenfeld}\ and\ \citenamefont
  {Schlichting}(2023)}]{Ochsenfeld:2023wxz}%
  \BibitemOpen
  \bibfield  {author} {\bibinfo {author} {\bibfnamefont {S.}~\bibnamefont
  {Ochsenfeld}}\ and\ \bibinfo {author} {\bibfnamefont {S.}~\bibnamefont
  {Schlichting}},\ }\href {\doibase 10.1007/JHEP09(2023)186} {\bibfield
  {journal} {\bibinfo  {journal} {JHEP}\ }\textbf {\bibinfo {volume} {09}},\
  \bibinfo {pages} {186} (\bibinfo {year} {2023})},\ \Eprint
  {http://arxiv.org/abs/2308.04491} {arXiv:2308.04491 [hep-th]} \BibitemShut
  {NoStop}%
\bibitem [{\citenamefont {Fotakis}\ \emph {et~al.}(2022)\citenamefont
  {Fotakis}, \citenamefont {Moln\'ar}, \citenamefont {Niemi}, \citenamefont
  {Greiner},\ and\ \citenamefont {Rischke}}]{Fotakis:2022usk}%
  \BibitemOpen
  \bibfield  {author} {\bibinfo {author} {\bibfnamefont {J.~A.}\ \bibnamefont
  {Fotakis}}, \bibinfo {author} {\bibfnamefont {E.}~\bibnamefont {Moln\'ar}},
  \bibinfo {author} {\bibfnamefont {H.}~\bibnamefont {Niemi}}, \bibinfo
  {author} {\bibfnamefont {C.}~\bibnamefont {Greiner}}, \ and\ \bibinfo
  {author} {\bibfnamefont {D.~H.}\ \bibnamefont {Rischke}},\ }\href {\doibase
  10.1103/PhysRevD.106.036009} {\bibfield  {journal} {\bibinfo  {journal}
  {Phys. Rev. D}\ }\textbf {\bibinfo {volume} {106}},\ \bibinfo {pages}
  {036009} (\bibinfo {year} {2022})},\ \Eprint
  {http://arxiv.org/abs/2203.11549} {arXiv:2203.11549 [nucl-th]} \BibitemShut
  {NoStop}%
\bibitem [{\citenamefont {Bass}\ \emph {et~al.}(1998)\citenamefont {Bass} \emph
  {et~al.}}]{Bass:1998ca}%
  \BibitemOpen
  \bibfield  {author} {\bibinfo {author} {\bibfnamefont {S.~A.}\ \bibnamefont
  {Bass}} \emph {et~al.},\ }\href {\doibase 10.1016/S0146-6410(98)00058-1}
  {\bibfield  {journal} {\bibinfo  {journal} {Prog. Part. Nucl. Phys.}\
  }\textbf {\bibinfo {volume} {41}},\ \bibinfo {pages} {255} (\bibinfo {year}
  {1998})},\ \Eprint {http://arxiv.org/abs/nucl-th/9803035}
  {arXiv:nucl-th/9803035} \BibitemShut {NoStop}%
\bibitem [{\citenamefont {Bleicher}\ \emph {et~al.}(1999)\citenamefont
  {Bleicher} \emph {et~al.}}]{Bleicher:1999xi}%
  \BibitemOpen
  \bibfield  {author} {\bibinfo {author} {\bibfnamefont {M.}~\bibnamefont
  {Bleicher}} \emph {et~al.},\ }\href {\doibase 10.1088/0954-3899/25/9/308}
  {\bibfield  {journal} {\bibinfo  {journal} {J. Phys. G}\ }\textbf {\bibinfo
  {volume} {25}},\ \bibinfo {pages} {1859} (\bibinfo {year} {1999})},\ \Eprint
  {http://arxiv.org/abs/hep-ph/9909407} {arXiv:hep-ph/9909407} \BibitemShut
  {NoStop}%
\bibitem [{\citenamefont {Weil}\ \emph {et~al.}(2016)\citenamefont {Weil} \emph
  {et~al.}}]{SMASH:2016zqf}%
  \BibitemOpen
  \bibfield  {author} {\bibinfo {author} {\bibfnamefont {J.}~\bibnamefont
  {Weil}} \emph {et~al.} (\bibinfo {collaboration} {SMASH}),\ }\href {\doibase
  10.1103/PhysRevC.94.054905} {\bibfield  {journal} {\bibinfo  {journal} {Phys.
  Rev. C}\ }\textbf {\bibinfo {volume} {94}},\ \bibinfo {pages} {054905}
  (\bibinfo {year} {2016})},\ \Eprint {http://arxiv.org/abs/1606.06642}
  {arXiv:1606.06642 [nucl-th]} \BibitemShut {NoStop}%
\bibitem [{\citenamefont {Rodrigues}\ \emph {et~al.}(2020)\citenamefont
  {Rodrigues} \emph {et~al.}}]{Rodrigues:2020syo}%
  \BibitemOpen
  \bibfield  {author} {\bibinfo {author} {\bibfnamefont {E.}~\bibnamefont
  {Rodrigues}} \emph {et~al.},\ }\href {\doibase 10.1051/epjconf/202024506028}
  {\bibfield  {journal} {\bibinfo  {journal} {EPJ Web Conf.}\ }\textbf
  {\bibinfo {volume} {245}},\ \bibinfo {pages} {06028} (\bibinfo {year}
  {2020})},\ \bibinfo {note} {\url{https://github.com/scikit-hep/particle}},\
  \Eprint {http://arxiv.org/abs/2007.03577} {arXiv:2007.03577
  [physics.comp-ph]} \BibitemShut {NoStop}%
\bibitem [{\citenamefont {Schenke}\ \emph {et~al.}(2011)\citenamefont
  {Schenke}, \citenamefont {Jeon},\ and\ \citenamefont
  {Gale}}]{Schenke:2010rr}%
  \BibitemOpen
  \bibfield  {author} {\bibinfo {author} {\bibfnamefont {B.}~\bibnamefont
  {Schenke}}, \bibinfo {author} {\bibfnamefont {S.}~\bibnamefont {Jeon}}, \
  and\ \bibinfo {author} {\bibfnamefont {C.}~\bibnamefont {Gale}},\ }\href
  {\doibase 10.1103/PhysRevLett.106.042301} {\bibfield  {journal} {\bibinfo
  {journal} {Phys. Rev. Lett.}\ }\textbf {\bibinfo {volume} {106}},\ \bibinfo
  {pages} {042301} (\bibinfo {year} {2011})},\ \Eprint
  {http://arxiv.org/abs/1009.3244} {arXiv:1009.3244 [hep-ph]} \BibitemShut
  {NoStop}%
\bibitem [{\citenamefont {Calzetta}\ and\ \citenamefont
  {Hu}(1988)}]{Calzetta:1986cq}%
  \BibitemOpen
  \bibfield  {author} {\bibinfo {author} {\bibfnamefont {E.}~\bibnamefont
  {Calzetta}}\ and\ \bibinfo {author} {\bibfnamefont {B.~L.}\ \bibnamefont
  {Hu}},\ }\href {\doibase 10.1103/PhysRevD.37.2878} {\bibfield  {journal}
  {\bibinfo  {journal} {Phys. Rev. D}\ }\textbf {\bibinfo {volume} {37}},\
  \bibinfo {pages} {2878} (\bibinfo {year} {1988})}\BibitemShut {NoStop}%
\bibitem [{\citenamefont {Jeon}\ and\ \citenamefont
  {Yaffe}(1996)}]{Jeon:1995zm}%
  \BibitemOpen
  \bibfield  {author} {\bibinfo {author} {\bibfnamefont {S.}~\bibnamefont
  {Jeon}}\ and\ \bibinfo {author} {\bibfnamefont {L.~G.}\ \bibnamefont
  {Yaffe}},\ }\href {\doibase 10.1103/PhysRevD.53.5799} {\bibfield  {journal}
  {\bibinfo  {journal} {Phys. Rev. D}\ }\textbf {\bibinfo {volume} {53}},\
  \bibinfo {pages} {5799} (\bibinfo {year} {1996})},\ \Eprint
  {http://arxiv.org/abs/hep-ph/9512263} {arXiv:hep-ph/9512263} \BibitemShut
  {NoStop}%
\bibitem [{\citenamefont {Arnold}\ \emph
  {et~al.}(2003{\natexlab{a}})\citenamefont {Arnold}, \citenamefont {Moore},\
  and\ \citenamefont {Yaffe}}]{Arnold:2002zm}%
  \BibitemOpen
  \bibfield  {author} {\bibinfo {author} {\bibfnamefont {P.~B.}\ \bibnamefont
  {Arnold}}, \bibinfo {author} {\bibfnamefont {G.~D.}\ \bibnamefont {Moore}}, \
  and\ \bibinfo {author} {\bibfnamefont {L.~G.}\ \bibnamefont {Yaffe}},\ }\href
  {\doibase 10.1088/1126-6708/2003/01/030} {\bibfield  {journal} {\bibinfo
  {journal} {JHEP}\ }\textbf {\bibinfo {volume} {01}},\ \bibinfo {pages} {030}
  (\bibinfo {year} {2003}{\natexlab{a}})},\ \Eprint
  {http://arxiv.org/abs/hep-ph/0209353} {arXiv:hep-ph/0209353} \BibitemShut
  {NoStop}%
\bibitem [{\citenamefont {Berges}\ and\ \citenamefont
  {Borsanyi}(2006)}]{Berges:2005md}%
  \BibitemOpen
  \bibfield  {author} {\bibinfo {author} {\bibfnamefont {J.}~\bibnamefont
  {Berges}}\ and\ \bibinfo {author} {\bibfnamefont {S.}~\bibnamefont
  {Borsanyi}},\ }\href {\doibase 10.1103/PhysRevD.74.045022} {\bibfield
  {journal} {\bibinfo  {journal} {Phys. Rev. D}\ }\textbf {\bibinfo {volume}
  {74}},\ \bibinfo {pages} {045022} (\bibinfo {year} {2006})},\ \Eprint
  {http://arxiv.org/abs/hep-ph/0512155} {arXiv:hep-ph/0512155} \BibitemShut
  {NoStop}%
\bibitem [{\citenamefont {Gagnon}\ and\ \citenamefont
  {Jeon}(2007)}]{Gagnon:2007qt}%
  \BibitemOpen
  \bibfield  {author} {\bibinfo {author} {\bibfnamefont {J.-S.}\ \bibnamefont
  {Gagnon}}\ and\ \bibinfo {author} {\bibfnamefont {S.}~\bibnamefont {Jeon}},\
  }\href {\doibase 10.1103/PhysRevD.76.105019} {\bibfield  {journal} {\bibinfo
  {journal} {Phys. Rev. D}\ }\textbf {\bibinfo {volume} {76}},\ \bibinfo
  {pages} {105019} (\bibinfo {year} {2007})},\ \Eprint
  {http://arxiv.org/abs/0708.1631} {arXiv:0708.1631 [hep-ph]} \BibitemShut
  {NoStop}%
\bibitem [{\citenamefont {Arnold}\ \emph
  {et~al.}(2003{\natexlab{b}})\citenamefont {Arnold}, \citenamefont {Moore},\
  and\ \citenamefont {Yaffe}}]{Arnold:2003zc}%
  \BibitemOpen
  \bibfield  {author} {\bibinfo {author} {\bibfnamefont {P.~B.}\ \bibnamefont
  {Arnold}}, \bibinfo {author} {\bibfnamefont {G.~D.}\ \bibnamefont {Moore}}, \
  and\ \bibinfo {author} {\bibfnamefont {L.~G.}\ \bibnamefont {Yaffe}},\ }\href
  {\doibase 10.1088/1126-6708/2003/05/051} {\bibfield  {journal} {\bibinfo
  {journal} {JHEP}\ }\textbf {\bibinfo {volume} {05}},\ \bibinfo {pages} {051}
  (\bibinfo {year} {2003}{\natexlab{b}})},\ \Eprint
  {http://arxiv.org/abs/hep-ph/0302165} {arXiv:hep-ph/0302165} \BibitemShut
  {NoStop}%
\bibitem [{\citenamefont {Baym}\ \emph {et~al.}(1990)\citenamefont {Baym},
  \citenamefont {Monien}, \citenamefont {Pethick},\ and\ \citenamefont
  {Ravenhall}}]{Baym:1990uj}%
  \BibitemOpen
  \bibfield  {author} {\bibinfo {author} {\bibfnamefont {G.}~\bibnamefont
  {Baym}}, \bibinfo {author} {\bibfnamefont {H.}~\bibnamefont {Monien}},
  \bibinfo {author} {\bibfnamefont {C.~J.}\ \bibnamefont {Pethick}}, \ and\
  \bibinfo {author} {\bibfnamefont {D.~G.}\ \bibnamefont {Ravenhall}},\ }\href
  {\doibase 10.1103/PhysRevLett.64.1867} {\bibfield  {journal} {\bibinfo
  {journal} {Phys. Rev. Lett.}\ }\textbf {\bibinfo {volume} {64}},\ \bibinfo
  {pages} {1867} (\bibinfo {year} {1990})}\BibitemShut {NoStop}%
\bibitem [{\citenamefont {Gradshteyn}\ and\ \citenamefont
  {Ryzhik}(2014)}]{gradshteyn2014table}%
  \BibitemOpen
  \bibfield  {author} {\bibinfo {author} {\bibfnamefont {I.~S.}\ \bibnamefont
  {Gradshteyn}}\ and\ \bibinfo {author} {\bibfnamefont {I.~M.}\ \bibnamefont
  {Ryzhik}},\ }\href@noop {} {\emph {\bibinfo {title} {Table of integrals,
  series, and products}}}\ (\bibinfo  {publisher} {Academic press},\ \bibinfo
  {year} {2014})\BibitemShut {NoStop}%
\bibitem [{\citenamefont {Struchtrup}(2004)}]{struchtrup2004stable}%
  \BibitemOpen
  \bibfield  {author} {\bibinfo {author} {\bibfnamefont {H.}~\bibnamefont
  {Struchtrup}},\ }\href@noop {} {\bibfield  {journal} {\bibinfo  {journal}
  {Physics of Fluids}\ }\textbf {\bibinfo {volume} {16}},\ \bibinfo {pages}
  {3921} (\bibinfo {year} {2004})}\BibitemShut {NoStop}%
\bibitem [{\citenamefont {Finazzo}\ \emph {et~al.}(2015)\citenamefont
  {Finazzo}, \citenamefont {Rougemont}, \citenamefont {Marrochio},\ and\
  \citenamefont {Noronha}}]{Finazzo:2014cna}%
  \BibitemOpen
  \bibfield  {author} {\bibinfo {author} {\bibfnamefont {S.~I.}\ \bibnamefont
  {Finazzo}}, \bibinfo {author} {\bibfnamefont {R.}~\bibnamefont {Rougemont}},
  \bibinfo {author} {\bibfnamefont {H.}~\bibnamefont {Marrochio}}, \ and\
  \bibinfo {author} {\bibfnamefont {J.}~\bibnamefont {Noronha}},\ }\href
  {\doibase 10.1007/JHEP02(2015)051} {\bibfield  {journal} {\bibinfo  {journal}
  {JHEP}\ }\textbf {\bibinfo {volume} {02}},\ \bibinfo {pages} {051} (\bibinfo
  {year} {2015})},\ \Eprint {http://arxiv.org/abs/1412.2968} {arXiv:1412.2968
  [hep-ph]} \BibitemShut {NoStop}%
\bibitem [{\citenamefont {Kanitscheider}\ and\ \citenamefont
  {Skenderis}(2009)}]{Kanitscheider:2009as}%
  \BibitemOpen
  \bibfield  {author} {\bibinfo {author} {\bibfnamefont {I.}~\bibnamefont
  {Kanitscheider}}\ and\ \bibinfo {author} {\bibfnamefont {K.}~\bibnamefont
  {Skenderis}},\ }\href {\doibase 10.1088/1126-6708/2009/04/062} {\bibfield
  {journal} {\bibinfo  {journal} {JHEP}\ }\textbf {\bibinfo {volume} {04}},\
  \bibinfo {pages} {062} (\bibinfo {year} {2009})},\ \Eprint
  {http://arxiv.org/abs/0901.1487} {arXiv:0901.1487 [hep-th]} \BibitemShut
  {NoStop}%
\bibitem [{\citenamefont {Olson}(1990)}]{Olson:1990rzl}%
  \BibitemOpen
  \bibfield  {author} {\bibinfo {author} {\bibfnamefont {T.~S.}\ \bibnamefont
  {Olson}},\ }\href {\doibase 10.1016/0003-4916(90)90366-V} {\bibfield
  {journal} {\bibinfo  {journal} {Annals Phys.}\ }\textbf {\bibinfo {volume}
  {199}},\ \bibinfo {pages} {18} (\bibinfo {year} {1990})}\BibitemShut
  {NoStop}%
\bibitem [{\citenamefont {Huang}\ \emph {et~al.}(2011)\citenamefont {Huang},
  \citenamefont {Kodama}, \citenamefont {Koide},\ and\ \citenamefont
  {Rischke}}]{Huang:2010sa}%
  \BibitemOpen
  \bibfield  {author} {\bibinfo {author} {\bibfnamefont {X.-G.}\ \bibnamefont
  {Huang}}, \bibinfo {author} {\bibfnamefont {T.}~\bibnamefont {Kodama}},
  \bibinfo {author} {\bibfnamefont {T.}~\bibnamefont {Koide}}, \ and\ \bibinfo
  {author} {\bibfnamefont {D.~H.}\ \bibnamefont {Rischke}},\ }\href {\doibase
  10.1103/PhysRevC.83.024906} {\bibfield  {journal} {\bibinfo  {journal} {Phys.
  Rev. C}\ }\textbf {\bibinfo {volume} {83}},\ \bibinfo {pages} {024906}
  (\bibinfo {year} {2011})},\ \Eprint {http://arxiv.org/abs/1010.4359}
  {arXiv:1010.4359 [nucl-th]} \BibitemShut {NoStop}%
\end{thebibliography}%

\end{document}